\documentclass[superscriptaddress,onecolumn,pre,nofootinbib]{revtex4}

\usepackage{amsmath}
\usepackage{graphicx,color}

\newcommand{\be}{\begin{equation}}
\newcommand{\ee}{\end{equation}}
\newcommand{\bea}{\begin{eqnarray}}
\newcommand{\eea}{\end{eqnarray}}

\begin{document}
\sloppy

%-title page-%

\title{Kinetic theory of collisionless relaxation for systems with long-range
interactions}

\author{Pierre-Henri Chavanis}
%\email{chavanis@irsamc.ups-tlse.fr}
\affiliation{Laboratoire de Physique Th\'eorique,  Universit\'e de Toulouse,
CNRS, UPS, Toulouse, France}

\begin{abstract}

We develop the kinetic theory of collisionless relaxation  for systems with
long-range interactions  in relation to the statistical
theory of Lynden-Bell. We
treat the multi-level case. We make the connection between the kinetic equation
obtained from the quasilinear theory of the Vlasov equation and the relaxation
equation obtained from  a maximum entropy production principle. We propose a
method to close the infinite hierarchy of kinetic equations for the 
phase level moments  and obtain a 
kinetic equation for the coarse-grained distribution function  in the form of a
generalized  Landau, Lenard-Balescu  or Kramers equation associated with a
generalized
form of entropy [P.H. Chavanis, Physica A {\bf 332}, 89 (2004)]. This allows us
to go beyond the two-level case associated with a Fermi-Dirac-type entropy. We
discuss the numerous analogies with
two-dimensional
turbulence. We also mention possible
applications of the present formalism to fermionic and
bosonic dark matter halos.
\end{abstract}

\maketitle

\section{Introduction}
\label{sec_introduction}

Recently, the dynamics and thermodynamics of systems with long-range
interactions has been a subject 
of considerable interest in statistical mechanics
\cite{houches,assise,oxford,cdr,campabook}. For systems with long-range
interactions, the
relaxation towards statistical equilibrium (Boltzmann
distribution) is governed by the
homogeneous \cite{lenard,balescu} or inhomogeneous \cite{heyvaerts,angleaction2}
Lenard-Balescu equation, which
is a generalization of the homogeneous \cite{landau} or inhomogeneous
\cite{aakin}
Landau equation taking into account collective effects. The collisional
relaxation time generically
scales with the number of particles $N$ as $t_{\rm relax}\sim N\, t_D$, where
$t_D$ is
the dynamical time.\footnote{The ``collisional'' evolution of systems with
long-range interactions described by the Lenard-Balescu equation is induced by
two-body correlations among the particles. The  Lenard-Balescu
equation is valid at the order
$1/N$ in a proper thermodynamic limit where $N\rightarrow +\infty$ with $m\sim
1/N$. Accordingly, the collisional
relaxation time generically scales as $t_{\rm relax}\sim N\, t_D$ \cite{epjp}.
For
self-gravitating systems, the Chandrasekhar relaxation time scales as $t_{\rm
relax}\sim (N/\ln N)\, t_D$ because of logarithmic corrections \cite{chandra}. A
similar scaling is
obtained in plasma physics where the particle number $N$ is replaced by
the plasma parameter $\Lambda$ giving the
number of electrons in the Debye sphere \cite{landau}. For spatially homogeneous
1D systems
with long-range interactions, the Lenard-Balescu collision term vanishes
\cite{epjp} and one has to account for higher order correlations among the
particles. A kinetic
equation, taking into account three-body correlations, has recently been derived
in Refs. \cite{n2a,n2b}. This equation is valid at the order
$1/N^2$, leading
to a collisional relaxation time scaling as $t_{\rm relax}\sim N^2\, t_D$.}
When
$N\rightarrow +\infty$, the relaxation time
diverges and the Boltzmann statistical equilibrium state is never reached. The
system is then governed by the Vlasov equation  \cite{jeans,vlasov} which
describes a collisionless evolution driven only by mean field effects. The
Vlasov
equation is reversible and conserves the Boltzmann entropy among an infinite
number of Casimir invariants. This seems to
preclude the relaxation towards an equilibrium state. However, systems governed
by the Vlasov
equation can experience a process of violent relaxation on the
coarse-grained
scale towards a metaequilibrium state. This process of collisionless relaxation
was first evidenced by King \cite{kingvrnewref} and H\'enon \cite{henonvr} 
in the case of stellar
systems
described by the Vlasov-Poisson equations. It can account for the structure and
regularity of galaxies whose collisional relaxation time exceeds the age of the
Universe by many orders of magnitude \cite{chandra}. Similarly, incompressible
and inviscid
flows in 2D hydrodynamics are described by the Euler-Poisson equations.
Systems governed by these equations also experience a process of violent
relaxation leading to the
formation of large-scale vortices like Jupiter's Great Red spot
\cite{houchesPH,bv}.

A statistical theory of violent relaxation has been developed by Lynden-Bell
\cite{lb} for stellar systems. The metaequilibrium state is obtained
by maximizing a mixing
entropy while accounting for all the constraints of the dynamics. This leads to
the Lynden-Bell distribution function (DF) which can be viewed as a
Fermi-Dirac-type DF in the two-level case, or a
superposition of Fermi-Dirac-type  DFs in the multi-level case. A similar
statistical theory has been developed by Miller \cite{miller} and Robert and
Sommeria \cite{rs} for 2D
incompressible and inviscid flows. The numerous analogies between
stellar systems and 2D vortices are described by Chavanis
\cite{phd,csr,houchesPH,ssvhmf,kindetail}. The
Lynden-Bell statistical theory has also been applied to other systems with
long-range interactions such as the HMF model
\cite{lbt,antoniazzi,antoniazzi2,antoniazzi3,cnfr,stan1,stan2,ccpoly}.
However, the power of
prediction of the Lynden-Bell statistical theory is limited by the problem of
incomplete relaxation \cite{lb,grand,incomplete,assisePH}. There are cases where
the
statistical prediction works well and cases where it fails. The
Lynden-Bell statistical theory relies on an assumption of ergodicity which is
not always fulfilled in practice. It is difficult to know {\it a priori} if the
system will mix efficiently, as required by the ergodicity assumption.

The Lynden-Bell theory is an equilibrium theory. It is then important to
develop a kinetic theory of (violent or quiescent) collisionless relaxation and
obtain an
evolution equation for the coarse-grained DF.\footnote{Lynden-Bell
\cite{lb} heuristically described the evolution of the coarse-grained DF by a
simple Fokker-Planck equation.} This project was first
considered 
by Kadomtsev and Pogutse (KP) \cite{kp} for the Vlasov equation of plasma
physics. They developed a quasilinear theory of collisionless relaxation and, in
the two-level case, obtained a fermionic-like Landau or Lenard-Balescu equation
which
relaxes towards the Lynden-Bell DF. Because of their assumptions, their kinetic
theory can only describe the late quiescent stages of the relaxation process
(gentle relaxation). Their approach was extended by Severne and Luwel
(SL) \cite{sl} in the multi-level case in an astrophysical context.
The quasilinear theory of  collisionless relaxation was further
discussed by Chavanis \cite{chavmnras,dubrovnik,kingen}. He
used it to derive a truncated model (a sort of fermionic King model) taking
into account the evaporation of high energy stars. This DF has a
finite mass in
contrast to the original Lynden-Bell DF which is not normalizable when coupled
to gravity. The
quasilinear theory of the Vlasov-Poisson equations has been exported to the case
of 2D incompressible and inviscid flows described by the Euler-Poisson equations
\cite{quasi,prep}.

A completely different approach has been developed by Chavanis, Sommeria and
Robert (CSR) \cite{csr} by using a maximum entropy production principle (MEPP)
previously introduced in 2D hydrodynamics \cite{rsmepp,rr,csmepp}. A relaxation
equation is
constructed
heuristically by maximizing the rate of production of Lynden-Bell's entropy
while accounting for all the constraints of the dynamics. This leads to a
generalized Fokker-Planck equation with a time-dependent temperature that
evolves so as to conserve the energy. In the two-level case, this relaxation
equation reduces to a fermionic-like Kramers equation which generalizes the
Fokker-Planck  equation for the coarse-grained DF introduced heuristically by
Lynden-Bell \cite{lb}. 

It is possible to obtain the CSR equation from the KP and SL equations by
making a sort of thermal bath approximation. This procedure provides
the explicit expression of the diffusion coefficient in the CSR equation which
is not given by the
MEPP \cite{csr}. The connection between the KP and the CSR equations was made in
Refs. \cite{chavmnras,dubrovnik,kingen} in the two-level case and is generalized
in the
present paper to the multi-level case.  Then, there
remains a complicated closure
problem: The SL and CSR equations yield an infinite hierarchy of equations for
the moments of the distribution. Following the  suggestion of Ref.
\cite{kingen},
we introduce a simple method to close this hierarchy of equations. We first
remark that
the equilibrium coarse-grained DF extremizes a ``generalized entropy'' at
fixed mass and energy. We also show that the variance of the
distribution
at equilibrium is related to the coarse-grained DF through this generalized
entropy. We then propose to close the hierarchy of kinetic equations at the
level of the coarse-grained DF by
extending this
relation out-of-equilibrium (this can be justified by a local
thermodynamical equilibrium assumption). This leads to a form of generalized
Landau or Lenard-Balescu equation associated with a generalized entropy. This
equation has been
studied in detail in \cite{kingen}. It conserves mass and  energy
and
satisfies an $H$-theorem for the generalized entropy. In the thermal bath
approximation, it reduces to a generalized Kramers equation and the diffusion
coefficient can be calculated explicitly.

We show how the kinetic theory is able
to account for the problem of incomplete relaxation through a space and
time-dependent diffusion coefficient.

We also discuss the nonlinear dynamical stability of stationary solutions of the
Vlasov
equation. We introduce an energy principle which provides the most refined
condition of dynamical stability. A stationary solution of the Vlasov equation
is dynamically stable if and only if it is a minimum of energy with respect to
symplectic perturbations (i.e. perturbations that conserve all the Casimirs). We
then propose a relaxation equation that
minimizes the energy while conserving all the Casimirs. This relaxation
equation can serve as a numerical algorithm to construct stable steady states
of the Vlasov equation. We also show that the maximization of a pseudo-entropy
at fixed mass and energy (thermodynamic-looking microcanonical principle)
provides a sufficient condition of dynamical
stability. As a result, the generalized Landau, Lenard-Balescu
and
Kramers equations of Refs. \cite{gen,kingen,nfp}
can be used as numerical algorithms to construct  stable steady states of
the Vlasov equation.

Finally, we discuss the notion of $H$-functions and ``selective decay''
following the work of Tremaine {\it et al.} \cite{thlb}. We
propose an evolution equation that monotonically increases all the $H$-functions
while
conserving the mass and the energy. It differs from the 
generalized Landau, Lenard-Balescu and Kramers equations of Refs.
\cite{gen,kingen,nfp}
which  monotonically increase only one particular $H$-function (the generalized
entropy) at
fixed mass and energy.

\section{Lynden-Bell's statistical theory}
\label{sec_stlb}

\subsection{Vlasov equation}
\label{sec_ve}

We consider a Hamiltonian system of $N$ particles  interacting
via a 
long-range binary potential $u(|{\bf r}-{\bf r}'|)$ which decays at large
distances as $r^{-\gamma}$ with $\gamma\le d$, where $d$ is the dimension of
space. Let $f({\bf r},{\bf v},t)$
denotes the DF defined such that $f \, d{\bf r}d{\bf v}$
gives the total
mass of particles with position ${\bf r}$ and velocity ${\bf v}$ at
time $t$. In the collisionless (or mean field) regime, the evolution of the
DF is governed by the Vlasov equation
\begin{equation}
\label{ve1}
{\partial f\over\partial t}+{\bf v}\cdot {\partial f\over\partial {\bf
r}}-\nabla\Phi\cdot {\partial f\over\partial {\bf v}}=0,
\end{equation}
where
\begin{eqnarray}
\Phi({\bf r},t)=\int u(|{\bf r}-{\bf r}'|)f({\bf r}',{\bf v}',t)\, d{\bf
r}'d{\bf v}'
\label{ve2}
\end{eqnarray}
is the  mean potential produced self-consistently by the particles.
The Vlasov equation can be obtained from the $N$-body Liouville equation
by making a mean-field approximation, i.e., by writing the $N$-body
DF as a product of $N$ one-body
DFs. The Vlasov equation is
valid when $t\ll t_{\rm relax}$, where $ t_{\rm relax}$ is the
collisional relaxation time.
Since $t_{\rm relax}$ grows algebraically with $N$,
the Vlasov equation becomes exact when $N\rightarrow +\infty$ in a proper
thermodynamic limit where the mass of the particles scales
as $m\sim 1/N$ \cite{campabook}. We note
that the individual mass of the
particles does not appear in the Vlasov equation. This implies that the
collisionless dynamics does not lead to a segregation by mass.

The Vlasov (or collisionless Boltzmann) equation simply
states that, in the absence of encounters, the DF $f$ is conserved by
the flow in phase space. This can be
written as $Df/Dt=0$ where $D/Dt=\partial/\partial t+{\bf v}\cdot\nabla_{\bf
r}-\nabla\Phi\cdot\nabla_{\bf v}$ is the material derivative (Stokes operator).
The conservation of the DF together with the incompressibility of the flow in
phase space imply that the total mass (or hypervolume)
of all phase elements with phase
density between $f$ and $f+\delta f$ is conserved. This is
equivalent to the conservation of an infinite number of invariants called
the Casimir integrals $I_{h}=\int h(f)\, d{\bf  r} d{\bf  v}$ for any
continuous
function $h(f)$. The conservation of the Casimirs
is equivalent to the conservation of all the moments of the
DF, denoted $M_n=\int f^{n} \, d{\bf r}d{\bf v}$,
which include the total mass $M=\int f \, d{\bf r}
d{\bf v}$ as a special case. The
Vlasov equation also conserves the total energy $E=\frac{1}{2}\int f v^2\,
d{\bf r}d{\bf v}+\frac{1}{2}\int f\Phi\, d{\bf r}d{\bf v}$ (kinetic +
potential), the
total impulse ${\bf P}=\int f {\bf v}\, d{\bf r}d{\bf v}$, and the total angular
momentum ${\bf L}=\int f {\bf r}\times {\bf v}\, d{\bf r}d{\bf v}$ (see
Appendices \ref{sec_b} and \ref{sec_propv}).  The
Vlasov
equation admits an infinite number of stationary solutions whose
general form is given by the Jeans theorem \cite{jeansth}. For example, any
DF of the
form $f=f(\epsilon)$, where $\epsilon=v^2/2+\Phi({\bf r})$ is the individual
energy of the particles by unit of mass, is a steady state of the Vlasov
equation. In astrophysics, it describes spherically symmetric and
isotropic stellar systems
\cite{bt}.

\subsection{Metaequilibrium state}
\label{sec_m}

The Vlasov-Poisson equations develop very complex filaments as a result
of a mixing process in phase space (collisionless mixing). In this sense, the
fine-grained DF ${f}({\bf
r},{\bf v},t)$ will never reach a stationary state but will rather
produce intermingled filaments at smaller and smaller scales. However, if we
introduce a coarse-graining
procedure, the coarse-grained DF $\overline{f}({\bf
r},{\bf v},t)$ is expected to reach a metaequilibrium state $\overline{f}({\bf
r},{\bf v})$ on a very short timescale, of the order of the dynamical
time $t_D$. This is because the evolution continues at 
scales smaller than the scale of observation (coarse-grained). This process is
known as ``phase mixing'' and ``violent relaxation'' (or collisionless
relaxation) \cite{bt}. Lynden-Bell \cite{lb} has tried to describe this
metaequilibrium state in terms of statistical mechanics. In the following, we
summarize his theory and provide some complements (see also 
\cite{csr,super,assisePH}).

Let $f_{0}({\bf r},{\bf v})$ denote the initial (fine-grained)
DF. We discretize $f_{0}({\bf r},{\bf v})$ in a
series of levels $\eta$ over which $f_{0}({\bf r},{\bf v})\simeq \eta$
is approximately constant. Thus, the levels $\lbrace \eta\rbrace$
represent all the values taken by the fine-grained DF. If the initial
condition is unsteady or unstable, the DF $f({\bf r},{\bf v},t)$ will
be stirred in phase space (phase
mixing) but will conserve its values $\eta$ and the corresponding
hypervolumes $\gamma(\eta)=\int \delta(f({\bf r},{\bf
v},t)-\eta)\, d{\bf r}d{\bf v}$ as a property of the Vlasov
equation (this is equivalent to the conservation of all the Casimirs).

Let us introduce the probability density $\rho({\bf r},{\bf v},\eta)$ of
finding the level of phase density $\eta$ in a small neighborhood of
the position $({\bf r},{\bf v})$ in phase space. This probability
density satisfies at each point the normalization
condition
\begin{equation}
\label{E1}
\int\rho({\bf  r,v},\eta)\, d\eta=1.
\end{equation}
The locally averaged (coarse-grained) DF is
then expressed in terms of the probability density as
\begin{equation}
\label{E2}
\overline{f}({\bf  r,v})=\int\rho({\bf  r,v}, \eta)\eta \, d \eta,
\end{equation}
and  the associated potential satisfies $\overline{\Phi}({\bf r},t)=\int u(|{\bf
r}-{\bf r}'|)\overline{f}({\bf r}',{\bf v}',t)\, d{\bf
r}'d{\bf v}'$. The conserved
quantities of the Vlasov equation can be decomposed into two
groups.\footnote{This distinction was first made in \cite{grand,kingen}.} The
mass and energy are called {\it robust integrals} because they are
(approximately) conserved by the  coarse-grained DF:
$\overline{M[f]}=M[\overline{f}]$ and $\overline{E[f]}\simeq
E[\overline{f}]$. Hence
\begin{equation}
\label{E4}
M=\int \overline{f}\, d{\bf  r} d{\bf  v},
\end{equation}
\begin{equation}
\label{E5}
E=\int{1\over 2}\overline{f} v^{2} \, d{\bf  r}
 d{\bf  v}+{1\over 2}\int \overline{f} \ \overline{\Phi} \, d{\bf  r} d{\bf  v}.
\end{equation}
The potential  is smooth since it is expressed as an integral
of the
DF, so we can express the energy in terms of the
coarse-grained fields $\overline{f}$ and $\overline{\Phi}$ neglecting
the internal energy of the fluctuations $\overline{\tilde f\tilde
\Phi}$. Therefore, the mass and the energy can be calculated at any time of the
evolution from the coarse-grained field $\overline{f}$. By contrast, the moments
$M_{n}$ with $n\ge 2$ are called
{\it fragile integrals} because they are altered on the coarse-grained
scale since $\overline{f^{n}}\neq \overline{f}^{n}$, where
$\overline{f^{n}}=\int \rho\eta^{n} \, d\eta$.
Therefore, only
the moments of the fine-grained DF,
$M^{f.g.}_{n}=\overline{M_{n}[f]}=\int \overline{f^{n}}\, d{\bf r}d{\bf v}$,
are conserved, i.e.
\begin{equation}
\label{E6}
M^{f.g.}_{n}=\int \rho({\bf  r,v},\eta)\eta^{n} \, d{\bf r}d{\bf v}d\eta.
\end{equation}
The moments of the coarse-grained DF, $M^{c.g.}_{n}[\overline{f}]=\int
\overline{f}^{n}\, d{\bf r}d{\bf v}$, are not conserved 
along the evolution since $M_{n}[\overline{f}]\neq
\overline{M_{n}[f]}$. Instead of conserving the fine-grained
moments, we can equivalently conserve the total hypervolume 
\begin{equation}
\gamma(\eta)=\int
\rho({\bf  r,v},\eta) \, d{\bf r}d{\bf v}
\end{equation}
of each phase level $\eta$. We note that $M^{f.g.}_{n}=\int \gamma(\eta)
\eta^{n} \, d\eta$.

After a complex evolution, we may expect the system to be in the most
probable, i.e. most mixed state, consistent with all the constraints
imposed by the dynamics.\footnote{This statement relies on an assumption of 
ergodicity
which may not always be realized in practice (see Refs.
\cite{lb,grand,incomplete,assisePH} for
a
discussion of the concept of incomplete relaxation).} Specifically, the
metaequilibrium state is obtained by maximizing the Lynden-Bell mixing
entropy\footnote{We take $k_B=1$ throughout the paper.}
 \begin{equation}
S_{\rm LB}[\rho]=-\int \rho({\bf r},{\bf v},\eta)\ln\rho({\bf r},{\bf
v},\eta)\, d{\bf r}d{\bf v}d\eta,
\label{E9}
\end{equation}
while conserving the mass $M$, the energy $E$ and all the Casimirs
(or the fine-grained moments $M_n^{f.g.}$). We also need to
account for the local normalization condition (\ref{E1}). The Lynden-Bell
entropy (\ref{E9}) can be obtained from a standard combinatorial analysis taking
into account the
specificities of the Vlasov equation (see Refs. \cite{lb,super,assisePH} for
details).

Introducing Lagrange multipliers, the first
variations satisfy
\begin{equation}
\delta S_{\rm LB}-\beta\delta E-\alpha\delta M-\sum_{n>1}\alpha_{n}\delta
M_{n}^{f.g.}-\int\zeta({\bf
r},{\bf v})\delta\biggl (\int \rho({\bf r},{\bf v},\eta)d\eta\biggr )\, d{\bf
r}d{\bf v}=0,
\label{E10}
\end{equation}
where ${ \beta}=1/T$ is the inverse temperature associated with the
conservation of
energy  and ${\alpha }_{n}$ are the
``chemical potentials" associated with the conservation of the
fine-grained moments $M_n^{f.g.}$ (the function $\zeta({\bf
r},{\bf v})$ accounts for the local
normalization condition).\footnote{For brevity, we
shall not consider here the conservation of linear impulse ${\bf P}$ and angular
momentum ${\bf
L}$. Actually, if we work in the barycentric frame of reference, the
Lagrange
multiplier ${\bf U}$ associated with the conservation of ${\bf P}$ vanishes. On
the other hand, the conservation of angular momentum implies that the
statistical equilibrium state has a solid rotation. When ${\bf L}={\bf 0}$,
it can be shown that the Lagrange multiplier ${\bf \Omega}$  associated with
the conservation of ${\bf L}$ also vanishes \cite{lb,csr}. We will assume that
we are in this situation, even though the general case can be treated
straightforwardly.} This variational principle
leads to the Gibbs state
\begin{equation}
\label{E11} \rho({\bf r},{\bf v},\eta) ={1\over
Z(\epsilon)}\chi(\eta)e^{-\eta(\beta\epsilon+\alpha)},
\end{equation}
where $\epsilon={v^2/2}+\overline{\Phi}({\bf r})$ is the energy of a particle by
unit of
mass.\footnote{We note that the Lynden-Bell
distribution (\ref{E11}) does not lead to a segregation by mass since
the individual mass of the particles does not appear in the Vlasov
equation on which the whole theory is based. However, it leads to a
segregation by phase levels $\eta$.} In writing Eq. (\ref{E11}), we have
distinguished the Lagrange
multipliers $\alpha$ and $\beta$ associated with the robust integrals
$M$ and $E$ from the Lagrange multipliers $\alpha_{n>1}$, associated
with the conservation of the fragile moments $M_{n>1}^{f.g.}$, which have
been regrouped in
the function 
\begin{equation}
\label{chi} \chi(\eta)\equiv {\rm
exp}(-\sum_{n>1}\alpha_{n}\eta^{n}).
\end{equation}
This distinction will make sense
in the following (see also Appendix \ref{sec_cano}). Under this form, we see
that the equilibrium
distribution of phase levels is a product of a universal Boltzmann
factor $e^{-(\beta\epsilon+\alpha)\eta}$ by a non-universal function
$\chi(\eta)$ which depends on the initial condition. The partition
function $Z$ is determined by the local normalization condition (\ref{E1})
yielding
\begin{equation}
\label{E12}
Z(\epsilon)=\int \chi(\eta)e^{-\eta(\beta\epsilon+\alpha)}\, d\eta.
\end{equation}
The
partition function $Z(\epsilon)$ can be
used as a generating
function for constructing the moments of the fine-grained
distribution (see Appendix \ref{sec_cgf}). We note
that the
Lynden-Bell statistics (\ref{E11}) has a
form similar to a
superstatistics  (see Ref. \cite{super} for the development of this analogy). 
The
equilibrium coarse-grained DF defined by Eq. (\ref{E2}) can be written as
\begin{equation}
\label{E14}
\overline{f}=\frac{1}{Z(\epsilon)}\int
\chi(\eta)\eta e^{-\eta (\beta\epsilon+\alpha)}\, d\eta=\frac{\int
\chi(\eta)\eta e^{-\eta(\beta\epsilon+\alpha)}\, d\eta}{\int
\chi(\eta)
e^{-\eta(\beta\epsilon+\alpha)}\, d\eta}.
\end{equation}
One can easily check that
\begin{equation}
\label{E13}
\overline{f}=-{1\over\beta}{\partial\ln Z\over\partial
\epsilon}=F(\beta\epsilon+\alpha)=\overline{f}(\epsilon).
\end{equation}
We note that
the coarse-grained DF predicted by Lynden-Bell depends
only on the individual energy $\epsilon$ of the particles. As such, it is a
particular stationary solution of the Vlasov equation. We also note
that $\overline{f}(\epsilon)$
is a monotonically decreasing function of energy. Indeed, from Eqs. (\ref{E11})
and
(\ref{E13}), it is easy to establish that (see Appendix \ref{sec_cgf})
\begin{equation}
\label{g1} \overline{f}'(\epsilon)=-\beta f_{2},
\end{equation}
where
\begin{equation}
\label{g1we} 
f_{2}\equiv \int \rho
(\eta-\overline{f})^{2}d\eta=\overline{f^2}-\overline{f}^2\ge 0
\end{equation}
is the  local centered variance
of
the distribution
$\rho({\bf r},{\bf v},\eta)$. Equation (\ref{g1}) is a form of
fluctuation-dissipation theorem. We note that $\overline{f}'(\epsilon)\le
0$ since $\beta\ge 0$ is required to make the velocity profile
normalizable. We can also easily show \cite{super} that
$\overline{f}({\bf r},{\bf v})\le f_{0}^{\rm max}$, where
$f_{0}^{\rm max}$ is the maximum value of the initial (fine-grained)
DF. The inequality $0\le \overline{f}\le
f_{0}^{\rm max}$ is clear from physical
considerations since the coarse-grained DF can only
decrease by mixing. Finally, one can show that the
coarse-grained DF predicted by Lynden-Bell is
nonlinearly dynamically Vlasov stable (see Sec. \ref{sec_nds}).

For a given initial condition, the
statistical theory of Lynden-Bell selects a particular stationary
solution of the Vlasov equation (the most probable -- most mixed -- one) among
an infinity of
stationary solutions. The Lynden-Bell equilibrium state is obtained by solving
the integral equation
\begin{eqnarray}
\overline{\Phi}({\bf r})=\int u(|{\bf r}-{\bf
r}'|)\overline{f}_{\alpha_{n},\beta}\left
\lbrack\frac{{v'}^2}{2}+\overline{\Phi}({\bf r}')\right \rbrack\, d{\bf
r}'d{\bf v}'
\label{ve2b}
\end{eqnarray}
and relating the Lagrange multipliers ($\alpha_{n}$, $\beta$) to the
constraints ($M_{n}^{f.g.}$, $E$). We also have to make sure that the
equilibrium state is an entropy maximum not a minimum or a saddle point (see
Appendix \ref{sec_cano}). We note that the coarse-grained DF
$\overline{f}(\epsilon)$ can take different forms
depending on the function $\chi(\eta)$ determined by the fragile
moments.  In the present context, the function
$\chi(\eta)$ is determined from the constraints {\it a
posteriori}. Indeed, we have to solve the full problem in order to get
the $\alpha_n$'s and obtain the expression of
$\chi(\eta)$.\footnote{In this sense, the
constraints
associated with the conservation of the fine-grained moments are
treated microcanonically. Following the approach of \cite{eht} in 2D turbulence,
we have suggested in \cite{super} that, when the system is forced by an external
medium, the fine-grained constraints may be
treated canonically.
In that case, the function $\chi(\eta)$ should be considered
as given {\it a priori} (it is determined by the external forcing). Treating
the Casimirs canonically also allows us to derive a sufficient condition of
 thermodynamical stability in the sense of Lynden-Bell (see Appendix
\ref{sec_cano}).} We
emphasize that the
Lynden-Bell statistical equilibrium state $\overline{f}(\epsilon)$ resulting
from a violent collisionless relaxation depends on the details of the initial
condition. This is different from the Boltzmann statistical equilibrium state
resulting from a collisional relaxation which depends only on the value of
the mass $M$ and the energy $E$. In the present case, we need to know the value
of
the fine-grained moments $M_{n}^{f.g.}$ which are accessible only in
the initial condition (or from the fine-grained field) since the {\it
observed} moments $M_{n}^{c.g.}$ are altered for $t>0$ by the coarse-graining as
the
system undergoes a mixing process ($M_{n}^{c.g.}\neq
M_{n}^{f.g.}$).

{\it Remark:} Similar results have been derived by Miller
\cite{miller} and Robert and Sommeria \cite{rs} in 2D turbulence. The analogy
between the Lynden-Bell theory and the Miller-Robert-Sommeria theory is
discussed in \cite{csr,houchesPH}.

\subsection{Two-level case}
\label{sec_tlc}

If the initial DF takes only two values
$f_0=\eta_{0}$ and  $f_0=0$ (vacuum), the Lynden-Bell entropy reduces to
\begin{eqnarray}
\label{lbent}
S=-\int \left\lbrace \frac{\overline{f}}{\eta_0}\ln
\frac{\overline{f}}{\eta_0}+\left (1-\frac{\overline{f}}{\eta_0}\right )\ln
\left (1-\frac{\overline{f}}{\eta_0}\right )\right\rbrace \, d{\bf r}d{\bf v},
\end{eqnarray}
which is similar to the Fermi-Dirac entropy. Furthermore, the constraints
reduce to the conservation of mass $M$ and energy $E$ since $M_{n>1}^{f.g.}=\int
\overline{f^n}\, d{\bf r}d{\bf v}=\int
\overline{\eta_0^{n-1}\times f}\, d{\bf r}d{\bf v}=
\eta_0^{n-1}M$. The metaequilibrium
state is then given by
\begin{equation}
\label{E15}
\overline{f}={\eta_{0}\over 1+ e^{\eta_{0}(\beta\epsilon+\alpha)}},
\end{equation}
which is similar to the Fermi-Dirac DF \cite{lb,csmnras}. Morphologically,
the Lynden-Bell
statistics corresponds to a $4^{\rm th}$ type of statistics since
the particles are distinguishable but subject to an exclusion
principle $\overline{f}\le \eta_0$ due to the incompressibility of the
flow in phase space \cite{lb}.
This constraint plays
a role similar to the Pauli exclusion principle in quantum
mechanics. In the dilute (nondegenerate) limit of the Lynden-Bell theory
$\overline{f}\ll
\eta_{0}$, the entropy (\ref{lbent}) and the DF (\ref{E15}) reduce to
\begin{equation}
\label{E17}
S=-\int 
\frac{\overline{f}}{\eta_0}\left
(\ln\frac{\overline{f}}{\eta_0}-1\right )\, d{\bf r}d{\bf
v},\qquad \overline{f}=\eta_0
e^{-\eta_{0}(\beta\epsilon+\alpha)},
\end{equation}
which are similar to the Boltzmann entropy and to the Boltzmann distribution. 

{\it Remark:} We note that the effective temperature $T=1/\beta$
in the DF (\ref{E17}) has not the dimension of a temperature. Indeed,
the mass $m$ of the
particles does not appear in the
Lynden-Bell theory since it is based on the Vlasov equation for collisionless
systems which is independent of the mass of the particles.
However, $T/\eta_0$ can  be
interpreted as a velocity dispersion (in the nondegenerate limit).  In this
sense, one
can say that the temperature in
Lynden-Bell's theory is proportional to the  mass of the particles
 (or more precisely to the ratio $m/\eta_0$) \cite{lb}.

\subsection{Generalized entropy}
\label{sec_ge}

Since the coarse-grained DF
$\overline{f}(\epsilon)$ predicted by the statistical theory of
Lynden-Bell depends only on the individual energy and is monotonically
decreasing, it extremizes a ``generalized entropy'' of the form
\cite{grand,kingen,super,assisePH}
\begin{equation}
\label{meta4} S[\overline{f}]=-\int C(\overline{f})\, d{\bf r}d{\bf v}
\end{equation}
at fixed mass $M$ and energy $E$, where
$C(\overline{f})$ is a convex function
(i.e. $C''>0$).  Indeed, introducing Lagrange multipliers  $\alpha$ and
$\beta$,
and
writing the
variational principle under the form
\begin{equation}
\label{meta5} \delta S-\beta \delta E-\alpha\delta M=0,
\end{equation}
we find that
\begin{equation}
\label{meta6} C'(\overline{f})=-\beta\epsilon-\alpha.
\end{equation}
Since $C'(\overline{f})$ is a monotonically increasing function of
$\overline{f}$, we can
inverse this relation to obtain
\begin{equation}
\label{meta7} \overline{f}=F(\beta\epsilon+\alpha)=\overline{f}(\epsilon),
\end{equation}
where the function $F(x)=(C')^{-1}(-x)$ is determined by the generalized
entropy $C(\overline{f})$. Inversely, for a given $F(x)$, the
generalized entropy is given by 
\begin{equation}
\label{g8q} C(\overline{f})=-\int^{\overline{f}}F^{-1}(x)\, dx.
\end{equation} From the identity
\begin{equation}
\label{meta7b} \overline{f}'(\epsilon)=-\frac{\beta}{C''(\overline{f})},
\end{equation}
obtained from Eq. (\ref{meta6}), we find that $\overline{f}(\epsilon)$ is a
monotonically decreasing function, i.e., $\overline{f}'(\epsilon)<0$ (since
$\beta>0$ as explained in \cite{cc}).

Therefore, for any Gibbs state of the form (\ref{E11}), there exists a
generalized entropy of the form (\ref{meta4}) that the
coarse-grained DF $\overline{f}$, given by Eq.
(\ref{E13}), extremizes at fixed mass $M$ and energy $E$. From the
statistical
theory of Lynden-Bell, we have $F(x)=-(\ln{Z})'(x)$ [see Eq. (\ref{E13})],
where $Z(x)$ depends only on $\chi(\eta)$ [see Eq. (\ref{E12})]. Substituting
this relation into Eq. (\ref{g8q}), we
find that the generalized entropy is given by \cite{super,assisePH}
\begin{equation}
\label{s3w} C(\overline{f})=-\int^{\overline{f}} \lbrack (\ln
Z)'\rbrack^{-1}(-x)\, dx.
\end{equation}
We expect that in many cases the coarse-grained
distribution
(\ref{E13}) {\it maximizes} the generalized entropy $S$ at fixed
mass $M$ and energy
$E$ (robust constraints) although this is not necessarily the case (see
Appendix \ref{sec_cano}). We emphasize that the generalized
entropy (\ref{s3w}) is a {\it non-universal} function which
depends on the
initial condition.  Indeed, it is determined by the function
$\chi(\eta)$ which depends indirectly on the initial condition
through the complicated procedure discussed at the end of Sec.
\ref{sec_m}.\footnote{In the case where the system experiences an
external
forcing (footnote 8), the function $\chi(\eta)$ and, consequently, the
generalized entropy $C(\overline{f})$ should be considered to be given {\it a
priori}, being determined by the forcing.} In general, the generalized
entropy (\ref{meta4})
with (\ref{s3w}) is
{\it not} the ordinary Boltzmann entropy $S_B[\overline{f}]=-\int
\overline{f}\ln \overline{f}\, d{\bf r}d{\bf v}$ because of the existence of
fine-grained constraints (Casimirs) that modify the form of the entropy
that we would naively expect. These constraints are sometimes refered to as
{\it hidden constraints} since they are not accessible from the
coarse-grained dynamics \cite{kingen,super,incomplete,assisePH}.

{\it Remark:} Similar results have been derived by Chavanis
\cite{vphydro,cnd,physicaD1,physicaD2} in 2D turbulence.

\subsection{$H$-functions and selective decay principle}
\label{sec_incomplete}

In order to quantify the importance of mixing during the process of violent
relaxation, Tremaine {\it et al.} \cite{thlb} have introduced the notion of
$H$-functions. They are
defined by
\begin{equation}
H[\overline{f}]=-\int C(\overline{f}) \, d{\bf r}d{\bf v}, \label{i1}
\end{equation}
where $C$ is any convex function (i.e. $C''>0$). It can
be shown that the
 $H$-functions $H[\overline{f}]$ calculated with the coarse-grained
DF increase during violent relaxation in the sense
that $H[\overline{f}({\bf r},{\bf v},t)]\ge H[\overline{f}({\bf
r},{\bf v},0)]$ for $t>0$ where it is assumed that, initially, the
system is not mixed so that $\overline{f}({\bf r},{\bf v},0)={f}({\bf
r},{\bf v},0)$. This is similar to the $H$-theorem in kinetic theory.
However, contrary to the Boltzmann equation, the Vlasov equation does
not single out a unique functional (the above inequality is true for
{\it all} $H$-functions) and the time evolution of the $H$-functions is not
necessarily monotonic (nothing is implied concerning the relative
values of $H(t)$ and $H(t')$ for $t,t'>0$). Yet, this observation
suggests a notion of generalized {\it selective decay
principle} (for $-H$):\footnote{A similar selective decay principle has been
advocated in 2D turbulence and magnetohydrodynamics (see, e.g., Refs.
\cite{km,matt,vphydro,chavjapon,ncd} and references
therein). It is either due to a small dissipation or to a coarse-graining (for
dissipationless systems). In 2D turbulence, it has been argued that the
enstrophy $\Gamma_2=\int
\omega^2\, d{\bf r}$ decreases while the circulation $\Gamma=\int\omega\,
d{\bf r}$ and the energy $E=\frac{1}{2}\int\omega\psi\, d{\bf r}$ are conserved
so that the system reaches a minimum enstrophy state
\cite{bretherton,leith,jfm1,ncd}. The minimization of enstrophy at fixed
circulation and energy (which is mathematically equivalent to the minimization
of energy at fixed circulation and enstrophy) leads to a linear relationship
$\omega=\lambda\psi+\mu$ between vorticity and
stream function. The minimization of ``generalized enstrophies'' has also been
considered. In magnetohydrodynamics
\cite{woltjer,taylor,mtv,leprovost,nmcd}, it has been argued that the magnetic
energy $E=\int {\bf B}^2\, d{\bf r}$ decreases while the helicity $H=\int {\bf
A}\cdot {\bf B}\, d{\bf r}$ is conserved so that the system reaches a minimum
energy state. The minimization of magnetic energy at fixed helicity (which is
mathematically equivalent to the maximization or minimization of helicity at
fixed energy) leads to
a linear relationship $\nabla\times {\bf B}=\lambda {\bf B}$ which
characterizes a force-free configuration. These variational principles ensure
the nonlinear dynamical stability of the system with respect to a
dissipationless evolution (we note that all the above functionals are conserved
for a purely inviscid and fine-grained evolution).}
Among all the invariants of the collisionless dynamics, the
$H$-functions
(fragile constraints) tend to increase ($-H$ tend to decrease) on the
coarse-grained scale while the mass and the energy (robust
constraints) are approximately conserved.  According to this
phenomenological principle, we may expect that the
metaequilibrium state reached
by the system as a result of violent relaxation will
maximize a certain $H$-function (non-universal), denoted $H^*[\overline{f}]$,
at fixed mass and
energy.\footnote{This is, however, not necessary: All the $H$-functions could
increase at fixed mass and energy without necessarily implying that the
coarse-grained DF reaches a
steady state that
maximizes one of them (see Sec. \ref{sec_cge}).} This would
guarantee that the
metaequilibrium state is nonlinearly dynamically stable (see Sec.
\ref{sec_nds}).
If the evolution is ergodic, the above-mentioned statement is presumably
correct. The
$H$-function $H^*[\overline{f}]$ that is effectively maximized at
metaequilibrium is the generalized entropy $S[\overline{f}]$ defined by Eqs.
(\ref{meta4})
and (\ref{s3w}), as obtained from
Lynden-Bell's theory (according to the comment that follows Eq. (\ref{s3w}) 
this is expected to be true in many cases but not in
all cases). Furthermore,
under
the assumption of Appendix \ref{sec_out}, it can be
shown that the generalized entropy  monotonically increases  during
violent
relaxation (see Sec. \ref{sec_chav}). In case of incomplete relaxation, it is
possible (but not necessary) that the metaequilibrium state maximizes a certain
$H$-function (sometimes also called a generalized entropy) which is different
from the one associated with the Lynden-Bell theory ($H^*[\overline{f}]\neq
S[\overline{f}]$). This discussion shows that the notion of selective decay for
collisionless systems with long-range interactions is quite subtle.

\subsection{Incomplete relaxation}
\label{sec_incompleteq}

The statistical theory of Lynden-Bell relies on the assumption that the
evolution is ergodic so that the equilibrium state maximizes the mixing
entropy (\ref{E9}) under the constaints of the dynamics. In reality, this
is
not
always the case. It has
been understood since the beginning \cite{lb} that violent
relaxation may be {\it incomplete} so that the mixing entropy
(\ref{E9}) is {\it not} maximized in the whole available phase
space. This is obvious in the case of 3D self-gravitating systems since there is
no maximum entropy state, even in theory (the Lynden-Bell DF has an infinite
mass). However,
even for simpler systems for which a maximum entropy state (in the sense of
Lynden-Bell) exists, there are
cases where this maximum entropy state is not
reached (see the discussion in \cite{incomplete} and in Sec. 6 of
\cite{assisePH}).
For example, in the context of the HMF model
\cite{lrt,cct,lbt,antoniazzi,antoniazzi2,antoniazzi3,cnfr,stan1,stan2,ccpoly}
and in the context of 2D
turbulence \cite{staquet,hd,boghosian,brands}, situations have been reported
where the Lynden-Bell prediction works
well and situations have been reported where the Lynden-Bell does not work well
(!).
Some authors have proposed to account for incomplete
relaxation by changing the form of entropy and by using for example the Tsallis
entropy \cite{tsallis}. Sometimes, the Tsallis
distribution provides a good fit of the
metaequilibrium state reached by the system (see the above-mentioned
references). However, this is not general.
Furthermore, this type of
appoach leads to some arbitrariness  since the generalized entropy depends on
unknown parameters (like, e.g., Tsallis' $q$ parameter) that are not predicted
by the theory.  In
practice, these parameters
have to be fitted to the observed distribution. In Sec. \ref{sec_incompletekin},
we shall
discuss an alternative approach to take
into account incomplete relaxation, based on kinetic theory, where there is no
such indetermination.

\section{Kinetic theory of quiescent collisionless relaxation}
\label{sec_whole}

In this section, we recall the kinetic theory of quiescent collisionless
relaxation, based on a quasilinear theory of the Vlasov equation,
initially developed by Kadomtsev and Pogutse \cite{kp}, Severne and Luwel
\cite{sl}, and Chavanis \cite{chavmnras,dubrovnik,kingen}. We give some details
of derivation
and complements.

\subsection{Quasilinear theory}

Basically, a collisionless system with long-range interactions is described in a
self-consistent mean field approximation by the Vlasov equation [see Eqs.
(\ref{ve1}) and (\ref{ve2})]. In principle, this equation completely 
determines the evolution of the DF
$f({\bf r},{\bf v},t)$. However, as discussed in Sec. \ref{sec_stlb},
we are not interested in practice by the finely striated structure of
the flow in phase space but only by its macroscopic,
i.e. smoothed-out, structure. Indeed, the observations and the
numerical simulations are always realized with a finite
resolution. Moreover, the coarse-grained DF
$\overline{f}({\bf r},{\bf v},t)$ is likely to converge towards a steady state
$\overline{f}({\bf r},{\bf v})$ (metaequilibrium state) contrary to the exact
distribution $f({\bf r},{\bf v},t)$
which
develops smaller and smaller scales for all times.

If we decompose the DF and the 
potential in a mean and fluctuating part
($f=\overline{f}+\delta f$,
$\Phi=\overline{\Phi}+\delta \Phi$) and take the local average of the
Vlasov equation (\ref{ve1}), we readily obtain an equation for the
coarse-grained DF of the
form
\begin{equation}
\label{E15b}
{\partial\overline{f}\over\partial t}+{\bf v}\cdot {\partial
\overline{f}\over\partial
{\bf r}}-\nabla
\overline{\Phi}\cdot {\partial \overline{f}\over\partial {\bf v}}=
-{\partial \over\partial {\bf v}}\cdot {\bf J}_{f}
\end{equation}
with a diffusion
current ${\bf J}_{f}=-\overline{\delta f\nabla \delta\Phi}$ related to
the correlations of the fine-grained fluctuations. The
right hand side of Eq. (\ref{E15b}) can be interpreted as an effective
``collision'' term. Any
systematic calculation of the diffusion current starting from the
Vlasov equation must necessarily introduce an
evolution equation for the fluctuation $\delta f$. This equation is simply
obtained by subtracting Eq. (\ref{E15b}) from Eq. (\ref{ve1}). This yields
\begin{equation}
\label{E29}
{\partial \delta{f}\over\partial t}+{\bf v}\cdot {\partial \delta{f}\over
\partial
{\bf r}}-\nabla\overline{\Phi}\cdot {\partial \delta {f}\over\partial
{\bf v}}=\nabla \delta{\Phi}\cdot {\partial \overline{f}\over\partial {\bf
v}}+\nabla\delta{\Phi}\cdot {\partial \delta
f\over\partial {\bf v}}-\overline{\nabla \delta{\Phi}\cdot {\partial \delta
f\over\partial
{\bf v}}}.
\end{equation}
Equations (\ref{E15b}) and (\ref{E29}) are exact since no approximation
has been made for the moment. To go further, we need to implement some
approximations. In the sequel, we shall develop a quasilinear theory 
which was
introduced by Kadomtsev and Pogutse \cite{kp} by analogy with the quasilinear
theory of collisional relaxation based on the Klimontovich equation. This will
provide a precise theoretical
framework to analyze the process of collisionless relaxation
in systems with long-range interactions.  The essence of the quasilinear theory
is to
assume that the fluctuations are weak and neglect the nonlinear
terms in Eq. (\ref{E29}) altogether. In that case, Eqs.
(\ref{E15b}) and (\ref{E29}) reduce to the coupled system
\begin{equation}
\frac{\partial \overline{f}}{\partial t}+{\bf v}\cdot\frac{\partial
\overline{f}}{\partial {\bf
r}}-\nabla\overline{\Phi}\cdot \frac{\partial \overline{f}}{\partial {\bf
v}}=\frac{\partial}{\partial
{\bf v}}\cdot \overline{\delta f\nabla\delta\Phi},
\label{pvm11}
\end{equation}
\begin{equation}
\frac{\partial\delta f}{\partial t}+{\bf v}\cdot \frac{\partial \delta
f}{\partial {\bf r}}-\nabla\overline{\Phi}\cdot \frac{\partial \delta
f}{\partial {\bf
v}}-\nabla\delta\Phi\cdot \frac{\partial \overline{f}}{\partial
{\bf v}}=0.
\label{lb2}
\end{equation}
Physically, these equations describe the coupling between a
subdynamics (played here by the small scale fluctuations $\delta f$) and a
macrodynamics (played by the coarse-grained DF
$\overline{f}$). Due to the strong simplifications implied by the
neglect of nonlinear terms in Eq. (\ref{E29}), the quasilinear theory
only describes the late quiescent stages of the violent relaxation
process, when the fluctuations have weaken (gentle
relaxation). Although this is essentially an asymptotic theory, it is
of importance to develop this theory in detail since it provides an
explicit expression of the effective ``collision operator'' which
appears on the coarse-grained scale. 

If we restrict
ourselves to spatially homogeneous distributions, the field
$-\nabla\overline{\Phi}$
vanishes. In that case, Eqs. (\ref{pvm11}) and (\ref{lb2}) reduce to the coupled
equations
\begin{equation}
\frac{\partial \overline{f}}{\partial t}=\frac{\partial}{\partial {\bf v}}\cdot
\overline{\delta f \nabla\delta\Phi},
\label{lb3}
\end{equation}
\begin{equation}
\frac{\partial\delta f}{\partial t}+{\bf v}\cdot \frac{\partial \delta
f}{\partial {\bf r}}
-\nabla\delta\Phi\cdot \frac{\partial \overline{f}}{\partial {\bf v}}=0.
\label{lb4}
\end{equation}
We shall assume that the fluctuations evolve rapidly compared
to the evolution of the coarse-grained fields, so that the time variation of
$\overline{f}$ and
$\overline{\Phi}$ can be
neglected in the calculation of the collision term. This is similar to
the Bogoliubov ansatz in the collisional theory.
Therefore, for the purpose of solving Eq. (\ref{lb4}) and obtaining the
correlation function $\overline{\delta f \delta\Phi}$, we shall
regard $\overline{f}({\bf v})$ as constant in time.  With this
approximation, Eqs. (\ref{lb3}) and (\ref{lb4}) can be solved with the aid of
Fourier-Laplace transforms and the collision term can be explicitly calculated.
The derivation proceeds similarly to the derivation of
the Lenard-Balescu equation from the Klimontovich equation in the collisional
theory (see, e.g., \cite{epjp}). The collisional and
collisionless kinetic theories are analogous because the
Klimontovich equation is formally similar to the Vlasov equation. However, 
the Klimontovich equation  involves a DF which is a sum of
$\delta$-functions while the Vlasov equation involves a continuous DF.

\subsection{Dielectric function}

The Fourier-Laplace transform of the fluctuations  of the DF 
$\delta f$ is defined by
\begin{equation}
\delta \tilde f({\bf k},{\bf v},\omega)=\int \frac{d{\bf r}}{(2\pi)^d}\int_{0}^{+\infty}dt\, e^{-i({\bf k}\cdot{\bf r}-\omega t)}\delta f({\bf r},{\bf v},t).
\label{lb5}
\end{equation}
The inverse transform is
\begin{equation}
\delta f({\bf r},{\bf v},t)=\int d{\bf k}\int_{\cal C}\frac{d\omega}{2\pi}\, e^{i({\bf k}\cdot{\bf r}-\omega t)}\delta\tilde f({\bf k},{\bf v},\omega),
\label{lb6}
\end{equation}
where the Laplace contour ${\cal C}$ in the complex $\omega$ plane must pass
above all 
poles of the integrand. Similar expressions hold for the fluctuations  of the
potential $\delta\Phi$. We note that, for periodic potentials, the integral over
${\bf k}$ is replaced by a discrete summation over the different modes. If we
take the  Fourier-Laplace transform of Eq. (\ref{lb4}), we find that
\begin{equation}
-\delta\hat{f}({\bf k},{\bf v},0)-i\omega\, \delta\tilde f({\bf k},{\bf
v},\omega)
+i {\bf k}\cdot{\bf v}\, \delta\tilde f({\bf k},{\bf v},\omega)-i {\bf
k}\cdot\frac{\partial \overline{f}}{\partial {\bf v}}\, \delta\tilde\Phi({\bf
k},\omega)=0,
\label{lb7}
\end{equation}
where the first term is the spatial Fourier transform of the initial value of
the fluctuations
\begin{equation}
\delta\hat{f}({\bf k},{\bf v},0)=\int\frac{d{\bf r}}{(2\pi)^d}\, e^{-i{\bf k}\cdot {\bf r}}\delta f({\bf r},{\bf v},0).
\label{lb8}
\end{equation}
Equation (\ref{lb7}) can be rewritten as
\begin{equation}
\delta\tilde f ({\bf k},{\bf v},\omega)=\frac{{\bf k}\cdot \frac{\partial
\overline{f}}{\partial {\bf v}}}{{\bf k}\cdot {\bf
v}-\omega}\delta\tilde\Phi({\bf k},\omega)+\frac{\delta\hat f({\bf k},{\bf
v},0)}{i({\bf k}\cdot {\bf v}-\omega)},
\label{lb9}
\end{equation}
where the first term takes into account ``collective effects''. The
fluctuations of the potential are related to the fluctuations of the
DF by a convolution
\begin{equation}
\delta\Phi({\bf r},t)=\int u(|{\bf r}-{\bf r}'|)\delta f({\bf r}',{\bf v}',t)\, d{\bf r}'d{\bf v}'.
\label{deltaconvol}
\end{equation}
Taking the Fourier-Laplace transform of this equation, we obtain
\begin{equation}
\delta\tilde\Phi({\bf k},\omega)=(2\pi)^d\hat{u}(k)\int\delta\tilde f({\bf k},{\bf v},\omega)\, d{\bf v}.
\label{lb10}
\end{equation}
Substituting Eq. (\ref{lb9}) into Eq. (\ref{lb10}), we find that the
Fourier-Laplace transform of the fluctuations of the potential is given by
\begin{equation}
\delta\tilde\Phi({\bf k},\omega)=(2\pi)^d\frac{\hat{u}(k)}{\epsilon({\bf k},\omega)}\int d{\bf v}\, \frac{\delta\hat f({\bf k},{\bf v},0)}{i({\bf k}\cdot {\bf v}-\omega)},
\label{lb11}
\end{equation}
where 
\begin{equation}
\epsilon({\bf k},\omega)=1-(2\pi)^d\hat{u}(k)\int  \frac{{\bf k}\cdot
\frac{\partial \overline{f}}{\partial {\bf v}}}{{\bf k}\cdot {\bf v}-\omega}\,
d{\bf v}
\label{lb12}
\end{equation}
is the dielectric function. The Fourier-Laplace transform of the
fluctuations of the DF is
then given by Eq. (\ref{lb9}) with Eq. (\ref{lb11}). The
dispersion
relation associated with the linearized Vlasov equation corresponds to
$\epsilon({\bf k},\omega)=0$. It determines the proper pulsations of the system.
If the DF is Vlasov stable,\footnote{The initial DF may be
dynamically unstable or even
unsteady but we focus on a regime of quiescent
relaxation where the DF is dynamically stable.} then ${\rm Im}[\omega]<0$ for
all modes $\omega$. In particular, $\epsilon({\bf k},\omega)$ does not vanish
when
$\omega$ is real so that  Eq. (\ref{lb11}) is well-defined. If collective
effects were neglected in Eq. (\ref{lb9}), we would obtain Eq. (\ref{lb11}) with
$\epsilon({\bf k},\omega)=1$. 
This shows that, because of collective effects, the bare potential of
interaction $\hat{u}(k)$ is replaced by a ``dressed'' potential
$\hat{u}(k)/\epsilon({\bf k},\omega)$ taking into account the polarization of
the medium.

We can use the foregoing equations to compute the effective collision term
appearing 
on the right hand side of Eq. (\ref{lb3}). One has
\begin{equation}
\overline{\delta f \nabla\delta\Phi}=\int d{\bf k}\int_{\cal C}
\frac{d\omega}{2\pi}\int d{\bf k'}\int_{\cal C} \frac{d\omega'}{2\pi} \, i {\bf
k}' e^{i({\bf k}\cdot {\bf r}-\omega t)}e^{i({\bf k}'\cdot {\bf r}-\omega'
t)}\overline{\delta\tilde f({\bf k},{\bf v},\omega)\delta\tilde\Phi({\bf
k}',\omega')}.
\label{lb13}
\end{equation}
Using Eq. (\ref{lb9}), we find that
\begin{equation}
\overline{\delta\tilde f({\bf k},{\bf v},\omega)\delta\tilde\Phi({\bf
k}',\omega')}=\frac{{\bf k}\cdot \frac{\partial \overline{f}}{\partial {\bf
v}}}{{\bf k}\cdot {\bf v}-\omega}\overline{\delta\tilde\Phi({\bf
k},\omega)\delta\tilde\Phi({\bf k}',\omega')}+\frac{\overline{\delta\hat
f({\bf k},{\bf v},0)\delta\tilde\Phi({\bf k}',\omega')}}{i({\bf k}\cdot
{\bf v}-\omega)}.
\label{lb14}
\end{equation}
As we shall see, the first term accounts for a diffusion and the second term
accounts for
a
friction.  Let us consider these two terms separately.

\subsection{Diffusion}

From Eq. (\ref{lb11}), we obtain
\begin{equation}
\overline{\delta\tilde\Phi({\bf k},\omega)\delta\tilde\Phi({\bf
k}',\omega')}=-(2\pi)^{2d}\frac{\hat{u}(k)\hat{u}(k')}{\epsilon({\bf
k},\omega)\epsilon({\bf k}',\omega')}\int d{\bf v}d{\bf v}'\, \frac{\overline{
\delta\hat f({\bf k},{\bf v},0)\delta\hat f({\bf k}',{\bf v}',0)}}{({\bf
k}\cdot {\bf v}-\omega)({\bf k}'\cdot {\bf v}'-\omega')}.
\label{lb15}
\end{equation}
To proceed further, we have to
evaluate the correlation function $\overline{\delta f({\bf r},{\bf
v},t)\delta f({\bf r'},{\bf v}',t)}$. Following Kadomtsev and Pogutse \cite{kp}
we shall assume that the mixing
in phase space is sufficiently efficient that the scale of the
kinematic correlations is small with respect to the coarse-graining
mesh size. In that case, we can write
\begin{equation}
\label{E40} \overline{\delta f({\bf r},{\bf v},0)\delta f({\bf
r'},{\bf v'},0)}=\epsilon_{r}^{d}\epsilon_{v}^{d} \delta({\bf
r}-{\bf r}') \delta ({\bf v}-{\bf v}') f_2({\bf v}),
\end{equation}
where $\epsilon_{r}$ and $\epsilon_{v}$ are the resolution scales
in position and velocity respectively and
\begin{equation}
\label{E41} f_2\equiv 
\overline{({\delta f})^{2}}=\overline{(f-\overline{f})^{2}}=\overline{f^{2}}
-\overline{f}^{2}
\end{equation}
is the local variance of the fine-grained fluctuations. Note that
$\epsilon_{r}^{d}\epsilon_{v}^{d}$ can be interpreted as the hypervolume
of a macrocell in Lynden-Bell's statistical theory.\footnote{The Lynden-Bell
entropy can be obtained from a combinatorial analysis by dividing the phase
space into  macrocells and microcells and by counting the number of microstates
associated with a given macrostate. The mixing entropy (\ref{E9}) is equal to
the
logarithm of this number \cite{lb,super,assisePH}.} Taking the
Fourier
transform of Eq. (\ref{E40}), we get
\begin{equation}
\overline{\delta\hat f({\bf k},{\bf v},0)\delta\hat f({\bf k}',{\bf
v}',0)}=\frac{1}{(2\pi)^d}\epsilon_{r}^{d}\epsilon_{v}^{d}\delta({\bf k}+{\bf
k}')\delta({\bf v}-{\bf v}') f_2({\bf v}).
\label{lb16}
\end{equation}
Substituting Eq.
(\ref{lb16}) into Eq. (\ref{lb15}), we find that
\begin{equation}
\overline{\delta\tilde\Phi({\bf k},\omega)\delta\tilde\Phi({\bf
k}',\omega')}=(2\pi)^{d}\epsilon_{r}^{d}\epsilon_{v}^{d}\frac{\hat{u}(k)^2}{
\epsilon({\bf k},\omega)\epsilon(-{\bf k},\omega')}\delta({\bf k}+{\bf k}')\int
d{\bf v}\, \frac{f_2({\bf v})}{({\bf k}\cdot {\bf v}-\omega)({\bf k}\cdot {\bf
v}+\omega')}.
\label{lb17}
\end{equation}
Considering only the contributions that do not decay in time, 
it can be shown \cite{pitaevskii} that $\lbrack ({\bf k}\cdot {\bf
v}-\omega)({\bf k}\cdot {\bf v}+\omega')\rbrack^{-1}$ can be substituted by
$(2\pi)^2\delta(\omega+\omega')\delta({\bf k}\cdot {\bf
v}-\omega)$.\footnote{See also Appendix A of \cite{bft2} for a more precise
justification of this procedure through a detailed calculation of the
integral (Laplace
transform) obtained by substituting Eq. (\ref{lb14}) with Eq. (\ref{lb17}) into
Eq. (\ref{lb13}).} Then, using
the property $\epsilon(-{\bf k},-\omega)=\epsilon({\bf k},\omega)^*$, one finds
that the correlations of the fluctuations of the potential are given by
\begin{equation}
\overline{\delta\tilde\Phi({\bf k},\omega)\delta\tilde\Phi({\bf
k}',\omega')}=(2\pi)^{d+2}\epsilon_{r}^{d}\epsilon_{v}^{d}\frac{\hat{u}(k)^2}{
|\epsilon({\bf k},\omega)|^2}\delta({\bf k}+{\bf k}')\delta(\omega+\omega')\int 
\delta({\bf k}\cdot {\bf v}-\omega)  f_2({\bf v})\, d{\bf v}.
\label{lb18}
\end{equation}
From Eq. (\ref{lb18}), we get the contribution to Eq. (\ref{lb13}) of the first
term of Eq. (\ref{lb14}). This yields the diffusion term  
\begin{equation}
(\overline{\delta f \nabla\delta\Phi})_i^{\rm Diff}=-i (2\pi)^{d+1}
\epsilon_{r}^{d}\epsilon_{v}^{d} \int d{\bf k}\int_{\cal
C}\frac{d\omega}{2\pi}\int  d{\bf v}'  \, {k}_i \frac{{\bf k}\cdot
\frac{\partial \overline{f}}{\partial {\bf v}}}{{\bf k}\cdot {\bf v}-\omega}
\frac{\hat{u}(k)^2}{|\epsilon({\bf k},\omega)|^2}f_2({\bf v}')\delta({\bf
k}\cdot {\bf v}'-\omega).
\label{lb20}
\end{equation}
Using the Landau prescription $\omega\rightarrow \omega+i0^+$ and the Plemelj
formula,
\begin{equation}
\frac{1}{x\pm i 0^+}={\cal P}\left (\frac{1}{x}\right )\mp i\pi\delta(x),
\label{lb21}
\end{equation}
where ${\cal P}$ denotes the principal value, we can replace $1/({\bf k}\cdot
{\bf v}-\omega-i0^+)$ by $+i\pi\delta({\bf k}\cdot {\bf v}-\omega)$. Then,
integrating over $\omega$,
we obtain
\begin{equation}
(\overline{\delta f \nabla\delta\Phi})_i^{\rm Diff}=\pi (2\pi)^{d}
\epsilon_{r}^{d}\epsilon_{v}^{d} \int d{\bf k}\, d{\bf v}'  \, k_ik_j  
\frac{\hat{u}(k)^2}{|\epsilon({\bf k},{\bf k}\cdot {\bf v})|^2}\delta\lbrack
{\bf k}\cdot ({\bf v}-{\bf v}')\rbrack f_2({\bf v}')\frac{\partial
\overline{f}}{\partial v_j}({\bf v}).
\label{lb23}
\end{equation}

\subsection{Friction}

Proceeding similarly, we obtain
\begin{equation}
\frac{\overline{\delta\hat f({\bf k},{\bf v},0)\delta\tilde\Phi({\bf
k}',\omega')}}{i({\bf k}\cdot {\bf
v}-\omega)}=(2\pi)^{d}
\frac{\hat{u}(k')}{\epsilon({\bf k}',\omega')}\frac{1}{i({\bf k}\cdot {\bf
v}-\omega)}\int d{\bf v}'\frac{\overline{\delta\hat f({\bf k},{\bf
v},0)\delta\hat f({\bf k}',{\bf v}',0)}}{i({\bf k}'\cdot {\bf
v}'-\omega')}
\label{lb19q1}
\end{equation}
and
\begin{equation}
\frac{\overline{\delta\hat f({\bf k},{\bf v},0)\delta\tilde\Phi({\bf
k}',\omega')}}{i({\bf k}\cdot {\bf
v}-\omega)}=\epsilon_r^d
\epsilon_v^d 
\frac{\hat{u}(k')}{\epsilon({\bf k}',\omega')} \delta({\bf k}+{\bf
k}')\frac{1}{i({\bf k}\cdot {\bf
v}-\omega)}\frac{f_2({\bf v})}{i({\bf k}'\cdot {\bf
v}-\omega')}.
\label{lb19q2}
\end{equation}
Considering only the
contributions that do not decay in time, 
it can be shown \cite{pitaevskii} (see footnote 14 with Eq.
(\ref{lb17}) replaced by Eq. (\ref{lb19q2})) that Eq.
(\ref{lb19q2}) can
be
substituted by
\begin{equation}
\frac{\overline{\delta\hat f({\bf k},{\bf v},0)\delta\tilde\Phi({\bf
k}',\omega')}}{i({\bf k}\cdot {\bf
v}-\omega)}=(2\pi)^{2}\epsilon_{r}^{d}\epsilon_{v}^{d} 
\frac{\hat{u}(k')}{\epsilon({\bf k}',\omega')}\delta({\bf k}+{\bf
k}')\delta(\omega+\omega')   \delta({\bf k}\cdot {\bf v}-\omega)f_2({\bf v}).
\label{lb19}
\end{equation}
From Eq. (\ref{lb19}), we get the contribution to Eq. (\ref{lb13}) of the second
term of Eq. (\ref{lb14}). This yields the friction term
\begin{equation}
(\overline{\delta f \nabla\delta\Phi})_i^{\rm Fric}=
\epsilon_{r}^{d}\epsilon_{v}^{d} \int d{\bf k} \, {k}_i
\frac{\hat{u}(k)}{|\epsilon({\bf k},{\bf k}\cdot {\bf v})|^2}{\rm Im}\,
\epsilon({\bf k},{\bf k}\cdot {\bf v})  f_2({\bf v}).
\label{lb24}
\end{equation}
Using the Landau prescription $\omega\rightarrow \omega+i0^+$ and the Plemelj
formula (\ref{lb21}), 
the imaginary part of the dielectric function (\ref{lb12}) reads
\begin{equation}
{\rm Im}\, \epsilon({\bf k},\omega)=-\pi(2\pi)^d\hat{u}(k)\int {\bf
k}\cdot\frac{\partial \overline{f}}{\partial {\bf v}}\delta({\bf k}\cdot {\bf
v}-\omega)\, d{\bf v}.
\label{lb25}
\end{equation} Substituting this expression into Eq. (\ref{lb24}), we obtain
\begin{equation}
(\overline{\delta f \nabla\delta\Phi})_i^{\rm Fric}=-\pi (2\pi)^{d}
\epsilon_{r}^{d}\epsilon_{v}^{d}   \int d{\bf k}\, d{\bf v}'  \, k_ik_j  
\frac{\hat{u}(k)^2}{|\epsilon({\bf k},{\bf k}\cdot {\bf v})|^2}\delta\lbrack
{\bf k}\cdot ({\bf v}-{\bf v}')\rbrack f_2({\bf v})\frac{\partial
\overline{f}}{\partial {v'}_j}({\bf v}').
\label{lb26}
\end{equation}

\subsection{Collision term: Kadomtsev-Pogutse
(1970) equation}

Regrouping Eqs. (\ref{lb3}), (\ref{lb23}) and (\ref{lb26}), we end up with
the kinetic equation
\begin{equation}
\frac{\partial \overline{f}}{\partial t}=\pi
(2\pi)^{d}\epsilon_{r}^{d}\epsilon_{v}^{d}  \frac{\partial}{\partial v_i} \int
d{\bf k} \, d{\bf v}'  \, k_ik_j  \frac{\hat{u}(k)^2}{|\epsilon({\bf k},{\bf
k}\cdot {\bf v})|^2}\delta\lbrack {\bf k}\cdot ({\bf v}-{\bf v}')\rbrack\left
(f'_2 \frac{\partial \overline{f}}{\partial {v}_{j}}-f_2
\frac{\partial \overline{f}'}{\partial {v'}_{j}}\right ),
\label{lb27}
\end{equation}
where $\overline{f}=\overline{f}({\bf v},t)$, $\overline{f}'=\overline{f}({\bf
v}',t)$, $f_2=f_2({\bf v},t)$, and $f'_2=f_2({\bf v}',t)$. This equation,
which takes collective effects into account, was first derived by Kadomtsev and
Pogutse \cite{kp}.\footnote{It is closely related to an
equation previously derived by Dupree \cite{dupree}.} It is formally
similar to
the Lenard-Balescu equation of the collisional theory (see, e.g., \cite{epjp}),
except that $m f$ in the
Lenard-Balescu equation is replaced by $\epsilon_{r}^{d}\epsilon_{v}^{d} f_2$ in
the KP equation. We also note that, contrary to the Lenard-Balescu equation, the
KP equation is not closed in the
general case since it involves the variance
$f_2$ of the fine-grained distribution. It can be closed exactly only in the
two-level case (see Sec. \ref{sec_two}).

If we neglect collective effects and take $|\epsilon({\bf k},{\bf k}\cdot {\bf
v})|=1$, Eq. (\ref{lb27}) reduces to
\begin{equation}
\frac{\partial \overline{f}}{\partial t}=\pi (2\pi)^{d}
\epsilon_{r}^{d}\epsilon_{v}^{d}\frac{\partial}{\partial v_i} \int d{\bf k} \,
d{\bf v}'  \, k_ik_j  \hat{u}(k)^2\delta\lbrack {\bf k}\cdot ({\bf v}-{\bf
v}')\rbrack \left
(f'_2 \frac{\partial \overline{f}}{\partial {v}_{j}}-f_2
\frac{\partial \overline{f}'}{\partial {v'}_{j}}\right ).
\label{lb28}
\end{equation}
The integral over ${\bf k}$ can be performed explicitly (see, e.g.,
\cite{kindetail}) and we obtain
\begin{equation}
\frac{\partial \overline{f}}{\partial t}=K_d\frac{\partial}{\partial v_i}\int
d{\bf v}' 
\frac{w^2\delta_{ij}-w_iw_j}{w^3}\left
(f'_2 \frac{\partial \overline{f}}{\partial {v}_{j}}-f_2
\frac{\partial \overline{f}'}{\partial {v'}_{j}}\right ),
\label{lb29}
\end{equation}
where ${\bf w}={\bf v}-{\bf v}'$ is the relative velocity and $K_d$ is a 
constant with value $K_3=8\pi^5
\epsilon_{r}^{3}\epsilon_{v}^{3}\int_0^{+\infty}k^3\hat{u}(k)^2\, dk$ in $d=3$
and  $K_2=8\pi^3
\epsilon_{r}^{2}\epsilon_{v}^{2}\int_0^{+\infty}k^2\hat{u}(k)^2\, dk$ in $d=2$.
This kinetic equation, that neglects collective effects, is formally similar
to the Landau equation 
\cite{landau} in the collisional theory (see, e.g., \cite{epjp}) with the
substitution $m f\rightarrow \epsilon_{r}^{d}\epsilon_{v}^{d} f_2$. For
a 3D plasma, using $(2\pi)^3\hat{u}(k)=4\pi
e^2/m^2k^2$, we get $K_3=(2\pi e^4/m^3)\ln\Lambda$ where
$\ln\Lambda=\int_0^{+\infty} dk/k$ is the Coulomb logarithm that has to be
regularized with appropriate cut-offs. The large-scale cut-off is the Debye
length $\lambda_D$ (the Debye length appears
naturally when we take into account collective effects) and the small-scale
cut-off is the
spatial resolution scale
$\epsilon_r$ which replaces the Landau
length $\lambda_{L}$ in the
collisional theory. This yields 
$\ln\Lambda=\ln({\lambda_D}/{\epsilon_r})$.  For a
2D plasma, using $(2\pi)^2\hat{u}(k)=2\pi e^2/m^2k^2$ and introducing a
large-scale cut-off at the Debye length, we obtain $K_2=2\pi e^4/m^3 k_D$.
There is no need to introduce a small-scale cut-off in that case but the
integration should be stoped at $\epsilon_r$ in principle. 
Returning to Eqs. (\ref{lb27}) and (\ref{lb28}), we note that collective effects
can be taken into account simply by replacing the bare potential $\hat{u}(k)$ in
the Landau equation by a ``dressed'' potential
$\hat{u}_{\rm d}(k)=\hat{u}(k)/|\epsilon({\bf k},{\bf k}\cdot {\bf v})|$,
including the dielectric function, without changing the overall structure of the
kinetic equation. Physically, this means that the particles are ``dressed'' by
their polarization cloud. In plasma physics, collective effects are important
because they account for screening effects and regularize, at the scale of the
Debye length, the logarithmic divergence that occurs in the Landau equation.
This avoids the introduction of {\it ad hoc} cut-offs at large scales.

We can write the kinetic equation (\ref{lb27}) in the compact form
\begin{equation}
\frac{\partial \overline{f}}{\partial t}=\frac{\partial}{\partial v_i}  \int
d{\bf v}'\, K_{ij} \, \left
(f_2'\frac{\partial \overline{f}}{\partial {v}_{j}}-f_2\frac{\partial
\overline{f}'}{\partial {v'}_{j}}\right )
\label{lb27a}
\end{equation}
by introducing the tensor 
\begin{equation}
K_{ij}=\pi (2\pi)^{d}\epsilon_{r}^{d}\epsilon_{v}^{d} \int
d{\bf k}  \, k_ik_j  \frac{\hat{u}(k)^2}{|\epsilon({\bf k},{\bf
k}\cdot {\bf v})|^2}\delta\lbrack {\bf k}\cdot ({\bf v}-{\bf v}')\rbrack. 
\label{lb27b}
\end{equation}
We note that it satisfies the identity $K_{ij}w_j=0$. If we neglect collective
effects, the tensor $K_{ij}$ is explicitly given by (see, e.g.,
\cite{kindetail})
\begin{eqnarray}
\label{E46}
K_{ij}^{\rm bare}=K_d{1\over w}\biggl
(\delta_{ij}-{w_{i}w_{j}\over w^{2}}\biggr ).
\end{eqnarray}
We also note that the kinetic equation (\ref{lb27a}) has the structure of a
generalized Fokker-Planck equation involving a diffusion term and a friction
term.

{\it Remark:} An equation similar to the KP equation has been
introduced in 2D turbulence \cite{quasi,prep} (see also Eq. (148) of
\cite{kinpre} and Eq. (88) of \cite{bbgkyvortex}). It reads
\begin{eqnarray}
\frac{\partial\overline{\omega}}{\partial
t}=2\pi^2\epsilon_r^2\frac{1}{r}\frac{\partial}{\partial
r}\sum_n\int_0^{+\infty}r'dr'\, |n|
|G(n,r,r',n\Omega)|^2\delta(\Omega-\Omega')\left
(\omega'_2\frac{1}{r}\frac{\partial\overline{\omega}}{\partial
r}-\omega_2\frac{1}{r'}\frac{\partial\overline{\omega}'}{\partial r'}\right ).
\end{eqnarray}

\subsection{3D self-gravitating systems}
\label{sec_3d}

Systems with attractive long-range interactions are generically
spatially inhomogeneous. It is important to develop a kinetic theory of
collisionless relaxation for such systems. If we implement a quasilinear
approximation and
neglect collective effects, we can derive
a generalized kinetic equation of the form \cite{chavmnras,dubrovnik,kingen}
\begin{eqnarray}
\frac{\partial \overline{f}}{\partial t}+{\bf v}\cdot \frac{\partial
\overline{f}}{\partial {\bf r}}- \nabla\overline{\Phi}\cdot \frac{\partial
\overline{f}}{\partial {\bf v}}=\epsilon_r^d\epsilon_v^d\frac{\partial}{\partial
{v}_{i}}\int_0^{t} ds\,  \int d{\bf
r}'d{\bf v}'\, {F}_{i}({\bf r}'\rightarrow {\bf r})_t
{F}_{j}({\bf r}'\rightarrow {\bf r})_{t-s}  \left
(f_2'\frac{\partial \overline{f}}{\partial {v}_{j}}-f_2\frac{\partial
\overline{f}'}{\partial {v'}_{j}}\right )_{t-s} 
\label{gkin}
\end{eqnarray}
that is valid for systems
that are not necessarily spatially homogeneous
and not necessarily Markovian. Here, $\overline{f}=\overline{f}({\bf r},{\bf
v},t)$,
$\overline{f}'=\overline{f}({\bf r}',{\bf
v}',t)$, $f_2=f_2({\bf r},{\bf v},t)$ and $f'_2=f_2({\bf r}',{\bf
v}',t)$. On the other hand, ${\bf F}({\bf r}'\rightarrow {\bf r})_t$ denotes the
force by unit
of mass exerted by a particle located in ${\bf r}'$ on a particle located in
${\bf r}$ at time $t$. In certain
cases, e.g. for 3D self-gravitating systems, we can make a {\it local
approximation} and proceed as if the system were spatially homogeneous. If we
also implement a {\it Markovian approximation}, the
kinetic equation (\ref{gkin}) is replaced by
\begin{eqnarray}
\frac{\partial \overline{f}}{\partial t}+{\bf v}\cdot \frac{\partial
\overline{f}}{\partial {\bf r}}- \nabla\overline{\Phi}\cdot \frac{\partial
\overline{f}}{\partial {\bf v}}=\epsilon_r^3\epsilon_v^3\frac{\partial}{\partial
{v}_{i}}\int_0^{+\infty} ds\,  \int d{\bf
r}'d{\bf v}'\, {F}_{i}({\bf r}'\rightarrow {\bf r})_t
{F}_{j}({\bf r}'\rightarrow {\bf r})_{t-s}  \left
(f_2'\frac{\partial \overline{f}}{\partial {v}_{j}}-f_2\frac{\partial
\overline{f}'}{\partial {v'}_{j}}\right )_{t}, 
\end{eqnarray}
where now $\overline{f}=\overline{f}({\bf r},{\bf
v},t)$,
$\overline{f}'=\overline{f}({\bf r},{\bf
v}',t)$, $f_2=f_2({\bf r},{\bf v},t)$ and $f'_2=f_2({\bf r},{\bf
v}',t)$. Passing in Fourier space, we obtain \cite{kindetail}
\begin{equation}
\frac{\partial \overline{f}}{\partial t}+{\bf v}\cdot \frac{\partial
\overline{f}}{\partial {\bf r}}- \nabla\overline{\Phi}\cdot \frac{\partial
\overline{f}}{\partial {\bf
v}}=\pi(2\pi)^3\epsilon_{r}^{3}\epsilon_{v}^{3}\frac{\partial}{\partial v_i} 
\int d{\bf
k} \, d{\bf v}'  \, k_ik_j \hat{u}(k)^2\delta\lbrack
{\bf k}\cdot ({\bf v}-{\bf v}')\rbrack\left (f'_2\frac{\partial
\overline{f}}{\partial {v}_{j}}-f_2\frac{\partial\overline{f}'}{\partial
{v'}_{j}}\right ).
\label{lb31}
\end{equation}
In this equation, the effects of spatial inhomogeneity are kept only in the 
advection (Vlasov) term, while the collision term is calculated as if the system
were spatially homogeneous. The integral over
${\bf k}$ can be performed explicitly  \cite{kindetail} and the foregoing
equation can be
rewritten as
\begin{equation}
\frac{\partial \overline{f}}{\partial t}+{\bf v}\cdot \frac{\partial
\overline{f}}{\partial {\bf r}}- \nabla\overline{\Phi}\cdot \frac{\partial
\overline{f}}{\partial {\bf v}}=K_3\frac{\partial}{\partial
v_i}\int d{\bf v}' \frac{w^2\delta_{ij}-w_iw_j}{w^3}\left (f'_2\frac{\partial
\overline{f}}{\partial {v}_{j}}-f_2\frac{\partial\overline{f}'}{\partial
{v'}_{j}}\right ).
\label{lb32}
\end{equation}
For a 3D self-gravitating system, using
$(2\pi)^3\hat{u}(k)=-4\pi G/k^2$, we get $K_3=2\pi
\epsilon_r^3\epsilon_v^3G^2\ln\Lambda$ where
$\ln\Lambda=\int_0^{+\infty} dk/k$ is the Coulomb factor that has to be
regularized with appropriate cut-offs. The large-scale cut-off is the Jeans
length $\lambda_J$ (which is of the order of the system size
$R$) and the small-scale cut-off is the spatial resolution scale
$\epsilon_r$ which replaces the gravitational Landau length
$\lambda_{L}$ in the
collisional theory. This yields 
$\ln\Lambda=\ln({\lambda_J}/{\epsilon_r})$.

{\it Remark:} We can also extend the KP equation to spatially inhomogeneous
systems by making a local approximation. In this manner, we can
heuristically take collective
effects into account. In Sec. \ref{sec_is}, we introduce more general kinetic
equations written with angle-action variables that take into account spatial
inhomogeneity and collective effects without relying on a local
approximation.

\section{Multi-level case}
\label{sec_mlc}

In the general case, the kinetic equation (\ref{lb27}) for the coarse-grained DF
$\overline{f}({\bf v},t)$ is not
closed because it depends on the local centered variance $f_2({\bf
v},t)$ of
the distribution $\rho({\bf v},\eta,t)$. It is therefore necessary to
extend the kinetic theory to the multi-level case and work in term of the DF
$\rho({\bf v},\eta,t)$ for each level.

\subsection{Severne-Luwel (1980) equation}
\label{sec_sl}

A kinetic equation for $\rho({\bf v},\eta,t)$ has been
derived by
Severne and Luwel \cite{sl} by generalizing the quasilinear theory of Kadomtsev
and Pogutse \cite{kp} to the multi-level
case.\footnote{The derivation is similar to the one detailed in
Sec. \ref{sec_whole}. In the
multi-level case, one has to work in terms of $\rho_e({\bf r},{\bf
v},\eta,t)=\delta(f({\bf r},{\bf v},t)-\eta)$ and $\rho({\bf
v},\eta,t)=\langle \delta(f({\bf r},{\bf v},t)-\eta)\rangle$ which generalize
$f({\bf r},{\bf v},t)$ and $\overline{f}({\bf v},t)$, respectively.  Equation
(\ref{E40}) is generalized into 
\begin{equation}
\overline{\delta\rho({\bf r},{\bf v},\eta,t)\delta\rho({\bf r}',{\bf
v}',\eta',t')}=\epsilon_r^d\epsilon_v^d \rho({\bf r},{\bf
v},\eta',t)\lbrack \delta(\eta-\eta')-\rho({\bf r},{\bf
v},\eta,t)\rbrack\delta({\bf r}-{\bf r}')\delta({\bf v}-{\bf v}'),
\label{sl}
\end{equation}
corresponding to Eq. (3.13) of
\cite{sl}.} They
obtained an equation of the
form\footnote{Severne and Luwel \cite{sl} developed their
theory for spatially inhomogeneous stellar systems. However, at the end of their
calculations, in order to obtain an explicit kinetic equation, they made a local
approximation which amounts to proceeding as if the system were infinite and
homogeneous (see Sec. \ref{sec_3d}).}
\begin{equation}
\frac{\partial \rho}{\partial t}=\frac{\partial}{\partial v_i}  \int d{\bf v}'\,
K_{ij} \, \left \lbrack f_2'\frac{\partial \rho}{\partial
{v}_{j}}-\rho(\eta-\overline{f})\frac{\partial \overline{f}'}{\partial
{v'}_{j}}\right \rbrack.
\label{sl1}
\end{equation}
We can check that the normalization condition ($\int \rho\, d\eta=1$) is
preserved with time and that the equation for the coarse-grained DF 
$\overline{f}=\int \rho \eta\,
d\eta$ returns Eq. (\ref{lb27a}). We can also show that the SL equation
conserves the
energy and all the Casimirs, that it satisfies and $H$-theorem for the
Lynden-Bell entropy (\ref{E9}), and that it relaxes towards the Gibbs state
(\ref{E11}) (see Appendix \ref{sec_propkpsl}).

From the SL equation, we can derive a hierarchy of equations for the moments
$\overline{f^{n}}=\int \rho \eta^{n} \, d\eta$ of the distribution. The general
term of this hierarchy is
\begin{equation}
\frac{\partial \overline{f^n}}{\partial t}=\frac{\partial}{\partial v_i}  \int
d{\bf v}'\,
K_{ij} \, \left \lbrack f_2'\frac{\partial \overline{f^n}}{\partial
{v}_{j}}-(\overline{f^{n+1}}-\overline{f^{n}}\,\,\, \overline{f})\frac{\partial
\overline{f}'}{\partial
{v'}_{j}}\right \rbrack.
\label{sl2}
\end{equation}
For $n=1$ we recover the KP equation [Eq. (\ref{lb27a})].

{\it Remark:} An equation similar to the SL equation can be
derived in 2D turbulence \cite{prep}. It reads 
\begin{eqnarray}
\frac{\partial\rho}{\partial
t}=2\pi^2\epsilon_r^2\frac{1}{r}\frac{\partial}{\partial
r}\sum_n\int_0^{+\infty}r'dr'\, |n|
|G(n,r,r',n\Omega)|^2\delta(\Omega-\Omega')\left
\lbrack \omega'_2\frac{1}{r}\frac{\partial\rho}{\partial
r}-\rho(\sigma-\overline{\omega})\frac{1}{r'}\frac{\partial\overline{\omega}'}{
\partial r'}\right \rbrack.
\end{eqnarray}

\subsection{Chavanis-Sommeria-Robert (1996) equation}
\label{sec_csr}

By using a different approach based on a Maximum Entropy Production
Principle (MEPP), Chavanis, Sommeria and Robert \cite{csr} have proposed
an equation for $\rho({\bf v},\eta,t)$ that relaxes towards the
Lynden-Bell distribution.\footnote{The MEPP can be formulated in
the inhomogeneous case, as discussed in Appendix \ref{sec_pcsr}, but in this
section we consider spatially homogeneous systems.} This
phenomenological equation is expected to describe
the whole process of violent relaxation, including the very nonlinear
early regime. In the late regime of
quiescent relaxation, it is possible to connect the CSR equation to the KP and
SL equations as follows.\footnote{This connection was made in
\cite{chavmnras,dubrovnik} in the
two-level case and is extended here to the multi-level case.}

The kinetic equation (\ref{sl1}) obtained from the quasilinear theory of
the Vlasov equation is an integrodifferential equation. The current in ${\bf v}$
depends on the value of $\overline{f}$ and  $f_2$ in ${\bf v}'$ through an
integral over ${\bf v}'$.  We can transform the integrodifferential equation
(\ref{sl1}) into a
differential equation by replacing $\overline{f}'=\overline{f}({\bf v}',t)$ and 
${f}'_2={f}_2({\bf v}',t)$ by their equilibrium values obtained from the
Gibbs state (\ref{E11}). This amounts to making a thermal bath
approximation. Using
the identity from Eq. (\ref{g1}), we find
that
\begin{equation}
\frac{\partial\overline{f}'}{\partial {\bf v}'}=-\beta f_2' {\bf v}'.
\label{csr1}
\end{equation}
Substituting Eq. (\ref{csr1}) into Eq. (\ref{sl1}) we get
\begin{equation}
\frac{\partial \rho}{\partial t}=\frac{\partial}{\partial v_i}  \int d{\bf v}'\,
K_{ij} \, \left \lbrack f_2'\frac{\partial \rho}{\partial
{v}_{j}}+\beta f_2' \rho(\eta-\overline{f}){v}_j'\right
\rbrack.
\label{csr2}
\end{equation}
In principle, $f'_2$ should be calculated at equilibrium. However,
in order to
be more general, we shall evaluate ${f}_2$ in Eq. (\ref{csr2}) at time $t$, not
at equilibrium. 
Using the identity $K_{ij}(v_j-v'_j)=0$, we can replace ${v}_j'$ by ${v}_j$
in the last term of Eq. (\ref{csr2}). In this manner, we obtain 
\begin{equation}
\frac{\partial \rho}{\partial t}=\frac{\partial}{\partial v_i} \left\lbrace
D_{ij} \left \lbrack \frac{\partial \rho}{\partial
{v}_{j}}+\beta \rho(\eta-\overline{f}){v}_j\right
\rbrack\right\rbrace
\label{csr3}
\end{equation}
with 
\begin{equation}
D_{ij}=\int d{\bf v}'\,
K_{ij} f_2'.
\label{csr4}
\end{equation}
This equation can be viewed as a generalized Fokker-Planck equation of
the Kramers type involving a
diffusion
term and a friction term. The friction term is the counterpart
of Chandrasekhar's dynamical friction \cite{chandra1} in the kinetic theory of
collisional
stellar systems. The friction coefficient is given by a form
of Einstein relation $\xi_{ij}=D_{ij}\beta\eta$.

By making the thermal bath approximation from
Eq. (\ref{csr1}), we have lost the conservation of energy. Indeed, we have
passed
from a microcanonical description (conservation of energy $E$ and $H$-theorem
for
the entropy $S$) to a canonical description (fixed temperature $T$ and
$H$-theorem for
the free energy $F=E-TS$). However, the conservation of energy can be
artificially restored by letting $\beta(t)$
depend on time in a suitable manner. To that purpose, using Eq. (\ref{csr3}), we
first write the
equation for the coarse-grained DF 
$\overline{f}=\int \rho \eta\,
d\eta$ which reads
\begin{equation}
\frac{\partial \overline{f}}{\partial t}=\frac{\partial}{\partial v_i}
\left\lbrace
D_{ij} \left \lbrack \frac{\partial \overline{f}}{\partial
{v}_{j}}+\beta f_2 {v}_j\right
\rbrack\right\rbrace.
\label{csr5}
\end{equation}
This equation can also be obtained by applying the thermal bath approximation
(\ref{csr1}) to
Eq. (\ref{lb27a}). Then, we compute
\begin{equation}
\dot E=\int \frac{\partial \overline{f}}{\partial t}
\frac{v^2}{2}\, d{\bf v}=-\int \frac{\partial {\bf J}_f}{\partial {\bf v}} 
\frac{v^2}{2}\, d{\bf v}=\int {\bf J}_f\cdot {\bf v}\, d{\bf v}=-\int
D_{ij} \left \lbrack \frac{\partial \overline{f}}{\partial
{v}_{j}}+\beta f_2 {v}_j\right
\rbrack v_i\,
d{\bf v}.
\label{csr6}
\end{equation}
We can enforce the conservation of energy ($\dot E=0$) by taking
\begin{equation}
\beta(t) =-\frac{\int  D_{ij} v_i \frac{\partial \overline{f}}{\partial
{v}_j}\, d{\bf v}}{\int D_{ij} f_2 v_i v_j \,
d{\bf v}}.
\label{csr7}
\end{equation}
This approach  returns the CSR equations formed by Eqs.
(\ref{csr3})
and (\ref{csr7}) above. It also provides the
explicit expression (\ref{csr4}) of the diffusion coefficient -- actually a
tensor $D_{ij}$ -- which was not given by the MEPP \cite{csr}. Equation
(\ref{csr3}) with
Eqs. (\ref{csr4}) and (\ref{csr7}) is an integrodifferential equation since
$\beta$ and
$D_{ij}$ are expressed as integrals of $\overline{f}$ and $f_2$, but it is
simpler than
the KP and SL equations. We can show  (see Ref. \cite{csr} and
Appendix \ref{sec_gcsr}) that the CSR equations 
conserve the energy and all the Casimirs, and that they
monotonically increase
the Lynden-Bell entropy
($H$-theorem). They usually relax towards the Lynden-Bell
DF except in the cases reported in Sec. \ref{sec_incompletekin} where 
the diffusion tensor $D_{ij}$ vanishes.

From the CSR equation (\ref{csr3}), we can derive a hierarchy of equations
for the moments
$\overline{f^{n}}=\int \rho \eta^{n} \, d\eta$ of the distribution. The general
term of this
hierarchy  is
\begin{equation}
\frac{\partial \overline{f^{n}}}{\partial t}=\frac{\partial}{\partial v_i}
\left\lbrace
D_{ij} \left \lbrack \frac{\partial \overline{f^{n}}}{\partial
{v}_{j}}+\beta (\overline{f^{n+1}}-\overline{f^n}\,\,\,  
\overline{f}){v}_j\right
\rbrack\right\rbrace.
\label{csr8}
\end{equation}
For $n=1$ we recover Eq. (\ref{csr5}). This hierarchy of equations can also be
obtained by applying the thermal bath approximation (\ref{csr1}) to
Eq. (\ref{sl2}).

{\it Remark:} In the context of 2D turbulence, the MEPP has been introduced by
Robert and Sommeria \cite{rsmepp}. It can be viewed as a variational
formulation of the linear thermodynamics of Onsager
\cite{onsager31a,onsager31b,om} (see \cite{nfp,entropy2}).
The relaxation equations for the
coarse-grained vorticity involves a diffusion term and a drift
term \cite{rsmepp,csr,csmepp}. The drift is the counterpart of Chandrasekhar's
dynamical friction \cite{drift,kinpre,houchesPH}. The moments of
the relaxation equation have been  derived in
\cite{rr,kazantsev,cnd,vphydro}. The
connection between the quasilinear theory of the 2D Euler equation and the MEPP
has been discussed by Chavanis \cite{quasi}. This connection allowed us to
compute the diffusion coefficient of 2D vortices (not given by the MEPP) and
recover the heuristic expression given in Ref. \cite{rr}. A more detailed
discussion of the quasilinear theory of the 2D Euler equation is
given in \cite{prep}. 

\subsection{Two-level case: Fermionic-like equations}
\label{sec_two}

If the initial condition in phase
space consists of patches of uniform DF
$f=\eta_{0}$ surrounded by vacuum $f=0$ (two-level
approximation), we can write $\overline{f^{2}}=\overline{\eta_{0}{\times} f}=
\eta_{0}\overline{f}$ yielding
\begin{equation}
f_{2}= \overline{f} (\eta_{0}-\overline{f}).
\label{lb27c}
\end{equation}
In that case, the local centered variance $f_2$ of the distribution can be
trivially related to the
coarse-grained DF $\overline{f}$. We see how the self-correlations and the
incompressibility of the flow in phase space give rise to an effective
``exclusion principle''. Substituting Eq. (\ref{lb27c}) into Eq.
(\ref{lb27a}), we obtain a kinetic equation of the form
\begin{equation}
\frac{\partial \overline{f}}{\partial t}=\frac{\partial}{\partial v_i}  \int
d{\bf v}'\, K_{ij} \, \left \lbrack\overline{f}'
(\eta_{0}-\overline{f}')\frac{\partial \overline{f}}{\partial
{v}_{j}}-\overline{f} (\eta_{0}-\overline{f})\frac{\partial
\overline{f}'}{\partial {v'}_{j}}\right \rbrack.
\label{lb27d}
\end{equation}
This equation is similar to the fermionic Landau (or Lenard-Balescu)
equation \cite{kp,sl,chavmnras,dubrovnik,kingen,kinquant}. It conserves the mass
and the 
energy and  monotonically increases the
Lynden-Bell (or Fermi-Dirac-like) entropy \cite{kinquant}. In the nondegenerate
limit
$\overline{f}\ll \eta_0$, it becomes similar to the classical  Landau (or
Lenard-Balescu) equation associated with the Boltzmann entropy.

On the
other hand, the CSR equations  (\ref{csr3}), (\ref{csr4}) and (\ref{csr7})
reduce to 
\begin{equation}
\frac{\partial \overline{f}}{\partial t}=\frac{\partial}{\partial v_i}
\left\lbrace
D_{ij} \left \lbrack \frac{\partial \overline{f}}{\partial
{v}_{j}}+ \beta(t)\overline{f} (\eta_{0}-\overline{f}) {v}_j\right
\rbrack\right\rbrace
\label{two1}
\end{equation}
with 
\begin{equation}
D_{ij}=\int d{\bf v}'\,
K_{ij} \overline{f}' (\eta_{0}-\overline{f}')
\label{two2}
\end{equation}
and
\begin{equation}
\beta(t) =-\frac{\int  D_{ij} v_i \frac{\partial \overline{f}}{\partial
{v}_j}\, d{\bf v}}{\int D_{ij} \overline{f}
(\eta_{0}-\overline{f}) v_i v_j \,
d{\bf v}}.
\label{two3}
\end{equation}
Equation (\ref{two1}) is similar to the fermionic Kramers equation. It can be
obtained from Eq. (\ref{lb27d}) by using the thermal bath approximation
(\ref{csr1}). This equation has been derived and studied in
\cite{chavmnras,dubrovnik,kingen,kinquant}. It conserves the mass and
the energy and  monotonically increases the
Lynden-Bell (or Fermi-Dirac-like) entropy.  In the nondegenerate limit
$\overline{f}\ll \eta_0$, it becomes similar to the classical Kramers
equation \cite{lb}.

\subsection{Chavanis (2004) equation}
\label{sec_chav}

For more complicated initial conditions
(multi-level case), we have to solve the 
equation for $\rho({\bf r},{\bf v},\eta,t)$, the
probability  density of finding the phase level $\eta$ in $({\bf r},{\bf v})$
at time $t$ [see Eq. (\ref{sl1}) or Eq. (\ref{csr3})]. The strategy is to
discretize the initial condition into ${\cal N}$ levels $\eta$. This approach
then leads us to a closed system of ${\cal N}$ coupled equations (one for each
level $\eta$). However, for generic initial conditions, we have to deal
with a great number of levels and these equations are not convenient to solve
when
${\cal N}\gg 1$. We can
alternatively try to solve the hierarchy of equations for the moments
$\overline{f^{n}}$ [see Eq. (\ref{sl2}) or Eq. (\ref{csr8})]  but
we then encounter a difficult closure problem. In practice, we are mainly
interested in
the evolution of the first moment, namely the coarse-grained DF $\overline{f}$.
As we have seen, the equation for  $\overline{f}$ [see Eq. (\ref{lb27a}) or
Eq. 
(\ref{csr5})] depends on the variance $f_{2}$. In order
to obtain a
self-consistent kinetic equation for $\overline{f}$, we need to
relate the variance $f_{2}$ to the coarse-grained DF $\overline{f}$. In Ref.
\cite{kingen}, we have
proposed a closure approximation that leads to a
simple kinetic equation. While not being exact, this equation
preserves the robust features of the process of violent relaxation and
is amenable to an easier numerical implementation. Its main interest
is to go beyond the two-level approximation while leaving the problem
tractable. The idea is to observe that Eqs. (\ref{g1}) and
(\ref{meta7b})
lead to the important relation \cite{gen,kingen}
\begin{equation}
\label{c1} 
f_{2}=\frac{1}{C''(\overline{f})}.
\end{equation}
This relation is valid at equilibrium but we propose to use it as a
closure approximation in Eqs.  (\ref{sl2}) and 
(\ref{csr8}). This is expected to be a
reasonable approximation if we are close to equilibrium, which is in
fact dictated by the quasilinear approximation. Of course, this
procedure assumes that we know the function $C(\overline{f})$ in advance. This
is
the case if we have already determined the equilibrium state by the procedure
discussed in Sec.
\ref{sec_ge} and we
want to describe the dynamics close to
equilibrium.\footnote{This is also the case when
the
generalized entropy $C(\overline{f})$ is determined by the
external forcing as discussed in footnotes 8 and 9.} A justification of the
closure
relation (\ref{c1}) is given in Appendix \ref{sec_out} following an argument
first given in Ref. \cite{physicaD1}.

If we close the hierarchy of equations (\ref{sl2})  with Eq. (\ref{c1}), we
obtain a
self-consistent kinetic equation of the form \cite{kingen}
\begin{equation}
\frac{\partial \overline{f}}{\partial t}=\frac{\partial}{\partial v_i}  \int
d{\bf v}'\, K_{ij} \, \left
\lbrack \frac{1}{C''(\overline{f}')}\frac{\partial \overline{f}}{\partial
{v}_{j}}-\frac{1}{C''(\overline{f})}\frac{\partial
\overline{f}'}{\partial {v'}_{j}}\right \rbrack.
\label{c2}
\end{equation}
It can be viewed as a generalized Landau (or Lenard-Balescu) equation.
It conserves the mass and the energy and monotonically increases the generalized
entropy
(\ref{meta4}) ($H$-theorem) \cite{kingen}.  In the two-level case, using
$f_2=1/C''(\overline{f})=\overline{f}
(\eta_{0}-\overline{f})$, we recover Eq.
(\ref{lb27d}). The generalized Landau equation (\ref{c2})
has been studied in detail in Ref. \cite{kingen}.

If we close the hierarchy of equations (\ref{csr8}) with Eq. (\ref{c1}), we
obtain a
self-consistent kinetic equation of the form \cite{gen}
\begin{equation}
\frac{\partial \overline{f}}{\partial t}=\frac{\partial}{\partial v_i}
\left\lbrace
D_{ij} \left \lbrack \frac{\partial \overline{f}}{\partial
{v}_{j}}+ \frac{\beta(t)}{C''(\overline{f})} {v}_j\right
\rbrack\right\rbrace
\label{c3}
\end{equation}
with 
\begin{equation}
D_{ij}=\int d{\bf v}'\,
K_{ij} \frac{1}{C''(\overline{f}')}
\label{c4}
\end{equation}
and
\begin{equation}
\beta(t) =-\frac{\int  D_{ij} v_i \frac{\partial \overline{f}}{\partial
{v}_j}\, d{\bf v}}{\int D_{ij} \frac{1}{C''(\overline{f})} v_i v_j \,
d{\bf v}}.
\label{c5}
\end{equation}
Equation
(\ref{c3})
can also be derived from Eq. (\ref{c2})  by making the
thermal bath approximation (\ref{csr1}). It can be viewed as a generalized
Kramers
equation. It conserves the mass and the energy and monotonically increases the
generalized
entropy
(\ref{meta4}) ($H$-theorem) \cite{gen}.  In two-level case, using
$f_2=1/C''(\overline{f})=\overline{f}
(\eta_{0}-\overline{f})$,  we recover Eqs.
(\ref{two1})-(\ref{two3}). The generalized Kramers equations
(\ref{c3})-(\ref{c5})
have been studied in detail in Ref. \cite{gen}.

{\it Remark:} Similar equations have been introduced in the context of 2D
turbulence in Refs. \cite{gen,cnd,vphydro,physicaD1,physicaD2,super}.

\subsection{One-dimensional systems}
\label{sec_oned}

For one-dimensional (1D) systems,
recalling the expression of $K_{ij}$ from Eq. (\ref{lb27b}), the SL equation
(\ref{sl1}) becomes
\begin{equation}
\frac{\partial \rho}{\partial
t}=2\pi^2\epsilon_{r}\epsilon_{v}\frac{\partial}{\partial v} 
\int d{v}'
d{k}  \, k^2  \frac{\hat{u}(k)^2}{|\epsilon({k},{k} {v})|^2}\delta\lbrack
k ({v}-{v}')\rbrack \left
\lbrack f_2'\frac{\partial \rho}{\partial
{v}}-\rho(\eta-\overline{f})\frac{\partial \overline{f}'}{\partial
{v'}}\right \rbrack.
\label{irk1b}
\end{equation}
Using the identity $\delta\lbrack k
({v}-{v}')\rbrack=\frac{1}{|k|}\delta(v-v')$, it can be rewritten as
\begin{equation}
\frac{\partial \rho}{\partial
t}=2\pi^2\epsilon_{r}\epsilon_{v}\frac{\partial}{\partial v} 
\int 
d{k}  \, |k|  \frac{\hat{u}(k)^2}{|\epsilon({k},{k}
{v})|^2} \left
\lbrack f_2\frac{\partial \rho}{\partial
{v}}-\rho(\eta-\overline{f})\frac{\partial \overline{f}}{\partial
{v}}\right \rbrack.
\label{irk1c}
\end{equation}
The corresponding hierarchy of moment equations takes the form
\begin{equation}
\frac{\partial \overline{f^n}}{\partial
t}=2\pi^2\epsilon_{r}\epsilon_{v}\frac{\partial}{\partial v} 
\int 
d{k}  \, |k|  \frac{\hat{u}(k)^2}{|\epsilon({k},{k}
{v})|^2} \left
\lbrack f_2\frac{\partial \overline{f^n}}{\partial
{v}}-(\overline{f^{n+1}}-\overline{f^n}\,\,\,  
\overline{f})\frac{\partial \overline{f}}{\partial
{v}}\right \rbrack.
\label{irk1d}
\end{equation}
For $n=1$ we get
\begin{equation}
\frac{\partial \overline{f}}{\partial
t}=0.
\label{irk1e}
\end{equation}
This result can be directly obtained from the KP equation (\ref{lb27}) applied
to 1D systems. It
implies that the coarse-grained DF does not change with time. However, the
higher moments evolve in
time. 

{\it Remark:} We note that the ``kinetic blocking'' of the coarse-grained DF in
1D occurs only for spatially homogeneous systems. For 1D
inhomogeneous systems, the  coarse-grained DF evolves in time according to the
inhomogeneous KP equation (\ref{is2}). On the other hand, for 1D
homogeneous systems, the CSR equation (\ref{csr5}) does not show such a
``kinetic blocking''. This may be related to the fact that
this equation is more justified in the phase of violent relaxation
than in the phase of quiescent relaxation.

\section{Incomplete violent relaxation}
\label{sec_incompletekin}

We have seen that the kinetic equation derived from the quasilinear theory
relaxes towards the Lynden-Bell distribution. In this sense, it provides a
justification of the 
maximum entropy principle and implies that the evolution is ergodic. However,  
direct numerical simulations of the Vlasov equation (or direct simulations of
the $N$-body problem performed in the collisionless regime) show that violent
relaxation is in general
incomplete \cite{lb,grand,incomplete,assisePH}. The fluctuations of the 
potential that
are
the engine of the collisionless relaxation can die out before the system has
reached the statistical equilibrium equilibrium state. How can we reconcile
these apparently contradictory results?

First, we have to recall that the  quasilinear theory, which is based on
the assumption that the nonlinear terms in the equation for the fluctuations can
be neglected,  describes only a regime of late quiescent
relaxation. Therefore, the relaxation toward the Lynden-Bell DF may be limited
to this ``gentle'' situation. In addition, we have assumed that the correlation
function is given by Eq. (\ref{E40}) in which  the resolution
scales $\epsilon_r$ and $\epsilon_v$ in position and velocity are constant in
time. This is also a strong assumption.
At the end of their paper, Kadomtsev and Pogutse \cite{kp} argue
that the scale of correlations may decrease in time as the variations of
the potential $\Phi$ decay. In that case,
the kinetic equation becomes
\begin{equation}
\frac{\partial \rho}{\partial t}=\pi
(2\pi)^{d}\epsilon_{r}(t)^{d}\epsilon_{v}(t)^{d}\frac{\partial}{\partial v_i} 
\int d{\bf v}'
d{\bf k}  \, k_ik_j  \frac{\hat{u}(k)^2}{|\epsilon({\bf k},{\bf
k}\cdot {\bf v})|^2}\delta\lbrack {\bf k}\cdot ({\bf v}-{\bf v}')\rbrack \left
\lbrack f_2'\frac{\partial \rho}{\partial
{v}_{j}}-\rho(\eta-\overline{f})\frac{\partial \overline{f}'}{\partial
{v'}_{j}}\right \rbrack,
\label{irk1}
\end{equation}
where $A(t)\equiv\epsilon_{r}(t)^{d}\epsilon_{v}(t)^{d}$ tends to zero for
$t\rightarrow
+\infty$. If the scale of correlations decreases rapidly in
time,\footnote{Kadomtsev and Pogutse \cite{kp} argue that $A(t)$
decreases like $t^{-1/2}$ or even more rapidly.} the
relaxation towards the Lynden-Bell distribution may be
inhibited. This
effect may account for incomplete relaxation.

The heuristic CSR approach  \cite{csr} aims at
describing
the very nonlinear regime of violent relaxation. It leads to a relaxation
equation of the form
\begin{equation}
{\partial\rho\over\partial t}+{\bf v}\cdot {\partial\rho\over\partial {\bf
r}}-\nabla\Phi\cdot {\partial\rho\over\partial {\bf v}}={\partial\over\partial
{\bf v}}\cdot \biggl\lbrace D({\bf r},{\bf v},t)\biggl\lbrack
{\partial\rho\over\partial {\bf v}}+\beta(t)(\eta-\overline{f})\rho {\bf
v}\biggr\rbrack\biggr\rbrace
\label{irk2}
\end{equation}
with a diffusion coefficient $D({\bf r},{\bf v},t)$ given by Eq. (\ref{csr4}).
The diffusion coefficient depends on the  local
centered variance $f_2$ of the distribution. Therefore, it vanishes in the
regions of phase space  where there are no fluctuations. Said
differently, the fluctuations $\delta\Phi$ of the potential
must be strong enough to provide an efficient mixing. The
vanishing of the diffusion coefficient can ``freeze'' the system in a
subdomain of phase space and account for incomplete relaxation and
non-ergodicity \cite{csr}.\footnote{This ``freezing'' has been observed
numerically in 2D
turbulence \cite{rr} and led to the concept of maximum entropy ``bubbles''
\cite{jfm2}. It is not clear if the vanishing of the diffusion coefficient in
certain regions of phase space completely stops the relaxation or simply slows
it down. However, if the relaxation is strongly slowed down (as observed in
\cite{rr}) the result is essentially the same on a practical point of view.}
For the same reason, the KP equation may also experience a process of incomplete
relaxation when
$f_2\rightarrow 0$.

This kinetic justification of incomplete relaxation is
interesting because
it is not based on a generalized entropy such as the Tsallis entropy (see Sec.
\ref{sec_incompleteq}), so it does not involve any free parameter like $q$
\cite{tsallis}.
However, it demands to solve a
dynamical equation [Eq. (\ref{irk1}) or Eq. (\ref{irk2}) with Eqs. (\ref{csr4})
and (\ref{csr7})] in
order to predict the incompletely mixed equilibrium
state reached by the system. The idea is that, in case of incomplete relaxation
(non-ergodicity), the prediction of the equilibrium state is
impossible without considering the dynamics \cite{incomplete}.

Finally, the kinetic theory allows us to take into account the evaporation of
high energy particles, which also prevents the relaxation of the system
towards a true (Lynden-Bell) statistical equilibrium state.\footnote{Evaporation
is particularly important for 3D self-gravitating systems since there is no
statistical equilibrium state in a strict sense. The Lynden-Bell DF coupled
to the Poisson equation has an infinite mass.} Truncated models
such as the fermionic King model have been derived in Refs.
\cite{chavmnras,kingen}.  These
truncated DFs differ from the Lynden-Bell DF and
have a finite mass contrary to the Lynden-Bell DF for 3D self-gravitating
systems.

{\it Remark:} Similar arguments have been developed in 2D turbulence to explain
the process of incomplete violent relaxation \cite{rr,csr,bbgkyvortex}.

\section{Collisionless versus collisional relaxation}
\label{sec_crt}

In this section, we discuss the collisionless relaxation time associated with
the KP equation (\ref{lb27}) and compare it with the collisional relaxation time
associated with the ordinary Lenard-Balescu equation \cite{epjp}. Our discussion
follows and
completes the discussion given in Refs. \cite{kp,sl,chavmnras}.

 The Lenard-Balescu equation describing the
evolution of the system sourced by finite $N$ effects (collisions) is (see,
e.g., \cite{epjp})
\begin{equation}
\frac{\partial f}{\partial t}=\pi
(2\pi)^{d} m  \frac{\partial}{\partial v_i} \int
d{\bf k} \, d{\bf v}'  \, k_ik_j  \frac{\hat{u}(k)^2}{|\epsilon({\bf k},{\bf
k}\cdot {\bf v})|^2}\delta\lbrack {\bf k}\cdot ({\bf v}-{\bf v}')\rbrack\left
(f'\frac{\partial f}{\partial {v}_{j}}-f
\frac{\partial f'}{\partial {v'}_{j}}\right ).
\label{crt1}
\end{equation}
As discussed previously, we can obtain the KP equation (\ref{lb27}) describing
the collisionless relaxation of the system by making the substitution $m
f\rightarrow \epsilon_r^d\epsilon_v^d f_2$ in the right hand side of Eq.
(\ref{crt1}). In the two-level case, this substitution
becomes
$mf\rightarrow \epsilon_r^d\epsilon_v^d (\eta_0-\overline{f})\overline{f}$. In
the nondegenerate limit $\overline{f}\ll\eta_0$, it reduces to $mf\rightarrow
\epsilon_r^d\epsilon_v^d \eta_0\overline{f}$, i.e., $m\rightarrow
\epsilon_r^d\epsilon_v^d
\eta_0$. In other words, we have to replace the mass $m$ of the particles by the
effective mass 
\begin{equation}
m_{\rm eff}=\epsilon_r^d\epsilon_v^d \eta_0,
\label{crt2}
\end{equation}
which is the mass of a completely filled macrocell. In the
terminology of
Dupree \cite{dupree} and Kadomtsev and Pogutse \cite{kp}, this can be viewed as
the effective mass of 
``macroparticles'' or ``clumps'', i.e., correlated regions. Using the
fact that $t_R\propto 1/m$ \cite{epjp}, the
ratio
between the collisionless relaxation time $t_R^*$ and the collisional relaxation
time $t_R$ is 
\begin{equation}
\frac{t_R^*}{t_R}=\frac{m}{m_{\rm eff}}.
\label{crt3}
\end{equation}
The quantity $m/m_{\rm eff}$ is the so-called reduction factor. In general
$m_{\rm eff}\gg m$ so that the reduction of the relaxation time can be quite
large. As a result, the collisionless
relaxation takes place on a timescale which is much smaller than the collisional
relaxation time in agreement with the Lynden-Bell concept of violent relaxation.

In $d\ge 2$, the collisional relaxation time scales as\footnote{In the case of
3D plasmas and 3D stellar systems we have to account for logarithmic corrections
yielding $t_R\sim [N/\ln(\lambda_{D,J}/\lambda_L)]\, t_D$. In $d=1$, the
KP and Lenard-Balescu operators vanish for spatially homogeneous
systems, implying that the relaxation is longer (see Sec. \ref{sec_oned}). In
that case, the collisional relaxation time scales as $N^2 t_D$
\cite{epjp,n2a,n2b}.} 
\begin{equation}
t_R\sim N t_D,
\label{crt4}
\end{equation}
where $t_D$ is the dynamical time. Using Eq. (\ref{crt3}), we find that the
collisionless relaxation time scales as 
\begin{equation}
t_R^*\sim \frac{Nm}{m_{\rm eff}} t_D\sim {\cal N}\, t_D,
\label{crt5}
\end{equation}
where ${\cal N}=Nm/m_{\rm eff}$ represents the number of completely filled
macrocells or, equivalently, the number of macroparticles. This number is
usually much smaller than the number of particles (${\cal N}\ll N$). This is
another manner to understand why the collisionless relaxation is much shorter
than the collisional relaxation. In general, the collisionless relaxation time
is equal to a
few dynamical times. We recall, however, that we need $t_R^*\gg t_D$ for the
validity of the Markovian description. This is typically the case when ${\cal
N}\gtrsim
10-100$.

The fact that the kinetic equation (\ref{lb27})  relaxes towards the
Lynden-Bell DF on a
few dynamical times is interpreted by Kadomtsev and Pogutse \cite{kp} in terms
of ``collisions'' between  macroparticles with a large effective mass $m_{\rm
eff}\sim \epsilon_r^d\epsilon_v^d \eta_0$. For partially degenerate systems,
the effective mass of the macroparticles scales as $m_{\rm
eff}=\epsilon_r^d\epsilon_v^d (\eta_0-\overline{f})$ or, more generally, as
$m_{\rm eff}=\epsilon_r^d\epsilon_v^d f_2/\overline{f}$. It is
therefore substantially reduced. Accordingly, the collisionless relaxation time
increases.\footnote{Similarly, the collisional relaxation time of  
self-gravitating fermions is larger than the collisional relaxation time of 
self-gravitating classical particles \cite{kinquant}.} This is because the
fluctuations that drive the collisionless relaxation are less effective in
establishing a statistical equilibrium state. As a result, the system can be
frozen for a long time in a
metaequilibrium state which is not the most mixed state. This kinetic
blocking can account for the process of incomplete relaxation as discussed in
Sec. \ref{sec_incompletekin}.

{\it Remark:} In the case of 3D plasmas and 3D stellar systems we have to
account for
logarithmic corrections. In that case, we get 
\begin{equation}
m_{\rm eff}=\epsilon_r^d\epsilon_v^d \eta_0 \frac{\ln\left
(\frac{\lambda_{D,J}}{\epsilon_r}\right )}{\ln\left
(\frac{\lambda_{D,J}}{\lambda_L}\right )}.
\label{crt6}
\end{equation}
Detailed estimates of the relaxation time are given
in \cite{kinquant,epjp,epjp2,epjp3}. Similar results are obtained in 2D
turbulence (see Sec. 4.2 of Ref. \cite{bbgkyvortex}).

\section{Kinetic theory of quiescent collisionless
relaxation for spatially inhomogeneous systems}
\label{sec_is}

We can easily extend the quasilinear theory of quiescent collisionless
relaxation to the case of spatially inhomogeneous systems by
introducing angle-action variables and using the formalism developed in Ref.
\cite{angleaction2}. The inhomogeneous Lenard-Balescu equation describing
the evolution of
the system sourced by finite $N$ effects (collisions) is
\cite{heyvaerts,angleaction2}
\begin{equation}
\frac{\partial f}{\partial t}=\pi (2\pi)^d m \frac{\partial}{\partial {\bf
J}}\cdot \sum_{{\bf k},{\bf k}'}\int d{\bf J}' \, {\bf k}\, |A^{d}_{{\bf
k},{\bf k}'}({\bf J},{\bf J}',{\bf k}\cdot {\bf \Omega})|^2\delta({\bf k}\cdot
{\bf \Omega}-{\bf k}'\cdot {\bf \Omega}') \left (f'{\bf k}\cdot \frac{\partial
f}{\partial {\bf J}}-f{\bf k}'\cdot \frac{\partial f'}{\partial {\bf J}'}\right
),
\label{is1}
\end{equation}
where $A^{d}_{{\bf
k},{\bf k}'}({\bf J},{\bf J}',\omega)$ is the dressed
potential of interaction (it is written as $-1/D_{{\bf
k},{\bf k}'}({\bf J},{\bf J}',\omega)$ in Refs. \cite{heyvaerts,angleaction2})
and ${\bf\Omega}({\bf J})$ is the pulsation of the orbit of a particle with
action ${\bf J}$. This equation can be obtained from a quasilinear
theory based on the Klimontovich equation \cite{angleaction2}. As we have
previously explained, the
quasilinear theory based on the Vlasov equation is similar to the quasilinear
theory based on the Klimontovich equation provided that we make the
substitution   $m
f\rightarrow \epsilon_r^d\epsilon_v^d f_2$. As a result, the inhomogeneous KP
equation describing
the quiescent collisionless relaxation of the system is
\begin{equation}
\frac{\partial \overline{f}}{\partial t}=\pi (2\pi)^d \epsilon_r^d\epsilon_v^d
\frac{\partial}{\partial {\bf
J}}\cdot \sum_{{\bf k},{\bf k}'}\int d{\bf J}' \, {\bf k}\, |A^{d}_{{\bf
k},{\bf k}'}({\bf J},{\bf J}',{\bf k}\cdot {\bf \Omega})|^2\delta({\bf k}\cdot
{\bf \Omega}-{\bf k}'\cdot {\bf \Omega}') \left (f'_2 {\bf k}\cdot
\frac{\partial
\overline{f}}{\partial {\bf J}}- f_2 {\bf
k}'\cdot\frac{\partial\overline{f}'}{\partial {\bf J}'}\right ).
\label{is2}
\end{equation}
In the multilevel case, we obtain the inhomogeneous SL equation
\begin{equation}
\frac{\partial \rho}{\partial t}=\pi (2\pi)^d \epsilon_r^d\epsilon_v^d
\frac{\partial}{\partial {\bf
J}}\cdot \sum_{{\bf k},{\bf k}'}\int d{\bf J}' \, {\bf k}\, |A^{d}_{{\bf
k},{\bf k}'}({\bf J},{\bf J}',{\bf k}\cdot {\bf \Omega})|^2\delta({\bf
k}\cdot
{\bf \Omega}-{\bf k}'\cdot {\bf \Omega}') \left (f'_2 {\bf k}\cdot
\frac{\partial\rho}{\partial {\bf J}}-\rho(\eta-\overline{f}) {\bf
k}'\cdot\frac{\partial\overline{f}'}{\partial {\bf J}'}\right ).
\label{is3}
\end{equation}
If we close the hierarchy of moment equations with the ansatz from Eq.
(\ref{c1}), we obtain the inhomogeneous Chavanis equation
\begin{equation}
\frac{\partial \overline{f}}{\partial t}=\pi (2\pi)^d \epsilon_r^d\epsilon_v^d
\frac{\partial}{\partial {\bf
J}}\cdot \sum_{{\bf k},{\bf k}'}\int d{\bf J}' \, {\bf k}\,|A^{d}_{{\bf
k},{\bf k}'}({\bf J},{\bf J}',{\bf k}\cdot {\bf \Omega})|^2\delta({\bf
k}\cdot
{\bf \Omega}-{\bf k}'\cdot {\bf \Omega}') \left
\lbrack\frac{1}{C''(\overline{f}')} {\bf k}\cdot
\frac{\partial
\overline{f}}{\partial {\bf J}}- \frac{1}{C''(\overline{f})} {\bf
k}'\cdot\frac{\partial\overline{f}'}{\partial {\bf J}'}\right \rbrack.
\label{is4}
\end{equation}
In the two-level case, the foregoing equations reduce to
\begin{equation}
\frac{\partial \overline{f}}{\partial t}=\pi (2\pi)^d \epsilon_r^d\epsilon_v^d
\frac{\partial}{\partial {\bf
J}}\cdot \sum_{{\bf k},{\bf k}'}\int d{\bf J}' \, {\bf k}\, |A^{d}_{{\bf
k},{\bf k}'}({\bf J},{\bf J}',{\bf k}\cdot {\bf \Omega})|^2\delta({\bf
k}\cdot
{\bf \Omega}-{\bf k}'\cdot {\bf \Omega}') \left
\lbrack \overline{f}'(\eta_0-\overline{f}') {\bf k}\cdot
\frac{\partial
\overline{f}}{\partial {\bf J}}- \overline{f}(\eta_0-\overline{f})  {\bf
k}'\cdot\frac{\partial\overline{f}'}{\partial {\bf J}'}\right \rbrack,
\label{is5}
\end{equation}
which can be viewed as a form of inhomogeneous fermionic
Lenard-Balescu
equation.

If we make a thermal bath approximation, using the identity from Eq. (\ref{g1})
and the relation ${\bf\Omega}({\bf J})=\partial\epsilon/\partial {\bf J}$, we
obtain
\begin{equation}
\frac{\partial\overline{f}'}{\partial {\bf J}'}=-\beta f'_2 {\bf\Omega}({\bf
J}').
\label{is6}
\end{equation}
Substituting Eq. (\ref{is6}) into Eq. (\ref{is3}) we get
\begin{equation}
\frac{\partial \rho}{\partial t}=\pi (2\pi)^d \epsilon_r^d\epsilon_v^d
\frac{\partial}{\partial {\bf
J}}\cdot \sum_{{\bf k},{\bf k}'}\int d{\bf J}' \, {\bf k}\, |A^{d}_{{\bf
k},{\bf k}'}({\bf J},{\bf J}',{\bf k}\cdot {\bf \Omega})|^2\delta({\bf
k}\cdot
{\bf \Omega}-{\bf k}'\cdot {\bf \Omega}') \left \lbrack f'_2 {\bf k}\cdot
\frac{\partial\rho}{\partial {\bf J}}+\beta f'_2 \rho(\eta-\overline{f}) {\bf
k}'\cdot {\bf\Omega}'\right \rbrack.
\label{is7}
\end{equation}
Using the properties of the $\delta$-function (resonance condition), we
can replace ${\bf k}'\cdot {\bf\Omega}'$ by ${\bf k}\cdot {\bf\Omega}$ in the
last term in brackets. We can
then rewrite the foregoing equation as
\begin{equation}
\frac{\partial \rho}{\partial t}=
\frac{\partial}{\partial {J}_i}\left \lbrace D_{ij} \left \lbrack
\frac{\partial\rho}{\partial {J_j}}+\beta  \rho(\eta-\overline{f})
{\Omega_j}\right \rbrack\right\rbrace
\label{is8}
\end{equation}
with
\begin{equation}
D_{ij}=\pi (2\pi)^d \epsilon_r^d\epsilon_v^d \sum_{{\bf k},{\bf k}'}\int d{\bf
J}' \, {k_i}{k_j}\, |A^{d}_{{\bf
k},{\bf k}'}({\bf J},{\bf J}',{\bf k}\cdot {\bf \Omega})|^2\delta({\bf
k}\cdot
{\bf \Omega}-{\bf k}'\cdot {\bf \Omega}') f_2({\bf J}'). 
\label{is9}
\end{equation}
The equation for the coarse-grained DF reads
\begin{equation}
\frac{\partial \overline{f}}{\partial t}=
\frac{\partial}{\partial {J}_i}\left \lbrace D_{ij} \left \lbrack
\frac{\partial\overline{f}}{\partial {J_j}}+\beta  f_2
{\Omega_j}\right \rbrack\right\rbrace.
\label{is10}
\end{equation}

As in Sec. \ref{sec_csr} we can enforce the conservation of energy $E=\int
f({\bf
J})\epsilon({\bf J})\, d{\bf J}$  by letting the
inverse temperature evolve in time according to
\begin{equation}
\beta(t) =-\frac{\int  D_{ij} \Omega_i \frac{\partial \overline{f}}{\partial
{J}_j}\, d{\bf J}}{\int D_{ij} f_2 \Omega_i \Omega_j \,
d{\bf J}}.
\label{is11}
\end{equation}
In this manner, we obtain another type of CSR equations for inhomogeneous
systems written with angle-action variables. Note that the diffusion
coefficient from Eq. (\ref{is9}) does not display a logarithmic divergence at
large scales for self-gravitating systems, contrary to the case where a local
approximation is made (see Sec. \ref{sec_3d}), since spatial inhomogeneity has
been properly accounted for. 

The above kinetic equations conserve the energy and all the Casimirs and
monotonically increase the Lynden-Bell entropy ($H$-theorem).

\section{Nonlinear dynamical stability and numerical algorithms}
\label{sec_nds}

In this section, we consider the nonlinear dynamical stability of  steady states
of the
Vlasov equation based on variational principles (see
\cite{assisePH,nyquist1,cc} and references therein for additional
discussions). We also introduce relaxation
equations that can serve
as numerical algorithms to compute stable steady states of the Vlasov equation.
Similar results obtained for the 2D Euler equation are given
in \cite{vphydro,cnd}.

\subsection{Energy principle}
\label{sec_ep}

The Vlasov equation conserves the energy and an infinite class of Casimirs.
It can be shown that a DF which is an extremum of energy ($\delta E=0$) with
respect to symplectic perturbations (i.e. perturbations that conserve all the
Casimirs) is a stationary solution of the Vlasov equation. Furthermore, this DF
is dynamically stable if and only if it is a minimum of energy ($\delta^2
E>0$)
with respect to symplectic perturbations  (see, e.g.,
\cite{cc} for a brief presentation of these results).
This energy
principle is the
most refined stability criterion because it takes into account all the
constraints of the Vlasov equation (an infinity of Casimirs). This
stability criterion has been introduced in astrophysics by Bartholomew
\cite{bartholomew} and
Kandrup \cite{kandrup91} for the Vlasov-Poisson equations. It is similar to
the Kelvin-Arnol'd energy principle for 2D inviscid
incompressible hydrodynamical 
flows
 governed by the Euler-Poisson equations  \cite{vphydro}. We are
led
therefore to considering the minimization problem
\begin{equation}
\min_{f} \left\lbrace E[f]\quad |\quad {\rm symplectic\,\,
perturbations}\right\rbrace
\label{ep1}
\end{equation}
or, equivalently,
\begin{equation}
\min_{f} \left\lbrace E[f]\quad |\quad M_{n\ge 1}[f]=M_{n\ge 1}\right\rbrace.
\label{ep2}
\end{equation}
Here, the perturbations must conserve all the Casimirs, which is equivalent
to the conservation of all the moments of the DF.\footnote{For
the
Newtonian gravitational
interaction,  it can be shown that all
the DFs of the form $f=f(\epsilon)$ with
$f'(\epsilon)<0$ are minima of energy with respect to symplectic
perturbations so, according to the stability criterion (\ref{ep1}), they are
dynamically Vlasov stable (see Refs.
\cite{doremus71,doremus73,gillon76,sflp,ks,kandrup91} 
for linear stability and
Ref. \cite{lmr} for nonlinear stability).  This is, however,
no more true in general relativity (see the discussion in
\cite{gr1}) nor for
other potentials of interaction.} If we restrict ourselves to DFs  of the form
$f=f(\epsilon)$
with $f'(\epsilon)<0$, it can be shown \cite{cc} that $f$ is a
local minimum of $E$ for isovortical perturbations if and only
if
\begin{eqnarray}
\delta^{2}{\cal E}[\delta f]&\equiv& -\frac{1}{2}\int
\frac{(\delta {f})^2}{f'(\epsilon)} \,
d{\bf
r}d{\bf v}+\frac{1}{2}\int \delta {f}\delta {\Phi}  \,
d{\bf
r}d{\bf v}>0,\nonumber\\
&\forall& \delta f\ | \ \delta E=\delta M_{n\ge 1}=0.
\label{sand1}
\end{eqnarray}
For the Coulombian potential of interaction in plasma physics, the second term
in Eq. (\ref{sand1}) is positive implying that all the DFs of the form
$f=f(\epsilon)$ with $f'(\epsilon)<0$ are stable ($\epsilon=v^2/2$ for
homogeneous plasmas). For
the Newtonian potential of interaction in astrophysics, the second term in Eq.
(\ref{sand1}) is negative. Still, it can be shown that all the DFs
of the form $f=f(\epsilon)$ with $f'(\epsilon)<0$ are stable (see
footnote 26).

{\it Numerical algorithm:} We can easily construct a modified dynamics for the
DF that conserves all the
Casimirs and that monotonically dissipates the energy. Let us
consider the equation\footnote{This equation
was suggested in footnote 6 of \cite{vphydro} based on similar results obtained
in
2D hydrodynamics.} 
\begin{equation}
\frac{\partial f}{\partial
t}+\lbrace f,\epsilon\rbrace=\alpha\lbrace f,\lbrace f,\epsilon\rbrace\rbrace,
\label{ep3}
\end{equation}
where $\lbrace f,g\rbrace$ is the Poisson bracket defined by Eq. (\ref{ep4}).
When $\alpha=0$, we recover
the Vlasov equation which conserves the energy and all the Casimirs (see
Appendix \ref{sec_propv}). When
$\alpha>0$, we show below that Eq. (\ref{ep3}) conserves all the Casimirs while
the
energy decreases monotonically. Therefore, it relaxes towards a minimum of
energy with respect to symplectic perturbations. By construction,
this is a dynamically stable steady state of the Vlasov equation. Therefore,
Eq. (\ref{ep3}) can be used as a numerical algorithm to construct stable steady
states of the Vlasov equation. This is interesting because it is generally
difficult to construct steady states of the Vlasov equation and be sure that
they are dynamically stable.

{\it Proof:} We first show that Eq. (\ref{ep3}) conserves all the
Casimirs. We have
\begin{equation}
\dot I_h=\int h'(f)\frac{\partial f}{\partial t}\, d{\bf r}d{\bf v}=\alpha\int
h'(f)\lbrace f,\lbrace f,\epsilon\rbrace\rbrace\, d{\bf r}d{\bf v}.
\label{ep5}
\end{equation}
Using the identity from Eq. (\ref{b1}), we get
\begin{equation}
\dot I_h=\alpha\int
\lbrace f,\epsilon\rbrace\lbrace h'(f),f\rbrace\, d{\bf r}d{\bf v}.
\label{ep6}
\end{equation}
Then, using the identity from Eq. (\ref{b2}) and the fact that $\lbrace
f,f\rbrace=0$, we
obtain
\begin{equation}
\dot I_h=\alpha\int h''(f)
\lbrace f,\epsilon\rbrace\lbrace f,f\rbrace\, d{\bf r}d{\bf v}=0.
\label{ep7}
\end{equation}
We now show that the energy decreases monotonically.  We have
\begin{equation}
\dot E=\int \epsilon\frac{\partial f}{\partial t}\, d{\bf r}d{\bf v}=\alpha\int
\epsilon\lbrace f,\lbrace f,\epsilon\rbrace\rbrace\, d{\bf r}d{\bf v}.
\label{ep8}
\end{equation}
Using the identity from Eq. (\ref{b1}), we get
\begin{equation}
\dot E=\alpha\int
\lbrace f,\epsilon\rbrace\lbrace \epsilon,f\rbrace\, d{\bf r}d{\bf v}.
\label{ep9}
\end{equation}
Then, using the identity from Eq. (\ref{b3}), we obtain
\begin{equation}
\dot E=-\alpha\int 
\lbrace f,\epsilon\rbrace^2\, d{\bf r}d{\bf v}\le 0.
\label{ep10}
\end{equation}
Therefore, the energy is non increasing. At equilibrium ($\dot E=0$), we have
$\lbrace f,\epsilon\rbrace=0$ implying that $f$ is a
stationary solution of the Vlasov equation.

\subsection{Sufficient conditions of dynamical stability}
\label{sec_s}

We have seen that a DF is a dynamically stable steady state  of the Vlasov
equation if and only if it is a minimum of energy for
perturbations that conserve all the Casimirs. Therefore, a {\it sufficient}
condition of dynamical stability is that $f$ is a minimum of energy for
perturbations
that conserve the mass $M$ and {\it one} Casimir of the form
\begin{equation}
\label{s1} S[f]=-\int C(f) d{\bf r}d{\bf v},
\end{equation}
where $C({f})$ is a convex function,
i.e. $C''>0$ \cite{ipser,ipserH}. In that case, it is {\it a fortiori} a minimum
of energy
for perturbations that conserve {\it all} the Casimirs (i.e. for symplectic
perturbations). We are therefore led to considering the two-constraint
minimization problem
\begin{equation}
\min_{f} \left\lbrace E[f]\quad |\quad M[f]=M,\quad S[f]=S\right\rbrace.
\label{s2}
\end{equation}
It is shown in \cite{cc} that this minimization problem is equivalent
to the maximization problem\footnote{If we view $f$ as the
coarse-grained DF $\overline{f}$, this maximization problem can be related to
the selective decay principle (for $-S$) of Sec. \ref{sec_incomplete}.}
\begin{equation}
\max_{f} \left\lbrace S[f]\quad |\quad M[f]=M,\quad E[f]=E\right\rbrace.
\label{s3}
\end{equation}
The first variations can be treated like in  Sec.
\ref{sec_ge} leading to the DF from Eq. (\ref{meta7}). It can
be shown \cite{cc} that $f$ is a
local minimum of $E$ at fixed $M$ and $S$ or a local maximum of $S$ at fixed
$E$ and $M$ if and only
if
\begin{eqnarray}
\delta^{2}{\cal E}[\delta f]&\equiv& -\frac{1}{2}\int
\frac{(\delta {f})^2}{f'(\epsilon)} \,
d{\bf
r}d{\bf v}+\frac{1}{2}\int \delta {f}\delta {\Phi}  \,
d{\bf
r}d{\bf v}>0,\nonumber\\
&\forall& \delta f\ | \ \delta E=\delta M=0.
\label{sand2}
\end{eqnarray}
Clearly, Eq. (\ref{sand2}) implies Eq. (\ref{sand1}). Indeed if $\delta^{2}{\cal
E}$ is positive for all perturbations that conserve mass and energy at first
order, it is {\it a fortiori} positive for all perturbations that conserve mass,
energy and all the Casimirs at first order. If we view
the functional (\ref{s1}) as a ``pseudo (or effective) entropy'' 
\cite{cst,assisePH,aaantonov} the maximization
problem (\ref{s3})
is similar to a condition of microcanonical stability in thermodynamics,
i.e., to the maximization of the entropy at fixed mass and
energy.\footnote{We stress that we are just making a ``thermodynamical
analogy'' \cite{cst,assisePH,aaantonov}. There is no thermodynamics involved in
the
dynamical stability
problem of the Vlasov equation. This thermodynamical
analogy (or effective thermodynamics) may provide an interpretation of the
Tsallis entropy $S_q=-\frac{1}{q-1}\int (f^q-f)\, d{\bf r}d{\bf v}$, leading to
the Tsallis distribution $f=(1/q)^{1/(q-1)}\lbrack
1-(q-1)(\beta\epsilon+\alpha)\rbrack^{1/(q-1)}$, in terms of a
``pseudo entropy'' \cite{cst,assisePH,aaantonov} in the sense given above. This
``Tsallis pseudo
entropy'' may be useful for dynamical
(not thermodynamical)
stability problems. The maximization of the Tsallis (pseudo) entropy at fixed
mass and energy ensures the dynamical stability of a particular class
of stationary solutions of the Vlasov equation known as polytropic DFs (see
\cite{cst,assisePH,cct,ccpoly,gr1} for a more detailed
discussion).} Therefore, a
maximum of pseudo entropy at fixed mass and energy is a dynamically
stable steady state of the Vlasov equation. In particular, considering
the Boltzmann entropy $S=-\int f\ln f\, d{\bf r}d{\bf v}$ of statistical
mechanics leading to the Boltzmann distribution $f=e^{-\beta
\epsilon-\alpha}$, we conclude that
microcanonical stability implies (Vlasov) dynamical stability. However, the
reciprocal is wrong: a dynamically stable steady state of the Vlasov equation is
not necessarily a maximum of pseudo entropy at fixed mass and energy. For
example, we have indicated in footnote 26 that, in the case of
Newtonian
self-gravitating systems, all the DFs of the form $f=f(\epsilon)$ with
$f'(\epsilon)<0$ are dynamically (Vlasov) stable, even those that do not
maximize a pseudo entropy at fixed mass and energy. The stability criteria
(\ref{s2}) and (\ref{s3}) are less refined than the stability
criterion (\ref{ep1}) because they do not take into account all the constraints
of the Vlasov equation. This is similar to a notion of ensemble inequivalence in
thermodynamics (see below).

An even less refined condition of dynamical stability is that
$f$ maximizes $J=S-\beta E$ at fixed mass or, equivalently, minimizes $F=E-TS$
at
fixed mass, where $J$ or $F$ is the Legendre transform of the pseudo entropy
with respect to the energy.  We are therefore led to considering
the one-constraint minimization problem
\begin{equation}
\min_{f} \left\lbrace F[f]=E[f]-TS[f] \quad |\quad M[f]=M\right\rbrace.
\label{sand3}
\end{equation}
The first variations return the results of Sec.
\ref{sec_ge} so that (\ref{s3}) and (\ref{sand3}) have the same critical points.
It can be shown
\cite{cc} that $f$ is a
local minimum of $F$ at fixed $M$ if
and only
if
\begin{eqnarray}
\delta^{2}{\cal E}[\delta f]&\equiv& -\frac{1}{2}\int
\frac{(\delta {f})^2}{f'(\epsilon)} \,
d{\bf
r}d{\bf v}+\frac{1}{2}\int \delta {f}\delta {\Phi}  \,
d{\bf
r}d{\bf v}>0,\nonumber\\
&\forall& \delta f\ | \ \delta M=0.
\label{sand4}
\end{eqnarray}
Clearly, Eq. (\ref{sand4}) implies Eq. (\ref{sand2}). Indeed if $\delta^{2}{\cal
E}$ is positive for all perturbations that conserve mass, it is {\it a
fortiori} positive for all perturbations that conserve mass and
energy at first order. If we view the functional $F$ as a ``pseudo (or
effective) free energy'' \cite{cst,assisePH,aaantonov} the minimization problem
(\ref{sand3}) is similar to a condition of canonical stability in
thermodynamics, i.e., to the minimization of the free energy at fixed mass.
Therefore,  a
minimum of pseudo free energy at fixed mass is a dynamically stable steady state
of the Vlasov equation.  The fact that (\ref{sand3}) implies (\ref{s3})
means that a minimum of free energy at fixed mass is necessarily a
maximum of entropy at fixed mass and energy. However, the reciprocal is
wrong: A maximum of entropy at fixed mass and energy is not necessarily a
minimum of
free energy at fixed mass. Therefore, canonical stability implies
microcanonical stability but not the converse \cite{cc}. This corresponds
the notion of ensemble inequivalence in thermodynamics for systems with
long-range interactions \cite{paddy,found,ijmpb,cc,campabook}. Transposed to
the present (dynamical) context, the minimization of
pseudo free energy at fixed mass (one-constraint problem) provides a sufficient
condition of dynamical
stability which is less refined than the maximization of pseudo entropy at fixed
mass and energy (two-constraint problem) which is itself less refined than the
minimization of energy
under symplectic perturbations (infinite-constraint
problem).\footnote{We can also introduce a no-constraint problem
by considering the maximization of the grand potential $G=S-\beta E-\alpha M$.
This is the least refined stability criterion (see Ref. \cite{vphydro}).} In
summary, we have the correspondances
\begin{eqnarray}
(\ref{sand3}) \quad \Rightarrow \quad (\ref{s3}) \quad \Rightarrow
\quad(\ref{ep1}) \quad  \Leftrightarrow \quad {\rm Vlasov \,\, stability}
\label{sand5}
\end{eqnarray}
The connection of these results with the so-called nonlinear Antonov first law
is discussed in detail in \cite{aaantonov,assisePH,gr1}.

{\it Numerical algorithms:} Let us consider  the relaxation
equation
\begin{equation}
\frac{\partial f}{\partial t}+{\bf v}\cdot \frac{\partial
f}{\partial {\bf r}}- \nabla\Phi\cdot \frac{\partial
f}{\partial {\bf v}}=\frac{\partial}{\partial v_i} 
\int
d{\bf v}'\, K_{ij} \, \left
\lbrack \frac{1}{C''({f}')}\frac{\partial {f}}{\partial
{v}_{j}}-\frac{1}{C''({f})}\frac{\partial
{f}'}{\partial {v'}_{j}}\right \rbrack,
\label{s4}
\end{equation}
or the relaxation equation
\begin{equation}
\frac{\partial f}{\partial t}+{\bf v}\cdot \frac{\partial
f}{\partial {\bf r}}- \nabla\Phi\cdot \frac{\partial
f}{\partial {\bf v}}=\frac{\partial}{\partial
{\bf v}}\cdot 
\left\lbrace
D \left \lbrack \frac{\partial {f}}{\partial
{\bf v}}+ \frac{\beta(t)}{C''({f})} {\bf v}\right
\rbrack\right\rbrace,
\label{s5}
\end{equation}
where  $D$ is a strictly positive constant
and
\begin{equation}
\beta(t) =-\frac{\int  \frac{\partial {f}}{\partial
{\bf v}}\cdot {\bf v}\, d{\bf r}d{\bf v}}{\int  \frac{v^2}{C''({f})} \,
d{\bf r}d{\bf v}}=\frac{3M}{\int 
\frac{v^2}{C''({f})} \,
d{\bf r}d{\bf v}}
\label{s6}
\end{equation}
is a time-dependent inverse temperature.\footnote{Here, we use the kinetic
equations (\ref{s4}) and (\ref{s5})
as numerical algorithms to compute stable steady states of the Vlasov equation,
not as parametrizations of the coarse-grained dynamics. As a result, we can
make the following simplifications: (i) we can write these equations for
spatially inhomogeneous systems and take $f'=f({\bf r},{\bf v}',t)$ in Eq.
(\ref{s4}) even if the local approximation is not justified for the true
evolution of the system; (ii) we can ignore
collective effects in Eq. (\ref{s4}) and define $K_{ij}$ by Eq. (\ref{E46})
instead of Eq.
(\ref{lb27b}); (iii) we can replace $D_{ij}$ by
$D_{ij}=D\delta_{ij}$ in Eq. (\ref{s5}), where $D$
is a strictly positive constant, in order to make the equation simpler and make
sure that it
relaxes towards a maximum  entropy state at fixed mass and energy without
experiencing a situation of
kinetic blocking (see Sec. \ref{sec_incompletekin}).} These equations conserve
the mass $M$ and the
energy $E$ and monotonically increase
the pseudo entropy $S$ ($H$-theorem).  They relax towards a maximum 
entropy state at fixed mass and energy.  This corresponds to a microcanonical
description. By construction, this equilibrium state is a stable
steady state of the Vlasov equation. 
Therefore, Eq. (\ref{s4}) or
Eqs. (\ref{s5}) and (\ref{s6})  can be used
as numerical algorithms to construct  stable steady states of the Vlasov
equation. If we fix $\beta$, Eq. (\ref{s5}) can be viewed as a generalized
Kramers equation. It conserves the mass $M$ and monotonically decreases
the pseudo free energy $F=E-TS$ \cite{gen,nfp}. It relaxes towards a minimum of
free energy  at fixed mass. This corresponds to a canonical description. By
construction, the equilibrium state of the generalized Kramers equation  is a
stable steady state of the Vlasov equation. Therefore, Eq. (\ref{s5}) with fixed
$\beta$ can be used as a numerical algorithm to construct  stable steady states
of the Vlasov equation. We can also obtain simpler numerical algorithms by
taking
the hydrodynamic moments of the generalized Landau and Kramers equations
(\ref{s4}) and (\ref{s5}), and closing the hierarchy of equations with a local
thermodynamic equilibrium assumption, leading to generalized Navier-Stokes,
Euler and Smoluchowski equations \cite{gen,nfp}. Similar
numerical
algorithms have been introduced in 2D turbulence \cite{gen,vphydro,cnd}.

\subsection{Dynamical and thermodynamical stability}

It can be shown that a thermodynamical equilibrium state in the sense of
Lynden-Bell is nonlinearly dynamically stable. Indeed, the coarse-grained DF
$\overline{f}$ obtained from the Gibbs state (\ref{E11}) which maximizes the
Lynden-Bell entropy (\ref{E9}) at fixed mass, energy and Casimir constraints
is a minimum of energy $E[\overline{f}]$ with respect to perturbations that
conserve the coarse-grained moments $M_{n\ge 1}^{c.g.}[\overline{f}]$ (see Sec.
7.8 of \cite{vphydro}). Therefore, according to Eq. (\ref{ep2}), this is a
nonlinearly dynamically stable steady state of the Vlasov
equation.  By contrast, the initial condition $f_0$, even
though it has the same energy as the metaequilibrium state, is generically not
a 
minimum of energy $E[f_0]$ with
respect to  perturbations that conserve the fine-grained moments $M_{n\ge
1}^{f.g.}[f_0]$, so it is dynamically unstable and relaxes towards the 
metaequilibrium state.

It can be shown  that a DF which maximizes the
generalized entropy  defined by Eqs. (\ref{meta4}) and
(\ref{s3w}) at fixed mass and energy is (i) nonlinearly dynamically stable
(see \ref{sec_s}) and
(ii) thermodynamically stable in the sense of Lynden-Bell (see
Appendix \ref{sec_cano}). We stress, however, that this is just a sufficient
condition of dynamical and thermodynamical stability. In particular, the DF
$\overline{f}$
associated with a thermodynamical equilibrium state in the sense of
Lynden-Bell does not
necessarily maximizes the
generalized entropy defined by Eqs. (\ref{meta4}) and (\ref{s3w}) at fixed
mass and energy (see Appendix \ref{sec_cano}).

We would be tempted to believe that Vlasov nonlinear dynamical stability implies
Lynden-Bell's thermodynamical stability. More precisely, we would be
tempted to believe that a DF which is a monotonically
decreasing function of
$\epsilon$ and which is a minimum of energy with respect to symplectic
perturbations is a thermodynamically equilibrium state in the sense of
Lynden-Bell. However, this is not
true as shown by the
following counter-example. For collisionless self-gravitating systems, all the 
DFs of the form $f=f(\epsilon)$ with $f'(\epsilon)<0$ are dynamically stable
(see footnote 26) even those that are not thermodynamically stable in the sense
of Lynden-Bell. In particular, all the extrema -- including saddle points -- of
Lynden-Bell's entropy at fixed
mass, energy and Casimir constraints are dynamically stable (since they are of
the form $\overline{f}=\overline{f}(\epsilon)$ with $\overline{f}'(\epsilon)<0$)
even if they are not maxima of Lynden-Bell's entropy  at fixed mass, energy and
Casimir constraints. This is because the dynamical stability criterion involves
the coarse-grained moments $M_{n\ge 1}^{c.g.}[\overline{f}]$ while the
Lynden-Bell thermodynamical criterion involves the fine-grained 
moments $M_{n\ge 1}^{f.g.}[\rho]$. This can be easily understood in the
two-level case. In that case, the Lynden-Bell statistical equilibrium state is
obtained by maximizing the Fermi-Dirac-like entropy (\ref{lbent}) at fixed mass
and energy (the fine-grained moments $M_{n\ge 1}^{f.g.}[\rho]$ are all
proportional to the mass) or, equivalently, by minimizing
the energy at fixed mass and Fermi-Dirac-like entropy. However, we have seen
that this optimization problem is just a
sufficient condition of dynamical stability. A more refined condition of
dynamical stability is that $\overline{f}$ is a minimum of energy
$E[\overline{f}]$ with respect to perturbations that
conserve all the coarse-grained moments $M_{n\ge 1}^{c.g.}[\overline{f}]$ (not
just the Fermi-Dirac-like entropy). For self-gravitating systems this
is the
case for all DFs of the form $f=f(\epsilon)$ with $f(\epsilon)<0$. Thus,
there exist DFs which are dynamically Vlasov
stable while they do not maximize the Fermi-Dirac-like entropy at fixed mass and
energy. Such DFs are dynamically stable but not thermodynamically stable in the
sense of Lynden-Bell.

\section{An equation that
conserves the mass and the energy and that  monotonically increases all
the
$H$-functions}
\label{sec_cge}

Using the same method as the one developed in Sec. \ref{sec_ep}, we can easily
construct a
modified dynamics for the DF that conserves the
mass and the energy and that monotonically increases all the
$H$-functions (see also Appendix \ref{sec_allh}).
Let us consider the equation
\begin{equation}
\frac{\partial \overline{f}}{\partial
t}+\lbrace \overline{f},\epsilon\rbrace=\alpha\lbrace \epsilon,\lbrace
\overline{f},\epsilon\rbrace\rbrace.
\label{cge1}
\end{equation}
When $\alpha=0$, we recover
the Vlasov equation which conserves the energy and all the Casimirs (see
Appendix \ref{sec_propv}). When
$\alpha>0$, we show below that Eq. (\ref{cge1}) conserves the energy while it
increases all the $H$-functions monotonically. We note that
all the stationary solutions of the Vlasov equation (satisfying $\lbrace
\overline{f},\epsilon\rbrace=0$) are
stationary solutions of Eq. (\ref{cge1}). Equation (\ref{cge1}) may admit
other stationary solutions (satisfying $\lbrace
\overline{f},\epsilon\rbrace=\alpha\lbrace \epsilon,\lbrace
\overline{f},\epsilon\rbrace\rbrace$) but, according to the result derived
below Eq. (\ref{cge6}), they are necessarily unstable. 

{\it Proof:} We first show that Eq. (\ref{cge1}) conserves the energy. We have
\begin{equation}
\dot E=\int \epsilon\frac{\partial \overline{f}}{\partial t}\, d{\bf r}d{\bf
v}=\alpha\int
\epsilon\lbrace \epsilon,\lbrace \overline{f},\epsilon\rbrace\rbrace\, d{\bf
r}d{\bf v}.
\label{cge2}
\end{equation}
Using the identity from Eq. (\ref{b1}) and the fact that $\lbrace
\epsilon,\epsilon\rbrace=0$,
we get
\begin{equation}
\dot E=\alpha\int
\lbrace \overline{f},\epsilon\rbrace\lbrace \epsilon,\epsilon\rbrace\, d{\bf
r}d{\bf v}=0.
\label{cge3}
\end{equation}
We now show that Eq. (\ref{cge1}) monotonically increases all the generalized
$H$-functions.
We have
\begin{equation}
\dot H=\int C'(\overline{f})\frac{\partial \overline{f}}{\partial t}\, d{\bf
r}d{\bf
v}=\alpha\int
C'(\overline{f})\lbrace \epsilon,\lbrace \overline{f},\epsilon\rbrace\rbrace\,
d{\bf r}d{\bf v}.
\label{cge4}
\end{equation}
Using the identity from Eq. (\ref{b1}), we get
\begin{equation}
\dot H=\alpha\int
\lbrace \overline{f},\epsilon\rbrace\lbrace C'(\overline{f}),\epsilon\rbrace\,
d{\bf r}d{\bf v}.
\label{cge5}
\end{equation}
Then, using the identity from Eq. (\ref{b2}), we obtain
\begin{equation}
\dot H=\alpha\int C''(\overline{f})
\lbrace \overline{f},\epsilon\rbrace^2\, d{\bf r}d{\bf v}\ge 0.
\label{cge6}
\end{equation}
Therefore, the $H$-functions are non decreasing. At equilibrium ($\dot H=0$),
Eq. (\ref{cge6}) implies 
$\lbrace \overline{f},\epsilon\rbrace=0$. Therefore, Eq. (\ref{cge1}) relaxes
towards a
stationary
solution of the Vlasov equation. Note that this stationary solution does not
necessarily maximize a particular $H$-function at fixed mass and energy. It
cannot be predicted {\it a priori}. One has to solve the kinetic equation
(\ref{cge1}) numerically to determine its equilibrium state.

\section{Conclusion}

In this paper, we have discussed the kinetic theory of collisionless relaxation
for systems with long-range interactions. We have recalled the basics of the
quasilinear theory of the Vlasov equation developed by Kadomtsev and Pogutse
\cite{kp}, Severne and Luwel \cite{sl}, and Chavanis
\cite{chavmnras,dubrovnik,kingen}. We
have established a connection between the kinetic equations derived from the 
quasilinear theory and the CSR relaxation equations obtained from a
phenomenological MEPP \cite{csr}. We have proposed a method to close the
hierarchy of
moment equations leading to a self-consistent kinetic equation for the
coarse-grained DF which is valid beyond the
two-level case \cite{kingen}. This equation [see Eq. (\ref{c2})] depends on a
generalized entropy $C(\overline{f})$ which can be obtained from
the equilibrium state and then used out-of-equilibrium, or which can be
obtained at any time of the dynamics by using the procedure explained in
Appendix \ref{sec_out}.  We have also discussed the nonlinear dynamical
stability of steady
states of the Vlasov equation and proposed
numerical
algorithms in the form of kinetic (relaxation) equations that can be used to
construct nonlinearly stable steady states. Similar results can be
obtained in 2D turbulence and vortex dynamics by exploiting the analogy between
the Vlasov and the 2D Euler equations. This will be discussed in a specific
paper \cite{prep}.

The statistical mechanics of violent relaxation was initiated by Lynden-Bell
\cite{lb} in the context of collisionless stellar systems. However, the present
paper was motivated by the possibility to apply these ideas to the context of
fermionic or bosonic dark matter
\cite{clm1,clm2,modeldmbosons,lbquant,modeldmfermions}. Indeed, these systems
also exhibit a process of violent relaxation (known as gravitational cooling
\cite{seidel94} in the case of boson stars). For these systems,
we have to take into account the
quantum nature of the particles. In the case of fermionic
dark matter, the quantum potential arising from the Heisenberg uncertainty
principle is negligible and we can use the
classical Vlasov equation (Thomas-Fermi approximation). The Lynden-Bell theory
of violent relaxation can justify the establishment of a Fermi-Dirac-like DF on
a timescale shorter than the age of the
universe \cite{clm1,clm2,modeldmfermions}.\footnote{For
self-gravitating fermions, gravitational encounters are completely negligible
and cannot establish a statistical equilibrium state on a relevant timescale.
However, a collisional relaxation may be relevant if the fermions are
self-interacting \cite{modeldmfermions}.} This leads to dark matter
halos with a ``core-halo'' structure. The quantum core (fermion ball) solves the
core-cusp problem of classical cold dark matter and the isothermal halo leads to
flat rotation curves in agreement with the observations \cite{modeldmfermions}.
In the case of bosonic dark matter the quantum potential is important and we
must replace the Vlasov equation by the Wigner equation. A generalization of the
Lynden-Bell theory of violent relaxation taking into account the specificities
of the Wigner equation has been recently proposed in \cite{lbquant}. This theory
also leads to dark matter halos with a ``core-halo'' structure where the quantum
core is a self-gravitating Bose-Einstein condensate (soliton) surrounded by a
halo made of quantum interferences. The collisional kinetic theory of fermions
and bosons has been studied in \cite{levkov2,bft,kinquant}. Fermions and
bosons behave antisymmetrically regarding their collisional relaxation.
The Pauli blocking $f(\eta_0-f)$ for fermions has the tendency to slow down the
relaxation and the Bose enhancement $f(\eta_0+f)$ for bosons, leading to the
formation of ``granules'' or ``quasiparticles'', has the tendency to
accelerate the relaxation.
Gravitational encounters (``collisions'') are completely negligible in fermionic
dark matter halos. In bosonic dark matter halos, they manifest themselves on a
(secular) timescale of the order of the age of the universe (see
\cite{kinquant} and references therein).

\appendix

\section{Basic properties of the Poisson bracket}
\label{sec_b}

The Poisson brackets are defined by
\begin{equation}
\lbrace f,g\rbrace=\nabla_{\bf r}f\cdot \nabla_{\bf v}g-\nabla_{\bf v}f\cdot
\nabla_{\bf r}g.
\label{ep4}
\end{equation}
We recall below some basic properties of the Poisson brackets that can be
established straightforwardly:
\begin{equation}
\int f \lbrace g,h\rbrace\, d{\bf r}d{\bf v}=\int h\lbrace f,g\rbrace\, d{\bf
r}d{\bf v},
\label{b1}
\end{equation}
\begin{equation}
\int  \lbrace h(f),g\rbrace\, d{\bf r}d{\bf v}=\int h'(f)\lbrace f,g\rbrace\,
d{\bf
r}d{\bf v},
\label{b2}
\end{equation}
\begin{equation}
\lbrace f,g\rbrace=-\lbrace g,f\rbrace.
\label{b3}
\end{equation}

\section{Basic properties of the Vlasov equation}
\label{sec_propv}

In this appendix, we establish some basic properties of the Vlasov equation
(\ref{ve1}).

(i) The conservation of the Casimirs can be established
as follows: 
\begin{eqnarray}
\dot I_h=\int h'(f)\frac{\partial f}{\partial t}\, d{\bf r}d{\bf v}&=&-\int
h'(f)\left ( {\bf v}\cdot {\partial f\over\partial {\bf
r}}-\nabla\Phi\cdot {\partial f\over\partial {\bf v}} \right )\,
d{\bf
r}d{\bf v}=-\int \left \lbrack {\bf v}\cdot {\partial h(f)\over\partial {\bf
r}}-\nabla\Phi\cdot {\partial h(f)\over\partial {\bf v}} \right \rbrack\,
d{\bf
r}d{\bf v}\nonumber\\
&=&-\int\left \lbrace\frac{\partial}{\partial {\bf r}}\cdot [h(f){\bf
v}]-\frac{\partial}{\partial {\bf v}}\cdot [h(f) \nabla\Phi] \right \rbrace\,
d{\bf
r}d{\bf v}=0.
\label{v1}
\end{eqnarray}

(ii) The conservation of the energy can be established
as follows: 
\begin{eqnarray}
\dot E=\int \epsilon\frac{\partial f}{\partial t}\, d{\bf r}d{\bf v}&=&-\int
\epsilon\left ( {\bf v}\cdot {\partial f\over\partial {\bf
r}}-\nabla\Phi\cdot {\partial f\over\partial {\bf v}} \right )\,
d{\bf
r}d{\bf v}=-\int \epsilon\left \lbrack\frac{\partial}{\partial {\bf r}}\cdot
(f{\bf
v})-\frac{\partial}{\partial {\bf v}}\cdot (f \nabla\Phi) \right \rbrack\,
d{\bf
r}d{\bf v}\nonumber\\
&=&\int f \nabla\Phi\cdot {\bf
v}\,d{\bf
r}d{\bf v}-\int f {\bf v}\cdot \nabla\Phi \,d{\bf
r}d{\bf v}  =0.  
\label{v1b}
\end{eqnarray}

(iii) The conservation of the impulse
can be
established
as follows: 
\begin{eqnarray}
\dot {\bf P}&=&\int {\bf v} \frac{\partial f}{\partial t}\, d{\bf r}d{\bf
v}=-\int {\bf v}\left ( {\bf v}\cdot {\partial f\over\partial {\bf
r}}-\nabla\Phi\cdot {\partial f\over\partial {\bf v}} \right )\,
d{\bf
r}d{\bf v}=-\int {\bf v} \left \lbrack\frac{\partial}{\partial {\bf r}}\cdot
(f{\bf
v})-\frac{\partial}{\partial {\bf v}}\cdot (f \nabla\Phi) \right \rbrack\,
d{\bf
r}d{\bf v}\nonumber\\
&=&\int  v_i \frac{\partial}{\partial
v_j}\cdot (f \partial_j\Phi)\,
d{\bf
r}d{\bf v}=-\int  \delta_{ij} f \partial_j\Phi
d{\bf
r}d{\bf v}=
-\int f \nabla\Phi \,d{\bf
r}d{\bf v}= -\int \rho \nabla\Phi \,d{\bf
r}={\bf 0}.  
\label{v1c}
\end{eqnarray}
The last equality results from the fact that the sum of the forces acting on
the system vanishes. Indeed, using Eq. (\ref{ve2}), we get
\begin{eqnarray}
-\int \rho \nabla\Phi \,d{\bf
r}=-\int d{\bf r}d{\bf r}' \rho({\bf r})\rho({\bf r}')\nabla u(|{\bf r}-{\bf
r}'|)=\int d{\bf r}d{\bf r}' \rho({\bf r})\rho({\bf r}')\nabla u(|{\bf r}-{\bf
r}'|)={\bf 0}.
\label{v1d}
\end{eqnarray}
To get the second equality, we have interchanged the dummy variables ${\bf r}$
and ${\bf r'}$, and to get the last equality we have added the half sum of the
two preceding expressions.

(iv) The conservation of the angular momentum 
can be
established
as follows: 
\begin{eqnarray}
\dot {\bf L}&=&\int ({\bf r}\times {\bf v}) \frac{\partial f}{\partial t}\,
d{\bf r}d{\bf
v}=-\int ({\bf r}\times {\bf v})\left ( {\bf v}\cdot {\partial
f\over\partial {\bf
r}}-\nabla\Phi\cdot {\partial f\over\partial {\bf v}} \right )\,
d{\bf
r}d{\bf v}=-\int ({\bf r}\times {\bf v}) \left
\lbrack\frac{\partial}{\partial {\bf r}}\cdot
(f{\bf
v})-\frac{\partial}{\partial {\bf v}}\cdot (f \nabla\Phi) \right \rbrack\,
d{\bf
r}d{\bf v}\nonumber\\
&=&-\int  \epsilon_{ijk}x_jv_k \left
\lbrack\frac{\partial}{\partial x_l}\cdot
(fv_l)-\frac{\partial}{\partial v_l}\cdot (f \partial_l\Phi) \right \rbrack\,
d{\bf
r}d{\bf v}=
\int  \epsilon_{ijk}\delta_{jl}v_k f v_l\,
d{\bf
r}d{\bf v}-\int  \epsilon_{ijk}x_j\delta_{kl} f\partial_l\Phi\,
d{\bf
r}d{\bf v}\nonumber\\
&=&\int  \epsilon_{ijk}v_k f v_j\,
d{\bf
r}d{\bf v}-\int  \epsilon_{ijk}x_j f\partial_k\Phi\,
d{\bf
r}d{\bf v}=
\int f {\bf v}\times {\bf v} \,d{\bf
r}d{\bf v}-\int f {\bf r}\times \nabla\Phi \,d{\bf
r}d{\bf v}=-\int \rho{\bf r}\times
\nabla\Phi \,d{\bf
r}={\bf 0}.  
\label{v1e}
\end{eqnarray}
The last equality results from the fact that the sum of
torques acting on
the system vanishes. Indeed, using Eq. (\ref{ve2}), we get
\begin{eqnarray}
&-&\int \rho {\bf r}\times \nabla\Phi \,d{\bf
r}=-\int d{\bf r}d{\bf r}' \rho({\bf r})\rho({\bf r}') {\bf r}\times\nabla
u(|{\bf r}-{\bf
r}'|)=\int d{\bf r}d{\bf r}' \rho({\bf r})\rho({\bf r}') {\bf
r}'\times\nabla u(|{\bf r}-{\bf
r}'|)\nonumber\\
&=&-\frac{1}{2}\int d{\bf r}d{\bf r}' \rho({\bf r})\rho({\bf
r}') ({\bf r}-{\bf
r}')\times\nabla u(|{\bf r}-{\bf
r}'|)=-\frac{1}{2}\int d{\bf r}d{\bf r}' \rho({\bf r})\rho({\bf
r}') u'(|{\bf r}-{\bf
r}'|) ({\bf r}-{\bf
r}')   \times \frac{{\bf r}-{\bf
r}'}{|{\bf r}-{\bf
r}'|}={\bf 0}.
\label{v1f}
\end{eqnarray}
To get the second equality, we have interchanged the dummy variables ${\bf r}$
and ${\bf r'}$, and to get the third equality we have added the half sum of the
two preceding expressions.

We can establish these results in a slightly different manner, by using the
properties of the Poisson brackets (see Appendix \ref{sec_b}). The Vlasov
equation can be written as
\begin{equation}
\frac{\partial f}{\partial
t}+\lbrace f,\epsilon\rbrace=0.
\label{v1g}
\end{equation}
The steady states of the Vlasov equation satisfy $\lbrace
f,\epsilon\rbrace=0$. 
The conservation of the Casimirs can be be established as
follows:
\begin{equation}
\dot I_h=\int h'(f)\frac{\partial f}{\partial t}\, d{\bf
r}d{\bf
v}=-\int h'(f)\lbrace f,\epsilon\rbrace\,
d{\bf r}d{\bf v}=-\int \epsilon\lbrace h'(f),f\rbrace\,
d{\bf r}d{\bf v}=-\int \epsilon h''(f) \lbrace f,f\rbrace\,
d{\bf r}d{\bf v}=0.
\label{cge4b}
\end{equation}
The conservation of the energy can be established as
follows:
\begin{equation}
\dot E=\int \epsilon\frac{\partial f}{\partial t}\, d{\bf r}d{\bf
v}=-\int
\epsilon\lbrace f,\epsilon\rbrace\, d{\bf
r}d{\bf v}=-\int f\lbrace \epsilon,\epsilon\rbrace\, d{\bf
r}d{\bf v}=0.
\label{cge2b}
\end{equation}

\section{Canonical treatment of the Casimir constraints}
\label{sec_cano}

In the  statistical theory of Lynden-Bell \cite{lb}, the Casimir constraints are
treated microcanonically. This is the correct approach of the problem for an
isolated system since these quantities are conserved by the Vlasov equation.
However, it makes the problem quite complicated to solve because we have to
relate a large number of Lagrange multipliers $\alpha_n$ (chemical potentials)
to the moments $M_n^{f.g.}$ of the fine-grained DF. For that reason, we may
consider a simpler problem where the Casimir constraints are treated
canonically (note that the energy and the mass are still treated
microcanonically). In that case, we assume that the
Lagrange multipliers $\alpha_n$ for $n>1$ are prescribed instead of the moments
$M_n^{f.g.}$.  

There are several justifications for treating the Casimir constraints
canonically:

1. If the system is not isolated, we may assume that forcing and
dissipation will destroy the conservation of the fine-grained moments $M_n^{\rm
f.g.}$ and fix the Lagrange multipliers  $\alpha_n$ (chemical potentials)
instead. While this is an interesting and convenient suggestion, it does not
rest on a firm solid basis. 

2. Treating the Casimir constraints canonically provides a simpler maximization
problem which determines a {\it sufficient} condition of thermodynamical
stability in the sense of Lynden-Bell (see Appendices
\ref{sec_canoa}-\ref{sec_loc}).

3. In the kinetic theory of collisionless relaxation, a canonical description of
the Casimir constraints is justified to close the hierarchy of moments equations
if the Lagrange multilpliers $\alpha_n$ do not differ too much from their
equilibrium value (see Appendix \ref{sec_out}). 

{\it Remark:} Similar results have been obtained for the Euler equation 
in 2D hydrodynamics \cite{eht,bouchet,cnd,vphydro,physicaD1,physicaD2}
 and their
adaptation to the Vlasov equation has been discussed in \cite{assisePH}.

\subsection{Sufficient condition of Lynden-Bell's thermodynamical stability}
\label{sec_canoa}

In the Lynden-Bell theory, the statistical equilibrium state is obtained
by maximizing the mixing entropy $S_{\rm LB}[\rho]$ at fixed mass $M$, energy
$E$, Casimirs $M_{n>1}^{f.g.}$, and normalization condition (see Sec.
\ref{sec_m}). This is a necessary and sufficient condition of thermodynamical
stability in the sense of Lynden-Bell. It determines the
{\it most probable} state of the system. We thus have to solve the maximization
problem
\begin{eqnarray}
\label{can1}
\max_{\rho}\quad \left\lbrace S_{\rm LB}[\rho]\quad |\quad M[\overline{f}]=M,
\quad E[\overline{f}]=E, \quad M_{n>1}^{f.g.}[\rho]=M_{n>1}^{f.g.}, \quad
\int \rho \, d\eta=1\right\rbrace.
\end{eqnarray}
The variational problem determining the extrema of $S_{\rm LB}$
at fixed $M$, $E$, $M_{n>1}^{f.g.}$ and normalization condition is
given by Eq. (\ref{E10}), leading to the Gibbs state (\ref{E11}).  This
equilibrium state is a local maximum of $S_{\rm LB}$ at fixed
$M$, $E$, $M_{n>1}^{f.g.}$ and normalization condition  if and only if 
\begin{eqnarray}
\delta^{2}J[\delta\rho]&\equiv& -\frac{1}{2}\int \frac{(\delta \rho)^{2}}{\rho}
\, d{\bf
r}d{\bf v}d\eta-\frac{\beta}{2}\int \delta\overline{f}\delta\overline{\Phi}  \,
d{\bf
r}d{\bf v}<0,\nonumber\\
&\forall& \delta\rho\ | \ \delta E=\delta M=\delta M_{n>1}^{f.g.}=\int
\delta\rho\,
d\eta=0.
\label{can2}
\end{eqnarray}

Let us now consider the maximization of the relative entropy 
\begin{equation}
S_{\chi}=S_{\rm LB}-\sum_{n>1}\alpha_n M_n^{f.g.}
\label{can3}
\end{equation}
at fixed mass $M$, energy $E$ and normalization condition. $S_{\chi}$ is the
Legendre transform of $S_{\rm LB}$ with respect to the fine-grained moments. As
compared to
the original maximization problem, this amounts to treating the Casimir
constraints canonically instead of microcanonically. We thus have to solve the
maximization problem
\begin{eqnarray}
\label{can4}
\max_{\rho}\quad \left \lbrace S_{\chi}[\rho]\quad |\quad M[\overline{f}]=M,
\quad E[\overline{f}]=E, \quad
\int \rho \, d\eta=1\right \rbrace.
\end{eqnarray}
The variational problem determining the extrema of $S_{\chi}$
at fixed $M$, $E$ and normalization is again given by Eq. (\ref{E10}),
leading to the same Gibbs state (\ref{E11}) as in the original problem. This
equilibrium state is a local maximum of $S_{\chi}$ at fixed
$M$, $E$ and normalization if and only if 
\begin{eqnarray}
\delta^{2}J[\delta\rho]&\equiv& -\frac{1}{2}\int \frac{(\delta \rho)^{2}}{\rho}
\, d{\bf
r}d{\bf v}d\eta-\frac{\beta}{2}\int \delta\overline{f}\delta\overline{\Phi}  \,
d{\bf
r}d{\bf v}<0,\nonumber\\
&\forall& \delta\rho\ | \ \delta E=\delta M=\int
\delta\rho\,
d\eta=0.
\label{can5}
\end{eqnarray}
The critical points (first variations) of (\ref{can1}) and
(\ref{can4}) are the same but the condition of stability (second
variations) is different. A maximum of $S_{\chi}$ at fixed $M$, $E$ and
normalization condition is always a maximum of $S$ at
fixed $M$, $E$, $M_{n>1}^{f.g.}$ and normalization condition, but the converse
is wrong. Indeed if inequality (\ref{can5}) is satisfied for all variations that
satisfy the conservation at first order of mass, energy and normalization
condition, it is {\it a fortiori} satisfied for all variations that satisfy the
conservation at first order of mass, energy, normalization condition {\it and}
Casimirs. Therefore (\ref{can5}) implies (\ref{can2}) but this is not
reciprocal. As a result, (\ref{can4}) provides just a {\it sufficient} condition
of thermodynamical stability (in the sense of Lynden-Bell). Making
the relative entropy explicit, we get
\begin{eqnarray}
S_{\chi}&=&-\int\rho({\bf r},{\bf v},\eta)\ln\rho({\bf r},{\bf v},\eta) \, d{\bf
r}d{\bf v}d\eta-\sum_{n>1}\alpha_n \int \rho({\bf r},{\bf v},\eta) \eta^n\,
d{\bf r}d{\bf v}d\eta\nonumber\\
&=&-\int\rho({\bf r},{\bf v},\eta) \left\lbrack \ln\rho({\bf r},{\bf v},\eta)
+\sum_{n>1}\alpha_n \eta^n\right\rbrack \, d{\bf r}d{\bf v}d\eta\nonumber\\
&=&-\int\rho({\bf r},{\bf v},\eta) \ln\left\lbrack
\frac{\rho({\bf r},{\bf v},\eta)}{\chi(\eta)}\right\rbrack \,
d{\bf r}d{\bf v}d\eta,
\label{can6}
\end{eqnarray}
where we have used Eq. (\ref{chi}) to get the last equality.

Let us finally consider the maximization of the generalized entropy
$S[\overline{f}]$ at fixed mass $M$ and energy $E$ (see Sec. \ref{sec_ge}).
We have to solve the maximization problem
\begin{eqnarray}
\label{can7}
\max_{\overline{f}}\quad \lbrace S[\overline{f}]\quad |\quad M[\overline{f}]=M,
\quad E[\overline{f}]=E\rbrace.
\end{eqnarray}
The variational problem determining the extrema of $S$
at fixed $M$ and  $E$ is
given by Eq. (\ref{meta5}), leading to the equilibrium  state (\ref{meta7})
corresponding to the Lynden-Bell coarse-grained DF. This
equilibrium state is a local maximum of $S$ at fixed
$M$ and $E$ if and only if\footnote{Using the identity from Eq. 
(\ref{meta7b}), we can check that Eq. (\ref{can8}) is equivalent to Eq.
(\ref{sand2}).}
\begin{eqnarray}
\delta^{2}J[\delta\overline{f}]&\equiv& -\frac{1}{2}\int
C''(\overline{f})(\delta \overline{f})^2 \,
d{\bf
r}d{\bf v}-\frac{\beta}{2}\int \delta\overline{f}\delta\overline{\Phi}  \,
d{\bf
r}d{\bf v}<0,\nonumber\\
&\forall& \delta\overline{f}\ | \ \delta E=\delta M=0.
\label{can8}
\end{eqnarray}
Below we show that the maximization of the relative entropy $S_{\chi}[\rho]$ at
fixed
mass $M$, energy $E$ and normalization condition is equivalent
to the maximization  of the generalized entropy  $S[\overline{f}]$ defined by
(\ref{s3w}) at fixed mass $M$ and energy $E$. As a result, (\ref{can7})
provides a {\it sufficient} condition
of thermodynamical stability (in the sense of Lynden-Bell). In summary
\begin{eqnarray}
(\ref{can7})\quad \Leftrightarrow \quad (\ref{can4})\quad  \Rightarrow \quad
(\ref{can1})
\label{can8b}
\end{eqnarray}

{\it Remark:} We may miss important solutions by maximizing
the relative entropy $S_{\chi}$ at fixed mass and energy instead of
maximizing the Lynden-Bell entropy $S_{\rm LB}$ at fixed mass, energy and
Casimirs. This is similar to the notion of ensemble
inequivalence for systems with
long-range interactions \cite{paddy,found,ijmpb,cc,campabook}. For example, for
systems with long-range interactions, equilibrium states with negative specific
heats are forbidden in the canonical ensemble (fixed $T$) while they are allowed
in the microcanonical ensemble (fixed $E$). Similarly, we may miss important
solutions by treating the Casimirs canonically instead of microcanonically.

\subsection{Equivalence for global maximization}
\label{sec_glob}

We first show the equivalence of (\ref{can4}) and (\ref{can7}) for global
maximization. To maximize $S_{\chi}[\rho]$ at fixed mass $M[\overline{f}]$,
energy
$E[\overline{f}]$ and normalization condition $\int\rho({\bf
r},{\bf v},\eta)\, d\eta=1$ we can proceed in two steps: 

(i) In a first step, we maximize $S_{\chi}[\rho]$ at fixed mass
$M[\overline{f}]$, energy $E[\overline{f}]$ and normalization
condition $\int\rho({\bf
r},{\bf v},\eta)\, d\eta=1$ for a given DF $\overline{f}({\bf r},{\bf v})$.
Since the specification of $\overline{f}({\bf r},{\bf v})$ determines
$M[\overline{f}]$ and $E[\overline{f}]$, this is equivalent to maximizing 
$S_{\chi}[\rho]$ at fixed normalization condition $\int\rho({\bf r},{\bf
v},\eta)\,
d\eta=1$ and with the constraint $\int\rho({\bf r},{\bf v},\eta)\eta\,
d\eta=\overline{f}({\bf r},{\bf v})$. Writing the variational
problem as
\begin{equation}
\delta S_{\chi}-\int\zeta({\bf
r},{\bf v})\delta\biggl (\int \rho({\bf r},{\bf v},\eta)d\eta\biggr )\, d{\bf
r}d{\bf v}-\int \Psi({\bf
r},{\bf v})\delta\biggl (\int \rho({\bf r},{\bf v},\eta)\eta d\eta\biggr )\,
d{\bf
r}d{\bf v}=0,
\label{can9}
\end{equation}
where $\zeta({\bf
r},{\bf v})$ and $\Psi({\bf
r},{\bf v})$ are Lagrange multipliers, we get
\begin{equation}
\rho_*({\bf r},{\bf v},\eta)=\frac{1}{Z[\Psi({\bf r},{\bf
v})]}\chi(\eta)e^{-\eta \Psi({\bf r},{\bf
v})}.
\label{can10}
\end{equation}
This is the global maximum of entropy with the
previous constraints since $\delta^2S_{\chi}=-\frac{1}{2}\int
[(\delta\rho)^2/\rho_*]\, d{\bf r}d{\bf v}d\eta<0$ (the constraints are linear
in $\rho$ so their second variations vanish). The functions $Z(\Psi)$ and $\Psi$
are
determined by
\begin{equation}
Z[\Psi({\bf r},{\bf
v})]=\int \chi(\eta)e^{-\eta \Psi({\bf r},{\bf
v})}\, d\eta,\qquad \overline{f}({\bf r},{\bf
v})=\frac{1}{Z[\Psi({\bf r},{\bf
v})]}\int \chi(\eta)\eta e^{-\eta \Psi({\bf r},{\bf
v})}\, d\eta
\label{can11}
\end{equation}
expressing the normalization condition and the specification of the DF
$\overline{f}({\bf r},{\bf v})$. 
These results are similar to those of Sec. \ref{sec_m} provided that we
replace
$\beta\epsilon+\alpha$ by $\Psi$. Then, we have 
\begin{equation}
\overline{f}=F(\Psi)=-(\ln Z)'(\Psi),\qquad \overline{f}'(\Psi)=-f_2(\Psi),
\label{can12}
\end{equation}
where $F$ and $f_2$ are defined in Sec. \ref{sec_m}.

We can then determine
$S[\overline{f}]\equiv S_{\chi}[\rho_*]$.
Substituting Eq. (\ref{can10}) into Eq. (\ref{can6}) we get
\begin{equation}
S[\overline{f}]=\int \overline{f}\Psi\, d{\bf
r}d{\bf v}+\int \ln Z\,  d{\bf
r}d{\bf v}.
\label{can13}
\end{equation}
This is of the form of Eq. (\ref{meta4}) with
\begin{equation}
C(\overline{f})=-\overline{f}\Psi-\ln Z.
\label{can14}
\end{equation}
Using Eq. (\ref{can12}) we find that
\begin{equation}
C'(\overline{f})=-\Psi-\overline{f}\frac{\partial
\Psi}{\partial\overline{f}}-\frac{\partial\ln Z}{\partial
\Psi}\frac{\partial \Psi}{\partial\overline{f}}=-\Psi-\overline{f}\frac{\partial
\Psi}{\partial\overline{f}}+\overline{f}\frac{\partial
\Psi}{\partial\overline{f}}=-\Psi=-[(\ln Z)']^{-1}(-\overline{f}).
\label{can15}
\end{equation}
Therefore,
\begin{equation}
C(\overline{f})=-\int^{\overline{f}}[(\ln Z)']^{-1}(-x)\, dx.
\label{can16}
\end{equation}
This returns the result from Eq. (\ref{s3w}) 
establishing the fact that $S[\overline{f}]$ is the
generalized entropy from Sec. \ref{sec_ge}. Therefore, the
generalized
entropy $S[\overline{f}]$ is equal to the relative entropy $S_{\chi}[\rho]$
calculated at $\rho_*$ when the Casimir
constraints are treated canonically (this is also true for the Lynden-Bell
entropy  $S_{\rm LB}[\rho]$ calculated at $\rho_*$ when the Casimir constraints
are treated microcanonically).

(ii) In a second step, we maximize $S[\overline{f}]\equiv S_{\chi}[\rho_*]$ at
fixed mass $M[\overline{f}]$ and energy $E[\overline{f}]$. Proceeding as in
Sec. \ref{sec_ge}, the cancellation of the first variations yields
\begin{equation}
C'(\overline{f})=-\beta\epsilon({\bf r},{\bf v})-\alpha.
\label{can17}
\end{equation}
Comparing Eqs. (\ref{can15}) and  (\ref{can17}) we find (at equilibrium) that
\begin{equation}
\Psi({\bf r},{\bf v})=\beta\epsilon({\bf r},{\bf v})+\alpha.
\label{can18}
\end{equation}
Substituting this relation into Eq. (\ref{can10}) we recover the Gibbs state
(\ref{E11}). However, we have proven more than that. The present approach shows
that $\rho({\bf r},{\bf v},\eta)$ is the global maximum of $S_{\chi}[\rho]$ 
at fixed $M$, $E$ and normalization condition
if and only if $\overline{f}({\bf r},{\bf v})$ is the global maximum of 
$S[\overline{f}]$ at fixed
$M$
and $E$ (this is where we need to treat the Casimir constraints canonically in
order to have a fixed shape of the generalized entropy).

{\it Remark:} Equation (\ref{can15})
implies
\begin{equation}
C''(\overline{f})=-\frac{1}{\overline{f}'(\Psi)}.
\label{can16b}
\end{equation}
Comparing this relation with Eq. (\ref{can12}) we obtain the important relation
\begin{equation}
f_2=\frac{1}{C''(\overline{f})}.
\label{can16c}
\end{equation}
We stress that this relation is valid even before maximizing 
$S[\overline{f}]\equiv S_{\chi}[\rho_*]$ at fixed mass $M[\overline{f}]$ and
energy $E[\overline{f}]$. In this sense, it is expected to remain valid (or
approximately valid) when the  coarse-grained DF
$\overline{f}({\bf r},{\bf v})$ is  out-of-equilibrium (see Appendix
\ref{sec_out}).

\subsection{Equivalence for local maximization}
\label{sec_loc}

We now show the
equivalence of (\ref{can4})
and (\ref{can7}) for local maximization,
i.e. $\rho({\bf r},{\bf v},\eta)$ is a (local) maximum of $S_{\chi}[\rho]$
at fixed $E$, $M$ and normalization condition if and only if the
corresponding coarse-grained DF $\overline{f}({\bf r},{\bf v})$ is
a (local) maximum of $S[\overline{f}]$ at fixed $E$ and
$M$. To that purpose,  we
show the equivalence between the stability criteria (\ref{can5}) and
(\ref{can8}).

Let us determine the perturbation $\delta\rho_{*}({\bf r},{\bf v},\eta)$
that maximizes $\delta^{2}J[\delta\rho]$ given by (\ref{can5})
with the
constraints $\delta\overline{f}=\int \delta\rho\eta\, d\eta$ and $\int
\delta\rho\, d\eta=0$, where $\delta\overline{f}({\bf r},{\bf v})$ is prescribed
(assumed to conserve energy and mass at first order). Since the
specification of  $\delta\overline{f}$ determines $\delta\overline{\Phi}$, hence
the second integral in Eq. (\ref{can5}), we can write the
variational problem under the form
\begin{eqnarray}
\label{eqa1}
\delta\left (-\frac{1}{2}\int\frac{(\delta\rho)^2}{\rho}\, d{\bf
r}d{\bf v}d\eta\right
)-\int\lambda({\bf r},{\bf v})\delta\left (\int\delta\rho\eta\, d\eta\right
)\,
d{\bf r}d{\bf v}
- \int\zeta({\bf r},{\bf v})\delta\left (\int\delta\rho\, d\eta\right )\,
d{\bf
r}d{\bf v}=0,\quad 
\end{eqnarray}
where $\lambda({\bf r},{\bf v})$ and $\zeta({\bf r},{\bf v})$ are Lagrange
multipliers. This gives
\begin{eqnarray}
\label{eqa2}
\delta\rho_*({\bf r},{\bf v},\eta)=-\rho({\bf r},{\bf v},\eta)[\lambda({\bf
r},{\bf v})\eta+\zeta({\bf r},{\bf v})],
\end{eqnarray}
which is the global maximum of $\delta^{2}J[\delta\rho]$ with the previous
constraints since $\delta^2(\delta^2 J)=-\int
\lbrace[\delta(\delta\rho)]^2/{2\rho}\rbrace\, d{\bf r}d{\bf v}d\eta< 0$ (the
constraints are linear in $\delta\rho$ so their second variations vanish). The
Lagrange multipliers are determined from the constraints
$\delta\overline{f}=\int \delta\rho\eta\, d\eta$ and $\int \delta\rho\,
d\eta=0$ yielding
$\delta\overline{f}=-\lambda\overline{f^2}-\zeta\overline{f}$ and
$0=-\lambda\overline{f}-\zeta$. Therefore, the optimal perturbation
(\ref{eqa2}) can finally be written \begin{eqnarray}
\label{eqa3}
\delta\rho_*=\frac{\delta\overline{f}}{f_2}\rho(\eta-\overline{f}).
\end{eqnarray}
Since it maximizes $\delta^{2}J[\delta\rho]$, we have
$\delta^{2}J[\delta\rho]\le \delta^{2}J[\delta\rho_*]$. Explicating
$\delta^{2}J[\delta\rho_*]$ using Eqs. (\ref{can5}) and
(\ref{eqa3}),
we obtain
\begin{eqnarray}
\label{eqa3by}
\delta^{2}J[\delta\rho]\le -\frac{1}{2}\int
\frac{(\delta\overline{f})^2}{f_2}\, d{\bf
r}d{\bf v}-\frac{1}{2}\beta\int\delta\overline{f}\delta\overline{\Phi}\, d{\bf
r}d{\bf v}.
\end{eqnarray}
Finally, using Eq. (\ref{c1}), which is rigorously valid at equilibrium, the
foregoing inequality can be rewritten as
\begin{eqnarray}
\label{eqa3b}
\delta^{2}J[\delta\rho]\le -\frac{1}{2}\int
C''(\overline{f})(\delta\overline{f})^2\, d{\bf
r}d{\bf v}-\frac{1}{2}\beta\int\delta\overline{f}\delta\overline{\Phi}\, d{\bf
r}d{\bf v}\equiv
\delta^{2}J[\delta\overline{f}],
\end{eqnarray}
where the r.h.s. is precisely the functional appearing in Eq. (\ref{can8}).
Furthermore, there is equality in Eq. (\ref{eqa3b}) if and only if
$\delta\rho=\delta\rho_*$. This proves that the stability criteria (\ref{can5})
and (\ref{can8})  are equivalent. Indeed: (i) if inequality (\ref{can8}) is
fulfilled for all perturbations $\delta\overline{f}$ that conserve
mass and energy at first order, then according to Eq. (\ref{eqa3b}), we
know that inequality (\ref{can5}) is fulfilled for all perturbations
$\delta\rho$ that conserve mass, energy, and normalization condition at first
order; (ii) if there exists a perturbation $\delta\overline{f}_c$ that
makes $\delta^{2}J[\delta\overline{f}_c]>0$, then the perturbation
$\delta\rho_c$ given by Eq. (\ref{eqa3}) with
$\delta\overline{f}=\delta\overline{f}_c$ makes
$\delta^{2}J[\delta\rho_c]=\delta^{2}J[\delta\overline{f}_c]>0$ (this is where
we need to treat the Casimir constraints canonically otherwise this
perturbation might not be allowed by the Casimir constraints). In conclusion,
the stability criteria (\ref{can5})
and (\ref{can8}) are equivalent.

{\it Remark:} We can also derive this result by using the method of
orthogonal perturbations \cite{frank} developed in
the Appendix of \cite{assisePH}.

\subsection{Out-of-equilibrium  distribution and justification of the closure
relation from Eq. (\ref{c1})}
\label{sec_out}

We can use the strategy developed above to propose a closure of the hierarchy
of moment equations (\ref{sl2}) describing the collisionless relaxation of
systems with long-range interactions.\footnote{This method was first introduced
in
Appendix
C of \cite{physicaD1} and in \cite{cnd} in the context of 2D turbulence.} The
idea is to
maximize, out-of-equilibrium, the relative entropy $S_{\chi}[\rho]$ at
fixed normalization $\int\rho({\bf r},{\bf v},\eta,t)\, d\eta=1$ and
coarse-grained DF
$\overline{f}({\bf r},{\bf v},t)=\int\rho({\bf r},{\bf v},\eta,t)\eta\,
d\eta$. This amounts  to
constructing a thermodynamical equilibrium distribution $\rho_*({\bf r},{\bf
v},\eta,t)$
corresponding to an out-of-equilibrium coarse-grained DF $\overline{f}({\bf
r},{\bf v},t)$, just like in the first step of Appendix
\ref{sec_glob}. This returns,  at
each time $t$, the equations of the first step of Appendix
\ref{sec_glob}. In
particular, one has
\begin{equation}
\rho_*({\bf r},{\bf v},\eta,t)=\frac{1}{Z[\Psi({\bf r},{\bf
v},t)]}\chi(\eta)e^{-\eta \Psi({\bf r},{\bf
v},t)},
\label{out1}
\end{equation}
where $Z[\Psi({\bf r},{\bf
v},t)]$ and $\Psi({\bf r},{\bf
v},t)$ are determined in terms of $\chi(\eta)$ and $\overline{f}({\bf r},{\bf
v},t)$ by Eq. (\ref{can11}). As a result, Eq. (\ref{can16c}) is valid at any
time
(under the previous assumption) yielding
\begin{equation}
f_2({\bf r},{\bf v},t)=\frac{1}{C''[\overline{f}({\bf r},{\bf v},t)]}.
\label{out2}
\end{equation}
As discussed in Sec.
\ref{sec_chav}, this important relation allows us to  close the hierarchy of
kinetic equations. This  leads to Eqs. (\ref{c2}) and (\ref{c3})-(\ref{c5}).
We have already indicated in Sec. \ref{sec_chav} that these equations conserve
mass and energy and satisfy an $H$-theorem for the generalized entropy
$S[\overline{f}]$. Since 
$S_{\chi}[\overline{\rho}_*]=S[\overline{f}]$, we conclude that the entropy 
$S_{\chi}(t)$ increases monotonically with time until the Gibbs state is
reached.

{\it Remark:} If we treat the Casimir constraints microcanonically, we find the
same results as above except that, at each time $t$, we have to relate
$\chi(\eta)$ to the Casimirs $M_{n>1}^{f.g.}$ and to the
coarse-grained DF $\overline{f}({\bf r},{\bf v},t)$. As a result,
$\chi_t(\eta)$ and $C_t(\overline{f})$ become functions of time. Therefore,
the shape of the generalized entropy changes with time. The kinetic
equation (\ref{c2}) remains valid except that we have to replace
$C(\overline{f})$
by $C_t(\overline{f})$.  A manner to justify treating the Casimir constraints
canonically is to assume that the function $\chi_t(\eta)$ is always close to
its equilibrium value so that it does not change substantially. Actually,
maximizing
out-of-equilibrium the relative entropy $S_{\chi}[\rho]$ at
fixed normalization  and coarse-grained DF to get Eq. (\ref{out1}) is only valid
close to equilibrium so the two assumptions are conditioned to each other.
In the canonical closure approach, we just have to solve the equilibrium
problem to get $\chi(\eta)$ and  $C(\overline{f})$ once for all. Then, Eq.
(\ref{c2})
determines the
dynamical evolution of the system for all times $t$ provided that we are
sufficiently close to equilibrium for the above assumptions to be valid.
Alternatively, in the microcanonical closure approach, we have to determine
$\chi_t(\eta)$ and $C_t(\overline{f})$ at each time in order to obtain
Eq. (\ref{c2}). Since 
$S_{\rm LB}[\overline{\rho}_*]=S_t[\overline{f}]$, we conclude that the
Lynden-Bell entropy 
$S_{\rm LB}(t)$ increases monotonically with time until the Gibbs state is
reached. This microcanonical closure approach is more precise, but it is also
much more
complicated.

\subsection{The equation for the distribution of phase levels} 
\label{sec_vd}

In the approach developed in the previous section, the coarse-grained
DF $\overline{f}({\bf v},t)$ evolves according to Eq.
(\ref{c2}) or Eqs. (\ref{c3})-(\ref{c5}). The distribution
$\rho_{*}({\bf v},\eta,t)$ is then given by Eq.
(\ref{out1}). It may be of interest to determine the relaxation equation
satisfied by $\rho_*({\bf v},\eta,t)$ explicitly. According to Eq. (\ref{out1}),
we have
\begin{equation}
\ln\rho_*=-\eta \Psi+\ln\chi(\eta)-\ln Z(\Psi), 
\label{vd1}
\end{equation}
where $\Psi({\bf v},t)$ is related to $\overline{f}({\bf v},t)$
according to Eq. (\ref{can11}). Differentiating Eq. (\ref{vd1}) with respect to
$t$  and using Eqs. (\ref{can12}) and (\ref{can16b}), we obtain
\begin{equation}
\frac{\partial\rho_*}{\partial
t}=-\rho_*(\eta-\overline{f})\frac{\partial\Psi}{\partial
t}=\rho_*(\eta-\overline
{ f } )C''(\overline { f})\frac{\partial \overline{f}}{\partial t}. 
\label{vd2}
\end{equation}
Similarly, we have
\begin{equation}
\frac{\partial\rho_*}{\partial
{\bf v}}=-\rho_*(\eta-\overline{f})\frac{\partial\Psi}{\partial
{\bf v}}=\rho_*(\eta-\overline
{ f } )C''(\overline { f})\frac{\partial \overline{f}}{\partial {\bf v}}.
\label{vd3}
\end{equation}
Combining Eq. (\ref{vd2}) with Eqs. (\ref{c2}) and (\ref{c3}), we get
\begin{equation}
\frac{\partial\rho_*}{\partial
t}=\rho_*(\eta-\overline{f})C''(\overline{
f}) \frac{\partial}{\partial v_i}  \int
d{\bf v}'\, K_{ij} \, \left
\lbrack \frac{1}{C''(\overline{f}')}\frac{\partial \overline{f}}{\partial
{v}_{j}}-\frac{1}{C''(\overline{f})}\frac{\partial
\overline{f}'}{\partial {v'}_{j}}\right \rbrack
\label{vd4}
\end{equation}
and
\begin{equation}
\frac{\partial\rho_*}{\partial
t}=\rho_*(\eta-\overline{f})C''(\overline{
f}) \frac{\partial}{\partial v_i}
\left\lbrace
D_{ij} \left \lbrack \frac{\partial \overline{f}}{\partial
{v}_{j}}+ \frac{\beta(t)}{C''(\overline{f})} {v}_j\right
\rbrack\right\rbrace.
\label{vd5}
\end{equation}
Using  Eqs. (\ref{can16c}) and (\ref{vd3}), the foregoing equations can be
rewritten as
\begin{equation}
\frac{\partial\rho_*}{\partial
t}=\frac{\rho_*(\eta-\overline{f})}{f_2} \frac{\partial}{\partial v_i}  \int
d{\bf v}'\, K_{ij} \frac{f_2}{\rho_*(\eta-\overline{f})} \, \left
\lbrack f'_2\frac{\partial \rho_*}{\partial
{v}_{j}}-\rho_*(\eta-\overline{f})\frac{\partial
\overline{f}'}{\partial {v'}_{j}}\right \rbrack
\label{vd6}
\end{equation}
and
\begin{equation}
\frac{\partial\rho_*}{\partial
t}=\frac{\rho_*(\eta-\overline{f})}{f_2}\frac{\partial}{\partial
v_i}
\left\lbrace
D_{ij} \frac{f_2}{\rho_*(\eta-\overline{f})} \left \lbrack \frac{\partial
\rho_*}{\partial
{v}_{j}}+\beta(t)\rho_*(\eta-\overline{f}){v}_j\right
\rbrack\right\rbrace.
\label{vd7}
\end{equation}
Under that form, we see some analogies (but also crucial differences)
with the SL and CSR equations (\ref{sl1}) and (\ref{csr3}).

{\it Remark:} Similar equations have been obtained in the context of 2D
turbulence \cite{cnd}. By proceeding similarly to Sec. 4.2
of \cite{cnd}, it is also possible to derive a relaxation for $\rho({\bf
r},{\bf v},t)$ associated with the maximization problem (\ref{can4}) where the
Casimir constraints are treated canonically. This equation can be used as a
numerical algorithm to solve the
maximization problem (\ref{can4}).

\subsection{Log-entropy} 
\label{sec_logo}

In the previous sections, we have treated the fine-grained moments
$M_{n>1}^{f.g.}$  canonically. If we do not take into account at all the
contribution of the fine-grained moments $M_{n>1}^{f.g.}$  in the
variational principle, the Gibbs state reduces to
\begin{equation}
\rho_*({\bf r},{\bf v},\eta)=\frac{1}{Z[\Psi({\bf r},{\bf
v})]}e^{-\eta \Psi({\bf r},{\bf
v})}.
\label{logo1}
\end{equation}
This amounts to writing $\chi(\eta)=1$ in Eq. (\ref{can10}). Using Eqs.
(\ref{can11}) and (\ref{can12}) it is easy to
establish that
 \begin{equation}
Z=\frac{1}{\Psi},\qquad \overline{f}=\frac{1}{\Psi},\qquad
f_2=\frac{1}{\Psi^2}=\overline{f}^2.
\label{logo2}
\end{equation}
We can then rewrite Eq. (\ref{logo1}) as
\begin{equation}
\rho_*({\bf r},{\bf v},\eta)=\frac{1}{\overline{f}({\bf r},{\bf
v})}e^{-\eta/\overline{f}({\bf r},{\bf
v})}.
\label{logo3}
\end{equation}
The generalized entropy associated with this distribution can be obtained from
the relation [see Eq. (\ref{out2})]
\begin{equation}
\frac{1}{C''(\overline{f})}=f_2=\overline{f}^2,
\label{logo4}
\end{equation}
leading to the functional
\begin{equation}
S=\int \ln\overline{f}\, d{\bf r}d{\bf v}.
\label{logo5}
\end{equation}
This is what we have called the log-entropy in Ref. \cite{super}. The
kinetic equation (\ref{c2}) associated with the log-entropy has been studied in
\cite{logo,kinquant}. Using Eq. (\ref{can18}), the equilibrium DF is given by 
\begin{equation}
\overline{f}=\frac{1}{\beta\epsilon+\alpha}.
\label{logo6}
\end{equation}
This is the Lorentzian DF. Note that this DF is not normalizable in $d=3$, so
there is no equilibrium state in that case.

\section{Cumulant generating function}
\label{sec_cgf}

In the multi-level case, the equilibrium distribution of the
statistical theory of Lynden-Bell is the Gibbs
state 
\begin{equation}
\label{cf1} \rho({\bf r},{\bf v},\eta) ={1\over
Z(\epsilon)}\chi(\eta)e^{-\eta(\beta\epsilon+\alpha)},
\end{equation}
where
\begin{equation}
\label{cf2}
Z(\epsilon)=\int \chi(\eta)e^{-\eta(\beta\epsilon+\alpha)}\, d\eta
\end{equation}
is the partition function. The coarse-grained DF
$\overline{f}=\int \rho\eta\, d\eta$  (first moment) is given by
\begin{eqnarray}
\overline{f}&=&\frac{1}{Z(\epsilon)}\int \eta \chi(\eta)
e^{-\eta(\beta\epsilon+\alpha)}\,
d\eta\nonumber\\
&=&-\frac{1}{\beta}\frac{1}{Z(\epsilon)}\frac{\partial}{\partial\epsilon}\int
\chi(\eta)
e^{-\eta(\beta\epsilon+\alpha)}\,
d\eta\nonumber\\
&=&-\frac{1}{\beta}\frac{1}{Z(\epsilon)}\frac{\partial Z}{\partial\epsilon}
\nonumber\\
&=&-\frac{1}{\beta}(\ln Z)'(\epsilon).
\label{cf3}
\end{eqnarray}
To the probability density (\ref{cf1}) we associate the cumulant generating
function
\begin{equation}
\kappa(\lambda,\epsilon)=\ln\overline{e^{-\lambda\beta\eta}}.
\label{cf3b}
\end{equation}
It satisfies
\begin{eqnarray}
\overline{e^{-\lambda\beta\eta}}&=&\frac{1}{Z(\epsilon)}\int
e^{-\lambda\beta\eta} \chi(\eta) e^{-\eta(\beta\epsilon+\alpha)}\,
d\eta\nonumber\\
&=&\frac{1}{Z(\epsilon)}\int \chi(\eta)
e^{-\eta[\beta(\epsilon+\lambda)+\alpha]}\,
d\eta\nonumber\\
&=&\frac{Z(\epsilon+\lambda)}{Z(\epsilon)}.
\label{cf4}
\end{eqnarray}
Therefore,
\begin{equation}
\kappa(\lambda,\epsilon)=\ln Z(\epsilon+\lambda)-\ln Z(\epsilon).
\label{cf4b}
\end{equation}
Taking the partial derivative of this expression with respect to $\lambda$, we
get
\begin{equation}
\frac{\partial\kappa(\lambda,\epsilon)}{\partial\lambda}=(\ln
Z)'(\epsilon+\lambda).
\label{cf5}
\end{equation}
Combined with Eq. (\ref{cf3}), we obtain
\begin{equation}
\frac{\partial\kappa(\lambda,\epsilon)}{\partial\lambda}=-\beta
\overline{f}(\epsilon+\lambda).
\label{cf6}
\end{equation}
Expanding both sides of Eq. (\ref{cf6}) in powers of $\lambda$, we find that the
cumulants $\kappa_n(\epsilon)=\kappa^{(n)}(0,\epsilon)$ are related to the
derivatives of $\overline{f}(\epsilon)$ by
\begin{equation}
\kappa_{n+1}(\epsilon)=-\beta
\frac{d^n \overline{f}}{d\epsilon^n}.
\label{cf7}
\end{equation}
For example,
\begin{equation}
\beta^2 (\overline{f^2}-\overline{f}^2)=-\beta \frac{d
\overline{f}}{d\epsilon},
\label{cf8}
\end{equation}
\begin{equation}
\beta^3 (\overline{f^3}-3\overline{f}\,\,\overline{f^2}+2\overline{f}^3)=\beta
\frac{d^2\overline{f}}{d\epsilon^2},
\label{cf9}
\end{equation}
\begin{equation}
\beta^4
(\overline{f^4}-3\overline{f^2}^2-4\overline{f^3}\,\,\overline{f}
+12\overline{f^2}\,\,\overline{f}^2-\overline{f}^4)=-\beta
\frac{d^3\overline{f}}{d\epsilon^3}.
\label{cf10}
\end{equation}

\section{Properties of the KP and SL equations}
\label{sec_propkpsl}

\subsection{Conservation of energy, linear impulse and Casimirs}

It is easy to show that the KP equation (\ref{lb27})
conserves the energy and the linear impulse. Indeed,
\begin{eqnarray}
\dot E&=&\int d{\bf v}\,\frac{v^2}{2}\frac{\partial \overline{f}}{\partial
t}=-\int d{\bf v}d{\bf v}'\, v_i K_{ij} \, \left
(f_2'\frac{\partial \overline{f}}{\partial {v}_{j}}-f_2\frac{\partial
\overline{f}'}{\partial {v'}_{j}}\right )=\int d{\bf v}d{\bf v}'\, v'_i K_{ij}
\, \left
(f_2'\frac{\partial \overline{f}}{\partial {v}_{j}}-f_2\frac{\partial
\overline{f}'}{\partial {v'}_{j}}\right )\nonumber\\
&=&-\frac{1}{2}\int d{\bf
v}d{\bf v}'\, w_i K_{ij}
\, \left
(f_2'\frac{\partial \overline{f}}{\partial {v}_{j}}-f_2\frac{\partial
\overline{f}'}{\partial {v'}_{j}}\right )=0,
\end{eqnarray}
where we have interchanged the dummy variables ${\bf v}$ and ${\bf v}'$ to
obtain the third equality and used the identity $K_{ij}w_j=0$ to obtain the last
equality. Similarly,
\begin{eqnarray}
\dot P_i&=&\int d{\bf v}\, v_i\frac{\partial \overline{f}}{\partial
t}=-\int d{\bf v}d{\bf v}'\, K_{ij} \, \left
(f_2'\frac{\partial \overline{f}}{\partial {v}_{j}}-f_2\frac{\partial
\overline{f}'}{\partial {v'}_{j}}\right )=\int d{\bf v}d{\bf v}'\,  K_{ij}
\, \left
(f_2'\frac{\partial \overline{f}}{\partial {v}_{j}}-f_2\frac{\partial
\overline{f}'}{\partial {v'}_{j}}\right )=0.
\end{eqnarray}
Since the KP equation (\ref{lb27}) is the first moment of the SL
equation (\ref{sl1}), the SL equation conserves the energy and the linear
impulse. The  SL equation also (trivially) conserves the hypersurface
$\gamma(\eta)=\int
\rho\, d{\bf v}$ of each level. This is equivalent to the conservation of
all the Casimirs.

\subsection{$H$-theorem}

We can also show that the  SL equation (\ref{sl1}) satisfies an $H$-theorem for
the  Lynden-Bell entropy (\ref{E9}).\footnote{This $H$-theorem was not derived
in \cite{sl}.} The SL equation (\ref{sl1}) can be rewritten in a more symmetric
form as
\begin{equation}
\frac{\partial \rho}{\partial t}=\frac{\partial}{\partial v_i}  \int d{\bf
v}'d\eta'\,
K_{ij} \, \eta' \left \lbrack \rho'(\eta'-\overline{f}')\frac{\partial
\rho}{\partial
{v}_{j}}-\rho(\eta-\overline{f})\frac{\partial \rho'}{\partial
{v'}_{j}}\right \rbrack.
\label{kpsl1}
\end{equation}
The rate of change of the Lynden-Bell entropy (\ref{E9}) is
\begin{eqnarray}
\dot S_{\rm LB}=-\int d{\bf v}d\eta\, (1+\ln\rho) \frac{\partial\rho}{\partial
t}.
\label{kpsl2}
\end{eqnarray}
Substituting Eq. (\ref{kpsl1}) into Eq. (\ref{kpsl2}), we get
\begin{eqnarray}
\dot S_{\rm LB}&=&\int d{\bf v}d{\bf v'}d\eta d\eta'\,
\frac{\eta'}{\rho}\frac{\partial\rho}{\partial
v_i}K_{ij}\left\lbrack \rho'(\eta'-\overline{f}')\frac{\partial\rho}{\partial
v_j}-\rho(\eta-\overline{f})\frac{\partial\rho'}{\partial
v'_j}\right\rbrack\nonumber\\
&=&-\int d{\bf v}d{\bf v'}d\eta d\eta'\,
\frac{\eta}{\rho'}\frac{\partial\rho'}{\partial
v'_i}K_{ij}\left\lbrack \rho'(\eta'-\overline{f}')\frac{\partial\rho}{\partial
v_j}-\rho(\eta-\overline{f})\frac{\partial\rho'}{\partial
v'_j}\right\rbrack\nonumber\\
&=&\frac{1}{2}\int d{\bf v}d{\bf v'}d\eta d\eta'\,
\frac{1}{\rho\rho'}\left (\rho'\eta'\frac{\partial\rho}{\partial
v_i}-\rho\eta\frac{\partial\rho'}{\partial
v'_i}\right )K_{ij}\left\lbrack
\rho'(\eta'-\overline{f}')\frac{\partial\rho}{\partial
v_j}-\rho(\eta-\overline{f})\frac{\partial\rho'}{\partial
v'_j}\right\rbrack.
\label{kpsl3}
\end{eqnarray}
To obtain the first line we have integrated by parts, to obtain the second line
we have interchanged the primed and unprimed variables, and to obtain the third
line we have taken the half-sum of the first and second lines. Equation
(\ref{kpsl3}) can be rewritten as
\begin{eqnarray}
\dot S_{\rm LB}=\frac{1}{2}\int d{\bf v}d{\bf v'}d\eta d\eta'\,
\frac{1}{\rho\rho'}\left\lbrack
\rho'(\eta'-\overline{f}')\frac{\partial\rho}{\partial
v_i}-\rho(\eta-\overline{f})\frac{\partial\rho'}{\partial
v'_i}\right\rbrack K_{ij}\left\lbrack
\rho'(\eta'-\overline{f}')\frac{\partial\rho}{\partial
v_j}-\rho(\eta-\overline{f})\frac{\partial\rho'}{\partial
v'_j}\right\rbrack+I,
\label{kpsl4}
\end{eqnarray}
where $I$ is the integral
\begin{eqnarray}
I=\frac{1}{2}\int d{\bf v}d{\bf v'}d\eta d\eta'\,
\frac{1}{\rho\rho'}\left \lbrack
\rho'\overline{f}'\frac{\partial\rho}{\partial
v_i}-\rho\overline{f}\frac{\partial\rho'}{\partial
v'_i}\right \rbrack K_{ij}\left\lbrack
\rho'(\eta'-\overline{f}')\frac{\partial\rho}{\partial
v_j}-\rho(\eta-\overline{f})\frac{\partial\rho'}{\partial
v'_j}\right\rbrack.
\label{kpsl5}
\end{eqnarray}
Expanding the terms in brackets, it can written as the sum of four integrals.
The first integral
\begin{eqnarray}
I_1=\frac{1}{2}\int d{\bf v}d{\bf v'}d\eta d\eta'\,
\frac{1}{\rho}\left (
\overline{f}'\frac{\partial\rho}{\partial
v_i}\right ) K_{ij}\left\lbrack
\rho'(\eta'-\overline{f}')\frac{\partial\rho}{\partial
v_j}\right\rbrack
\label{kpsl6}
\end{eqnarray}
vanishes because $\int d\eta'\,
\rho'(\eta'-\overline{f}')=\overline{f}'-\overline{f}'=0$. The second integral
\begin{eqnarray}
I_2=\frac{1}{2}\int d{\bf v}d{\bf v'}d\eta d\eta'\,\left (
\overline{f}'\frac{\partial\rho}{\partial
v_i}\right ) K_{ij}\left\lbrack
(\eta-\overline{f})\frac{\partial\rho'}{\partial
v'_j}\right\rbrack=0
\label{kpsl7}
\end{eqnarray}
vanishes because $\int d\eta'\, (\partial\rho'/{\partial
v'_j})=0$ (recall that $\int  d\eta'\, \rho'=1$).
The two other integrals $I_3$ and
$I_4$ vanish for the same reasons. As a result, we find
that $I=0$. The rate of change of the Lynden-Bell entropy (\ref{kpsl4}) can
therefore be written as
\begin{eqnarray}
\dot S_{\rm LB}=\frac{1}{2}\int d{\bf v}d{\bf v'}d\eta d\eta'\,
\frac{1}{\rho\rho'}X_i K_{ij}X_j
\label{kpsl4a}
\end{eqnarray}
with
\begin{eqnarray}
{\bf X}=\rho'(\eta'-\overline{f}')\frac{\partial\rho}{\partial
{\bf v}}-\rho(\eta-\overline{f})\frac{\partial\rho'}{\partial
{\bf v}'}.
\label{kpsl4b}
\end{eqnarray}
Since 
\begin{equation}
X_iK_{ij}X_j=\pi (2\pi)^{d}\epsilon_{r}^{d}\epsilon_{v}^{d} \int
d{\bf k}  \, ({\bf k}\cdot {\bf X})^2  \frac{\hat{u}(k)^2}{|\epsilon({\bf
k},{\bf
k}\cdot {\bf v})|^2}\delta\lbrack {\bf k}\cdot ({\bf v}-{\bf v}')\rbrack,
\label{lb27ba}
\end{equation}
we conclude that $\dot S_{\rm LB}\ge 0$ with equality if and only if ${\bf
X}$ is parallel to ${\bf v}'-{\bf v}$. Therefore, the Lynden-Bell entropy
increases
monotonically ($H$-theorem).

\subsection{Gibbs state}

Let us check that
the Gibbs state (\ref{E11}) is a stationary solution of the SL
equation (\ref{sl1}). From
Eq. (\ref{E11}) we have
\begin{equation}
\ln\rho=-\eta(\beta\epsilon+\alpha)+\ln\chi(\eta)-\ln Z.
\label{nsl1}
\end{equation}
Taking the derivative of Eq. (\ref{nsl1}) and using Eq. (\ref{E13}) we get
\begin{equation}
\frac{\partial\rho}{\partial {\bf
v}}=-\beta\rho(\eta-\overline{f}){\bf v}.
\label{nsl2}
\end{equation}
On the other hand, according to Eq. (\ref{g1}), we have
\begin{equation}
\frac{\partial\overline{f}}{\partial {\bf
v}}=\overline{f}'(\epsilon){\bf v}=-\beta f_2 {\bf v}.
\label{nsl3}
\end{equation}
Therefore, at statistical equilibrium,
\begin{equation}
f_2'\frac{\partial \rho}{\partial
{\bf v}}-\rho(\eta-\overline{f})\frac{\partial \overline{f}'}{\partial
{\bf v'}}=-f'_2\beta\rho(\eta-\overline{f}){\bf w}.
\label{nsl4}
\end{equation}
Since $K_{ij}w_j=0$, we find that the current in Eq. (\ref{sl1}) vanishes
implying that $\partial\rho/\partial t=0$.

Inversely, the condition  that ${\bf X}$ must be
parallel to ${\bf v}'-{\bf
v}$ at equilibrium  (this condition results from the $H$-theorem as shown 
above) can be written as 
\begin{equation}
\frac{1}{\eta-\overline{f}}\frac{\partial
\ln\rho}{\partial
{\bf v}}-\frac{1}{\eta'-\overline{f}'}\frac{\partial \ln\rho'}{\partial
{\bf v'}}=-A(\eta,\eta',{\bf v},{\bf v}') ({\bf v}-{\bf v}').
\label{kpsl9}
\end{equation}
From the symmetry of the left hand side of Eq. (\ref{kpsl9}) it can be shown
\cite{lenard} that $A(\eta,\eta',{\bf v},{\bf v}')$ is a constant that we shall
denote $\beta$. This then implies that
\begin{equation}
\frac{\partial
\ln\rho}{\partial
{\bf v}}+\beta (\eta-\overline{f})({\bf v}-{\bf u})=0,
\label{kpsl10}
\end{equation}
where ${\bf u}$ is another constant. At that stage, we can  repeat the argument
of
\cite{csr} (see also Appendix \ref{sec_gcsr}) to show that Eq. (\ref{kpsl10})
leads to the Gibbs
state (\ref{E11}). In conclusion, the SL equation relaxes towards the
Lynden-Bell distribution.

\section{Interpretation of the global temperature in the CSR equations}
\label{sec_pcsr}

In this Appendix, we provide a physical interpretation of the inverse
temperature $\beta(t)$ in the CSR equation which was introduced in \cite{csr}
as a Lagrange multiplier associated with the conservation of energy.

\subsection{Spatially inhomogeneous systems}

For spatially inhomogeneous systems, the CSR equations can be written as
\begin{equation}
\frac{\partial \rho}{\partial t}+{\bf
v}\cdot \frac{\partial \rho}{\partial
{\bf r}}-\nabla\overline{\Phi}\cdot \frac{\partial \rho}{\partial
{\bf v}}=\frac{\partial}{\partial v_i}
\left\lbrace
D_{ij} \left \lbrack \frac{\partial \rho}{\partial
{v}_{j}}+\beta(t) \rho(\eta-\overline{f})v_j\right
\rbrack\right\rbrace
\label{csr8b}
\end{equation}
with
\begin{equation}
D_{ij}=\int d{\bf v}'\,
K_{ij} f'_2
\label{two2w}
\end{equation}
and
\begin{equation}
\beta(t) =-\frac{\int  D_{ij} v_i \frac{\partial \overline{f}}{\partial
{v}_j}\, d{\bf r}d{\bf v}}{\int D_{ij} f_2 v_i v_j \,
d{\bf r}d{\bf v}},
\label{csr7b}
\end{equation}
where we have made a local approximation $f'_2=f_2({\bf r},{\bf v}',t)$ in Eq.
(\ref{two2w}).\footnote{See another possible expression of $D_{ij}$ in Appendix
B of \cite{csr}. More generally, we can leave $D_{ij}$
unspecified provided that the quadratic form $D_{ij}X_iX_j\ge 0$ for any ${\bf
X}$ is definite positive.} If we consider a
simplified model where $D_{ij}=D\delta_{ij}$ with $D$ constant, we obtain after
an
integration by parts
\begin{equation}
\beta(t) =\frac{d\int \overline{f}\, d{\bf r}d{\bf v}}{\int 
f_2
v^2 \, d{\bf r}d{\bf v}}.
\label{csr7d}
\end{equation}

In the two-level case, and in the nondegenerate limit,  the CSR
equations 
reduce to 
\begin{equation}
\frac{\partial \overline{f}}{\partial t}+{\bf
v}\cdot \frac{\partial \overline{f}}{\partial
{\bf r}}-\nabla\overline{\Phi}\cdot \frac{\partial \overline{f}}{\partial
{\bf v}}=\frac{\partial}{\partial v_i}
\left\lbrace
D_{ij} \left \lbrack \frac{\partial \overline{f}}{\partial
{v}_{j}}+ \beta(t)\eta_{0}\overline{f} {v}_j\right
\rbrack\right\rbrace
\label{two1b}
\end{equation}
with 
\begin{equation}
D_{ij}=\int d{\bf v}'\,
K_{ij} \eta_{0} \overline{f}' 
\label{two2b}
\end{equation}
and
\begin{equation}
\beta(t) =-\frac{\int  D_{ij} v_i \frac{\partial \overline{f}}{\partial
{v}_j}\, d{\bf r}d{\bf v}}{\int D_{ij}\eta_{0} \overline{f} v_i v_j \,
d{\bf r}d{\bf v}}.
\label{two3b}
\end{equation}
Equation (\ref{two1b}) is similar to the classical Kramers
equation, except that it involves a time-dependent temperature. If we
consider a
simplified model where $D_{ij}=D\delta_{ij}$ with $D$ constant, we obtain after
an integration by parts
\begin{equation}
\beta(t) =\frac{d\int \overline{f}\, d{\bf r}d{\bf v}}{\eta_0
\int 
\overline{f}
v^2 \, d{\bf r}d{\bf v}}=\frac{dM}{2\eta_0K(t)},
\label{csr7e}
\end{equation}
where $M=\int \overline{f}\, d{\bf r}d{\bf v}$ is the total mass and
$K(t)=\frac{1}{2}\int 
\overline{f}
v^2 \, d{\bf r}d{\bf v}$ is the total kinetic energy. Writing $\beta=1/T$, we
get
\begin{equation}
T(t) =\frac{2\eta_0K(t)}{dM} \qquad \Leftrightarrow\qquad 
K(t)=\frac{d}{2}\frac{M}{\eta_0}T(t).
\label{csr7g}
\end{equation}
This relation shows that $T(t)$ can be interpreted as a global 
kinetic temperature. It is, however, different from the spatial average value of
the
local kinetic temperature. The local kinetic temperature is defined
by
\begin{equation}
T_{\rm kin}({\bf r},t)=\frac{\eta_0\int\overline{f}[{\bf v}-{\bf u}({\bf
r},t)]^2 \,
d{\bf v}}{d\int
\overline{f}\,d{\bf
v}},
\label{csr7h}
\end{equation}
where ${\bf u}({\bf r},t)=\frac{1}{\rho}\int \overline{f}{\bf v}\, d{\bf
v}$ is the local velocity. The spatial average of the
kinetic temperature is
\begin{equation}
\langle  T_{\rm kin} \rangle(t)=\frac{\int \rho T_{\rm kin}({\bf
r},t)\, d{\bf r}}{\int \rho \, d{\bf r}}=\frac{\eta_0}{dM}\left
(\int \overline{f}v^2\, d{\bf r}d{\bf v}-\int \rho {\bf u}^2\, d{\bf r}\right ).
\label{csr7ha}
\end{equation}
We have the following relation
\begin{equation}
T(t)=\langle  T_{\rm kin} \rangle(t)+\frac{\eta_0}{dM}\int \rho {\bf u}^2\,
d{\bf r}
\label{csr7hb}
\end{equation}
between the global kinetic temperature (\ref{csr7g}) and  the
spatial average value of the local kinetic temperature (\ref{csr7ha}).

\subsection{Spatially homogeneous systems}

For spatially homogeneous systems, the energy reduces to the kinetic energy 
($K=E$) implying that the inverse temperature defined by Eq. (\ref{csr7e}) is
constant
\begin{equation}
\beta=\frac{1}{T}=\frac{dM}{2\eta_0E}.
\label{csr7j}
\end{equation}
In that case, the CSR equation (\ref{two1b}) becomes
\begin{equation}
\frac{\partial \overline{f}}{\partial t}=D\frac{\partial}{\partial {\bf
v}}\cdot \left (\frac{\partial \overline{f}}{\partial
{\bf v}}+\beta\eta_0\overline{f}{\bf v}\right ),
\label{csr8bc}
\end{equation}
which is similar to the usual Kramers (or Klein-Kramers-Chandrasekhar) equation
\cite{klein,kramersbrown,chandra1}. For the initial condition
$\overline{f}_0({\bf v})=M\delta({\bf v}-{\bf v}_0)$ it has
the analytical solution
\begin{equation}
\overline{f}({\bf v},t)=M\left\lbrack
\frac{\beta\eta_0}{2\pi(1-e^{-2D\beta\eta_0t})}\right\rbrack^{d/2}e^{-\frac{
\beta\eta_0({\bf v}-e^{-D\beta\eta_0 t}{\bf v}_0)^2}{2(1-e^{-2D\beta\eta_0
t})}}.
\label{lord1}
\end{equation}
We can check that this solution relaxes towards the Boltzmann DF
$\overline{f}({\bf
v})=M(\beta\eta_0/2\pi)^{d/2}e^{-\beta\eta_0v^2/2}$. The solution (\ref{lord1})
was first found by Lord Rayleigh \cite{lr} long before 
the seminal paper of Einstein \cite{einstein} on Brownian motion (see
\cite{sst} for more details). Taking the time derivative of $E=\frac{1}{2}\int
\overline{f} v^2\,
d{\bf v}$ and using Eq. (\ref{csr8bc}), we get\footnote{If we make the
correspondance ${\bf v}\leftrightarrow{\bf r}$, the Kramers equation
(\ref{csr8bc}) is equivalent to the Smoluchowski equation \cite{smoluchowski}
for a Brownian particle in a harmonic potential. In that case, the kinetic
energy is equivalent to the moment of inertia and Eq. (\ref{lord2}) can be
interpreted as a form of virial theorem.}
\begin{equation}
\dot E+2D\beta\eta_0E=dDm.
\label{lord2}
\end{equation}
This equation can be integrated into 
\begin{equation}
E(t)=\left (E_0-\frac{dM}{2\beta\eta_0}\right )e^{-2D\beta\eta_0
t}+\frac{dM}{2\beta\eta_0}.
\label{lord3}
\end{equation}
This result can also be directly obtained from Eq. (\ref{lord1}). The Kramers
equation (\ref{csr8bc}) satisfies an $H$-theorem for the free energy
$F=E-TS$ where $S=-\int (\overline{f}/\eta_0)\ln(\overline{f}/\eta_0)\, d{\bf
v}$ is the Boltzmann entropy. Indeed,
\begin{equation}
\dot F=-\int \frac{DT}{\eta_0\overline{f}}\left (\frac{\partial
\overline{f}}{\partial
{\bf v}}+\beta\eta_0\overline{f}{\bf v}\right )^2\le 0.
\label{lord4}
\end{equation}
Equations (\ref{lord1})-(\ref{lord4}) are valid for arbitrary values of
$\beta$. In general, the energy is not conserved since the Kramers equation is
associated with the canonical ensemble (thermal bath). However, when $\beta$ is
exactly given
by Eq. (\ref{csr7j}) it turns out that $E(t)=E_0$ is constant. In that case, Eq.
(\ref{csr8bc}) satisfies an
$H$-theorem for the  Boltzmann entropy $S$.

\section{Generalized CSR equations}
\label{sec_gcsr}

In the CSR equations \cite{csr} the energy, the linear impulse and the angular
momentum are conserved globally thanks to uniform
time-dependent Lagrange multipliers (inverse temperature $\beta(t)$, linear
velocity ${\bf U}(t)$ and angular
velocity ${\bf\Omega}(t)$). It is possible to introduce more general
relaxation equations that conserve
the energy, the linear impulse and the angular momentum locally. The equation
for $\rho({\bf r},{\bf v},\eta,t)$ 
reads\footnote{This equation can be obtained from the SL equation (\ref{sl}) by
first extending it to spatially inhomogeneous systems, making a local
approximation (see Sec. \ref{sec_3d}), then by computing the term
$\partial\overline{f}'/\partial {\bf v}'$ with the distribution 
\begin{equation}
\rho({\bf r},{\bf v}',\eta,t)=\frac{1}{Z({\bf r},{\bf
v}',t)}\chi(\eta)e^{-\eta\left\lbrack \beta({\bf r},t)\frac{({\bf v}-{\bf
u}({\bf r},t))^2}{2}+\alpha({\bf r},t)\right\rbrack},
\end{equation}
which relies  on a local thermodynamic equilibrium approximation. In that case,
the
diffusion tensor in Eq. (\ref{csr3b}) is given by Eq. (\ref{csr4}) with
$f'_2=f_2({\bf r},{\bf v}',t)$. The usual CSR equations \cite{csr}
are recovered for $\beta({\bf r},t)=\beta(t)$ and ${\bf u}({\bf r},t)={\bf
U}(t)-{\bf\Omega}(t)\times {\bf r}$.}
\begin{equation}
\frac{\partial \rho}{\partial t}+{\bf
v}\cdot \frac{\partial \rho}{\partial
{\bf r}}-\nabla\overline{\Phi}\cdot \frac{\partial \rho}{\partial
{\bf v}}=\frac{\partial}{\partial v_i}
\left\lbrace
D_{ij} \left \lbrack \frac{\partial \rho}{\partial
{v}_{j}}+\beta({\bf r},t) \rho(\eta-\overline{f})({\bf v}-{\bf u}({\bf
r},t))_j\right
\rbrack\right\rbrace.
\label{csr3b}
\end{equation}
Multiplying Eq. (\ref{csr3b}) by $\eta$ and integrating over the phase levels,
we get
\begin{equation}
\frac{\partial \overline{f}}{\partial t}+{\bf
v}\cdot \frac{\partial \overline{f}}{\partial
{\bf r}}-\nabla\overline{\Phi}\cdot \frac{\partial
\overline{f}}{\partial
{\bf v}}=\frac{\partial}{\partial v_i}
\left\lbrace
D_{ij} \left \lbrack \frac{\partial \overline{f}}{\partial
{v}_{j}}+\beta({\bf r},t) f_2 ({\bf v}-{\bf u}({\bf
r},t))_j\right
\rbrack\right\rbrace.
\label{csr5b}
\end{equation}
These relaxation equation can be written as
\begin{equation}
\frac{\partial \rho}{\partial t}+{\bf
v}\cdot \frac{\partial \rho}{\partial
{\bf r}}-\nabla\overline{\Phi}\cdot \frac{\partial \rho}{\partial
{\bf v}}=-\frac{\partial}{\partial {\bf
v}}\cdot {\bf J}\qquad {\rm and}\qquad \frac{\partial \overline{f}}{\partial
t}+{\bf
v}\cdot \frac{\partial \overline{f}}{\partial
{\bf r}}-\nabla\overline{\Phi}\cdot \frac{\partial
\overline{f}}{\partial
{\bf v}}=-\frac{\partial}{\partial {\bf
v}}\cdot {\bf J}_f,
\label{csr3c}
\end{equation}
where ${\bf J}$ is the current of the phase levels $\eta$ and ${\bf J}_f$
is the current of the coarse-grained DF given by
\begin{equation}
J_i=-
D_{ij} \left \lbrack \frac{\partial \rho}{\partial
{v}_{j}}+\beta({\bf r},t) \rho(\eta-\overline{f})({\bf v}-{\bf u}({\bf
r},t))_j\right
\rbrack\qquad {\rm and}\qquad J_f^i=-D_{ij} \left \lbrack
\frac{\partial \overline{f}}{\partial
{v}_{j}}+\beta({\bf r},t) f_2 ({\bf v}-{\bf u}({\bf
r},t))_j\right
\rbrack.
\label{csr3cb}
\end{equation}
We note that $\int {\bf J}\, d\eta={\bf 0}$ (according to the
normalization condition) and ${\bf J}_f=\int {\bf J}\eta\, d\eta$.
The local conservation of linear impulse and
energy imposes that
\begin{equation}
\int {\bf J}_f\, d{\bf v}={\bf 0}\qquad {\rm and}\qquad \int {\bf J}_f\cdot {\bf
v}\, d{\bf
v}=0.
\label{csr6b}
\end{equation}
Substituting the current ${\bf J}_f$  from Eq. (\ref{csr3cb}) into the
constraints from Eq. (\ref{csr6b}), we obtain a set of two linear equations
\begin{equation}
\int D_{ij} \frac{\partial \overline{f}}{\partial
{v}_{j}}\, d{\bf v}+\beta({\bf r},t) \int  D_{ij}f_2 ({\bf v}-{\bf u}({\bf
r},t))_j\, d{\bf v}=0,
\label{csr5d}
\end{equation}
\begin{equation}
\int D_{ij} v_i \frac{\partial \overline{f}}{\partial
{v}_{j}}\, d{\bf v}+\beta({\bf r},t) \int  D_{ij} v_i f_2 ({\bf v}-{\bf u}({\bf
r},t))_j\, d{\bf v}=0,
\label{csr5e}
\end{equation}
which determine $\beta({\bf r},t)$ and ${\bf u}({\bf
r},t)$.

The $H$-theorem can be derived as follows. First we note that the
Lynden-Bell entropy, and more generally all the functionals of $\rho$, are
conserved by the advection term of Eq. (\ref{csr3b}). The proof
is similar to the one given in Appendix \ref{sec_propv} for the Vlasov equation:
\begin{eqnarray}
\dot I_h=\int h'(\rho)\frac{\partial \rho}{\partial t}\, d{\bf r}d{\bf
v}d\eta&=&-\int
h'(\rho)\left ( {\bf v}\cdot {\partial \rho\over\partial {\bf
r}}-\nabla\Phi\cdot {\partial \rho\over\partial {\bf v}} \right )\,
d{\bf
r}d{\bf v}d\eta=-\int \left \lbrack {\bf v}\cdot {\partial h(\rho)\over\partial
{\bf
r}}-\nabla\Phi\cdot {\partial h(\rho)\over\partial {\bf v}} \right \rbrack\,
d{\bf
r}d{\bf v}d\eta\nonumber\\
&=&-\int\left \lbrace\frac{\partial}{\partial {\bf r}}\cdot [h(\rho){\bf
v}]-\frac{\partial}{\partial {\bf v}}\cdot [h(\rho) \nabla\Phi] \right \rbrace\,
d{\bf
r}d{\bf v}d\eta=0.
\label{v1ba}
\end{eqnarray}
Therefore, the change of entropy is only due to the current ${\bf J}$. It is
given by
\begin{equation}
\dot S_{\rm LB}=-\int (\ln\rho+1)\frac{\partial \rho}{\partial t}\, d{\bf
r}d{\bf v}d\eta=\int (\ln\rho+1) \frac{\partial}{\partial {\bf v}}\cdot {\bf
J} \, d{\bf r}d{\bf v}d\eta=-\int \frac{\partial \ln\rho}{\partial {\bf v}}
\cdot {\bf
J} \, d{\bf r}d{\bf v}d\eta=-\int \frac{{\bf
J}}{\rho}\cdot \frac{\partial\rho}{\partial {\bf v}} \, d{\bf r}d{\bf v}d\eta.
\label{csr5h}
\end{equation}
The last term of this equation can be rewritten as
\begin{equation}
\dot S_{\rm LB}=-\int \frac{{\bf
J}}{\rho}\cdot \left\lbrack \frac{\partial\rho}{\partial {\bf
v}}+\beta\rho(\eta-\overline{f})({\bf v}-{\bf u})\right\rbrack \, d{\bf r}d{\bf
v}d\eta+\int  \beta\rho(\eta-\overline{f})\frac{{\bf
J}}{\rho}\cdot({\bf v}-{\bf u})\,
d{\bf r}d{\bf
v}d\eta.
\label{csr5i}
\end{equation}
Integrating over $\eta$ and using the normalization condition and the local
conservation of impulse and
energy from Eq. (\ref{csr6b}) we see that the second term in Eq.
(\ref{csr5i})
vanishes:
\begin{equation}
\int  \beta(\eta-\overline{f}){\bf
J}\cdot({\bf v}-{\bf u})\,
d{\bf r}d{\bf
v}d\eta=\int  \beta {\bf
J}_f\cdot({\bf v}-{\bf u})\,
d{\bf r}d{\bf
v}=0.
\label{csr5j}
\end{equation}
As a result, there remains
\begin{equation}
\dot S_{\rm LB}=-\int \frac{{\bf
J}}{\rho}\cdot \left\lbrack \frac{\partial\rho}{\partial {\bf
v}}+\beta\rho(\eta-\overline{f})({\bf v}-{\bf u})\right\rbrack \, d{\bf r}d{\bf
v}d\eta,
\label{csr5k}
\end{equation}
which, using Eq. (\ref{csr3cb}), can be written as
\begin{equation}
\dot S_{\rm LB}=\int \frac{1}{\rho}\left\lbrack
\frac{\partial\rho}{\partial {\bf
v}}+\beta\rho(\eta-\overline{f})({\bf v}-{\bf u})\right\rbrack_i  D_{ij}
\left\lbrack \frac{\partial\rho}{\partial {\bf
v}}+\beta\rho(\eta-\overline{f})({\bf v}-{\bf u})\right\rbrack_j \, d{\bf
r}d{\bf
v}d\eta.
\label{csr5l}
\end{equation}
Assuming that the quadratic form $X_iD_{ij}X_j\ge 0$ for any ${\bf
X}$ is positive definite (we
can check that this is the case with the
expression of $D_{ij}$ from Eqs. (\ref{lb27b}), (\ref{E46}) and (\ref{csr4}))
we conclude
that $\dot
S_{\rm LB}\ge 0$. At equilibrium, the current ${\bf J}$ vanishes
leading to the Gibbs state (\ref{E11}). This can be proven as
follows.  The condition ${\bf J}={\bf 0}$
can be written as
\begin{equation}
\frac{\partial\ln\rho}{\partial {\bf
v}}+\beta(\eta-\overline{f})({\bf v}-{\bf u})={\bf 0}.
\label{ek1}
\end{equation}
Applying this relation to a reference level $\eta_0$, we get
\begin{equation}
\frac{\partial\ln\rho_0}{\partial {\bf
v}}+\beta(\eta_0-\overline{f})({\bf v}-{\bf u})={\bf 0},
\label{ek2}
\end{equation}
where $\rho_0=\rho({\bf r},{\bf v},\eta_0)$. Subtracting Eqs. (\ref{ek1}) and
(\ref{ek2}), we obtain
\begin{equation}
\frac{\partial}{\partial {\bf
v}}\ln\left (\frac{\rho}{\rho_0}\right )+\beta(\eta-\eta_0)({\bf v}-{\bf
u})={\bf 0}.
\label{ek3}
\end{equation}
This equation can be integrated into
\begin{equation}
\ln\left (\frac{\rho}{\rho_0}\right
)+\frac{1}{2}\beta(\eta-\eta_0)({\bf v}-{\bf
u})^2=A({\bf r},\eta),
\label{ek4}
\end{equation}
where $A({\bf r},\eta)$ is a constant of integration. At equilibrium, the
advection term in Eq. (\ref{csr3b}) must also vanish yielding
\begin{equation}
{\bf
v}\cdot \frac{\partial \rho}{\partial
{\bf r}}-\nabla\overline{\Phi}\cdot \frac{\partial \rho}{\partial
{\bf v}}=0.
\label{ek5}
\end{equation}
Repeating the same procedure as above, we get
\begin{equation}
{\bf
v}\cdot \frac{\partial}{\partial
{\bf r}}\ln\left (\frac{\rho}{\rho_0}\right ) -\nabla\overline{\Phi}\cdot
\frac{\partial}{\partial
{\bf v}}\ln\left (\frac{\rho}{\rho_0}\right )=0.
\label{ek6}
\end{equation}
One can show from the combination of Eqs. (\ref{ek1}) and (\ref{ek5}) that, at
equilibrium, $\beta$ must be uniform and ${\bf u}$ must vanish (this
can be viewed as a consequence of the Jeans
theorem \cite{jeansth}). Therefore, $\lim_{t\rightarrow
+\infty}\beta({\bf r},t)=\beta$ and $\lim_{t\rightarrow
+\infty}{\bf u}({\bf r},t)={\bf 0}$.\footnote{Note that in the
CSR approach \cite{csr}, at each time $t$, the inverse temperature  $\beta(t)$
is uniform and ${\bf u}={\bf
0}$. One then have $\lim_{t\rightarrow +\infty}\beta(t)=\beta$.} Then,
Eq. (\ref{ek4}) reduces to
\begin{equation}
\ln\left (\frac{\rho}{\rho_0}\right
)+\frac{1}{2}\beta(\eta-\eta_0)v^2=A({\bf r},\eta).
\label{ek7}
\end{equation}
Taking its gradient with respect to ${\bf r}$, we get
\begin{equation}
\frac{\partial}{\partial {\bf r}}\ln\left (\frac{\rho}{\rho_0}\right
)=\nabla A({\bf r},\eta).
\label{ek8}
\end{equation}
Substituting Eq. (\ref{ek3})  and 
Eq. (\ref{ek8}) into Eq. (\ref{ek6}) we get ${\bf v}\cdot[\nabla
A+\beta(\eta-\eta_0)\nabla\Phi]=0$.
This equality must be true for all ${\bf v}$, implying that $\nabla
A+\beta(\eta-\eta_0)\nabla\Phi={\bf 0}$, which can be integrated into
$A({\bf r},\eta)=-\beta(\eta-\eta_0)\Phi({\bf r})-B(\eta)$,
where $B(\eta)$ is a constant of integration. Finally, Eq. (\ref{ek7}) can be
rewritten as 
\begin{equation}
\ln\left (\frac{\rho}{\rho_0}\right
)=-\beta(\eta-\eta_0)\epsilon-B(\eta),
\label{ek11}
\end{equation}
which is equivalent to the Gibbs state (\ref{E11}) with
$1/Z({\bf r},{\bf v})=\rho({\bf r},{\bf v},\eta_0) e^{\beta\eta_0\epsilon({\bf
r},{\bf v})}$ and $\chi(\eta)e^{-\eta\alpha}=e^{-B(\eta)}$. Inversely,
starting from the Gibbs state (\ref{E11}) and using Eqs. (\ref{nsl1}) and
(\ref{nsl2}), we get ${\bf J}={\bf 0}$.

{\it Remark:} If we assume that $D_{ij}=D\delta_{ij}$ with $D$ constant, the
linear equations
(\ref{csr5d}) and (\ref{csr5e}) reduce to
\begin{equation}
{\bf u}({\bf r},t)=\frac{\int f_2 {\bf v}\, d{\bf v}}{\int f_2 \, d{\bf
v}},\qquad T({\bf r},t)=\frac{\int f_2 ({\bf v}-{\bf
u}({\bf
r},t))^2\, d{\bf
v}}{d\int \overline{f}\, d{\bf v}}.
\label{csr5o}
\end{equation}
In the two-level case and in the nondegenerate limit, we obtain
\begin{equation}
\frac{\partial \overline{f}}{\partial t}+{\bf
v}\cdot \frac{\partial \overline{f}}{\partial
{\bf r}}-\nabla\overline{\Phi}\cdot \frac{\partial
\overline{f}}{\partial
{\bf v}}=D\frac{\partial}{\partial {\bf v}}\cdot
\left \lbrack \frac{\partial \overline{f}}{\partial
{\bf v}}+\beta({\bf r},t) \overline{f}\eta_0 ({\bf v}-{\bf u}({\bf
r},t))\right
\rbrack
\label{ek12}
\end{equation}
with
\begin{equation}
{\bf u}({\bf r},t)=\frac{\int \overline{f} {\bf v}\, d{\bf v}}{\int
\overline{f} \, d{\bf
v}},\qquad T({\bf r},t)=\frac{\int \overline{f} ({\bf v}-{\bf
u}({\bf
r},t))^2\, d{\bf
v}}{d\int \overline{f}\, d{\bf v}}.
\label{ek13}
\end{equation}
In that case, we recover the usual expressions of the local velocity and local
kinetic temperature. Equations (\ref{ek12}) and (\ref{ek13}) are similar to the
kinetic equations introduced by Dougherty \cite{dougherty} for collisional
systems.

\section{Another equation that conserves the mass and the energy and that
monotonically increases all
the
$H$-functions}
\label{sec_allh}

In Sec. \ref{sec_cge} we have introduced an equation that conserves the mass and
the energy and that monotonically increases all the $H$-functions. In this
Appendix, we introduce another equation that satisfies the same properties.

\subsection{Anisotropic diffusion equation}

The CSR equation is given by Eq.
(\ref{csr3}) with Eq. (\ref{csr7}). If we assume for simplicity that the
diffusion tensor is isotropic and constant, so that $D_{ij}=D\delta_{ij}$, this
equation reduces to
\begin{equation}
\frac{\partial \rho}{\partial t}=D\frac{\partial}{\partial {\bf v}}\cdot
\left \lbrack \frac{\partial \rho}{\partial
{\bf v}}+\beta(t) \rho(\eta-\overline{f}){\bf v}\right
\rbrack
\label{hg1}
\end{equation}
with
\begin{equation}
\beta(t) =-\frac{\int  {\bf v}\cdot \frac{\partial \overline{f}}{\partial
{\bf v}}\, d{\bf v}}{\int  f_2 v^2 \,
d{\bf v}}.
\label{h4}
\end{equation}
If we get rid of the integrals in Eq. (\ref{h4}), we get
\begin{equation}
\beta({\bf v},t) =-\frac{{\bf v}\cdot \frac{\partial \overline{f}}{\partial
{\bf v}}}{f_2 v^2}.
\label{h5}
\end{equation}
Substituting this relation into Eq. (\ref{hg1}) we obtain
\begin{equation}
\frac{\partial \rho}{\partial t}=D\frac{\partial}{\partial {\bf v}}\cdot
\left \lbrack \frac{\partial \rho}{\partial
{\bf v}}-\frac{{\bf v}\cdot \frac{\partial \overline{f}}{\partial
{\bf v}}}{f_2 v^2}\rho(\eta-\overline{f}){\bf v}\right
\rbrack.
\label{hg1b}
\end{equation}
This equation can also be obtained by applying the MEPP with a local
conservation of energy ${\bf J}_{f}\cdot {\bf v}=0$ in velocity
space.\footnote{An equation similar to Eq. (\ref{hg1b}),
but acting in
position space instead of velocity space, has been obtained in the
context of 2D turbulence in Ref. \cite{chavjapon}.} It
conserves
the normalization condition, the Casimirs (or the total hypervolume of each
phase level $\eta$) and the energy. It also increases the mixing entropy
(\ref{E9}) monotonically ($H$-theorem). The proof is
essentially the same as for the CSR equations (see Ref. \cite{csr} and
Appendix \ref{sec_gcsr}). However,
it does {\it not} relax towards the Gibbs state (\ref{E11}).\footnote{It is not
clear if this property is a drawback of this equation or if it can account for
situations of incomplete relaxation where the quasistationary state is different
from the Lynden-Bell
statistical equilibrium state.} To see that, let us
consider the equation for the coarse-grained DF (\ref{csr5}) which, for an
isotropic and constant diffusion tensor, can be written as
\begin{equation}
\frac{\partial \overline{f}}{\partial t}=D\frac{\partial}{\partial {\bf
v}}\cdot\left \lbrack \frac{\partial \overline{f}}{\partial
{\bf v}}+\beta(t) f_2 {\bf v}\right
\rbrack.
\label{h3}
\end{equation}
Replacing $\beta(t)$ by Eq. (\ref{h5}) we obtain
\begin{equation}
\frac{\partial \overline{f}}{\partial t}=D\frac{\partial}{\partial {\bf
v}}\cdot \left ( \frac{\partial \overline{f}}{\partial
{\bf v}}-\frac{{\bf v}\cdot \frac{\partial \overline{f}}{\partial
{\bf v}}}{v^2}{\bf v}\right ).
\label{h6}
\end{equation}
This equation can also be obtained by multiplying Eq. (\ref{hg1b}) by $\eta$
and
integrating over $\eta$. We note that, unlike Eq. (\ref{h3}), this equation is
closed since the second
moment $f_2$ has cancelled
out. Equation (\ref{h6}) can be rewritten as
\begin{equation}
\frac{\partial \overline{f}}{\partial t}=D\frac{\partial}{\partial {v}_i}
\left \lbrack \left (\delta_{ij}-\frac{v_iv_j}{v^2}\right )\frac{\partial
\overline{f}}{\partial
{v}_j}\right
\rbrack.
\label{h7}
\end{equation}
This is an anisotropic diffusion equation of the form
\begin{eqnarray}
\frac{\partial \overline{f}}{\partial t}=\frac{\partial}{\partial
v_i}\left
(D_{ij}\frac{\partial \overline{f}}{\partial v_j}\right )
\label{h8}
\end{eqnarray}
with a diffusion tensor 
\begin{eqnarray}
D_{ij}=D\frac{v^2\delta_{ij}-v_iv_j}{v^2}.
\label{h9}
\end{eqnarray}
The diffusion tensor $D_{ij}$ has
the property that
$D_{ij}v_j=0$. As
a result, all isotropic DFs are stationary solutions of
Eq. (\ref{h6}). Indeed, for a DF of the form
$\overline{f}=\overline{f}(v)$ with $v=|{\bf v}|$, we
have
$\partial \overline{f}/\partial
v_j=\overline{f}'(v)v_j/v$. Since $D_{ij}v_j=0$, we obtain $D_{ij}\partial
\overline{f}/\partial
v_j=0$, hence $\partial \overline{f}/\partial t=0$. When the initial
DF $\overline{f}_0({\bf v})$ is anisotropic, the system evolves until
$\overline{f}({\bf v},t)$ becomes isotropic. Therefore, the effect of the
diffusion
equation
(\ref{h6}) is to ``isotropize'' an initially anisotropic
DF.\footnote{For simplicity, we have considered spatially 
homogeneous systems. However, Eq. (\ref{h6}) remains valid for
spatially inhomogeneous systems provided that we introduce an advection term in
the left hand side. In that case, it relaxes towards an isotropic 
DF of the form $f(\epsilon)$ where $\epsilon=v^2/2+\Phi({\bf r})$ which
cancels both the advection term and the ``collision'' term.}

{\it Remark:} For 1D systems, like the HMF model, Eq. (\ref{h6}) reduces to 
\begin{equation}
\frac{\partial \overline{f}}{\partial t}=0
\label{h10}
\end{equation}
so there is no evolution in that case.

\subsection{Properties of Eq. (\ref{h6})}

Let us write Eq. (\ref{h6}) under the conservative form
\begin{eqnarray}
\frac{\partial \overline{f}}{\partial t}=-\frac{\partial}{\partial
{\bf v}}\cdot {\bf J}_f
\label{h11}
\end{eqnarray}
with the diffusion current
\begin{eqnarray}
{\bf J}_f=- D_{ij}\frac{\partial \overline{f}}{\partial v_j}=-D\left\lbrack
\frac{\partial \overline{f}}{\partial {\bf v}}-\left ({\bf v}\cdot
\frac{\partial
\overline{f}}{\partial {\bf v}}\right )\frac{\bf
v}{v^2}\right\rbrack=-D\left (
\frac{\partial \overline{f}}{\partial {\bf v}}-\frac{\partial
\overline{f}}{\partial v}\frac{\bf v}{v}\right ).
\label{h12}
\end{eqnarray}
We note that the diffusion current is normal to the
velocity:
\begin{eqnarray}
{\bf J}_f\cdot {\bf v}=0.
\label{h12b}
\end{eqnarray}
As a result, Eq. (\ref{h6}) trivially conserves the energy (\ref{E5}). Indeed
\begin{eqnarray}
\dot E=\int \frac{v^2}{2}\frac{\partial \overline{f}}{\partial t}\, d{\bf
v}=-\int \frac{v^2}{2}\frac{\partial}{\partial
{\bf v}}\cdot {\bf J}_f\, d{\bf v}=\int
{\bf J}_f\cdot {\bf v}\, d{\bf v}=0.
\label{h13}
\end{eqnarray}
We can also show that Eq. (\ref{h6})  monotonically increases all the
$H$-functions (\ref{i1}). We have
\begin{eqnarray}
\dot H=-\int C'(\overline{f})\frac{\partial \overline{f}}{\partial t}\, d{\bf
v}=\int
C'(\overline{f})\frac{\partial}{\partial
{\bf v}}\cdot {\bf J}_f\, d{\bf v}=-\int
C''(\overline{f}) {\bf J}_f\cdot \frac{\partial \overline{f}}{\partial {\bf
v}}\,
d{\bf v}.
\label{h14}
\end{eqnarray}
Using Eq. (\ref{h12}), the last equality of this equation can be rewritten as
\begin{eqnarray}
\dot H=\int
C''(\overline{f})\frac{1}{D} {\bf J}_f\cdot \left\lbrack {\bf J}_f-D \left ({\bf
v}\cdot \frac{\partial
\overline{f}}{\partial {\bf v}}\right )\frac{\bf
v}{v^2}\right\rbrack\,
d{\bf v}.
\label{h15}
\end{eqnarray}
Using Eq. (\ref{h12b}), we get
\begin{eqnarray}
\dot H=\int
C''(\overline{f}) \frac{{\bf J}_f^2}{D}\, d{\bf v}\ge 0.
\label{h16}
\end{eqnarray}
Therefore, all the $H$-functions increase monotonically. At equilibrium, we have
${\bf J}_f={\bf 0}$. This determines an isotropic DF of the form
$\overline{f}=\overline{f}(v)$.

{\it Remark:} We note that Eqs. (\ref{cge1}) and (\ref{h6}) share
similar general properties (conservation of energy and
monotonic increase of all the $H$-functions). However, these two equations are
very different. In particular, Eq. (\ref{cge1}) reduces to
$\partial_t\overline{f}=0$ for spatially homogeneous systems contrary to Eq.
(\ref{h6}), and Eq. (\ref{h6}) reduces to $\partial_t\overline{f}=0$ for 1D
systems contrary to Eq. (\ref{cge1}).

\subsection{Analytical solution of Eq. (\ref{h6})}

It turns out that Eq. (\ref{h6}) can be solved analytically. Taking the
divergence
of
the current from Eq. (\ref{h12}), we can rewrite Eq. (\ref{h6}) as
\begin{eqnarray}
\frac{\partial \overline{f}}{\partial t}=D\left (\Delta_{{\bf
v}} \overline{f}-\Delta_v \overline{f}\right )=\frac{D}{v^2}\Delta_S
\overline{f},
\label{h19}
\end{eqnarray}
where $\Delta_{{\bf v}}$ is the Laplacian operator in velocity space,
$\Delta_v=\frac{\partial^2}{\partial v^2}+\frac{d-1}{v}\frac{\partial}{\partial
v}$ is the part of the Laplacian operator which involves derivatives with
respect to the modulus of ${\bf v}$, and $\Delta_S$ is the part of the
Laplacian
operator which involves derivatives with respect to the orientation of the
vector ${\bf v}$ (on the unit sphere).

In $d=2$, introducing a polar system of coordinates, we have
\begin{eqnarray}
\Delta_S=\frac{\partial^2}{\partial\theta^2}.
\label{h20}
\end{eqnarray}
The solution of Eq. (\ref{h19}) is then
\begin{eqnarray}
\overline{f}(v,\theta,t)=\sum_{n=-\infty}^{+\infty} c_n(v)e^{i n\theta} e^{-D
n^2 t/v^2}
\label{h21}
\end{eqnarray}
with
\begin{eqnarray}
c_n(v)=\frac{1}{2\pi}\int_0^{2\pi}\overline{f}_0(v,\theta)e^{-i n \theta}\,
d\theta.
\label{h22}
\end{eqnarray}
For $t\rightarrow +\infty$, we get
\begin{eqnarray}
\overline{f}(v,\theta,t)\rightarrow
c_0(v)=\frac{1}{2\pi}\int_0^{2\pi}\overline{f}_0(v,\theta)\,d\theta=\langle
\overline{f}_0(v,\theta)
\rangle_{\theta}.
\label{h23}
\end{eqnarray}
Therefore, $\overline{f}({\bf v},t)$  tends to an isotropic DF which
is equal
to the average over the angle $\theta$ of the initial DF
$\overline{f}_0({\bf v})=f_0(v,\theta)$.

In $d=3$,  introducing a spherical system of coordinates, we have
\begin{eqnarray}
\Delta_S=\frac{1}{\sin^2\theta}\frac{\partial^2}{\partial\phi^2}+\frac{1}{
\sin\theta}\frac{\partial}{\partial\theta}\left
(\sin\theta\frac{\partial}{\partial \theta}\right ).
\label{h24}
\end{eqnarray}
In that case, the solution of Eq. (\ref{h19}) is
\begin{eqnarray}
\overline{f}(v,\theta,\phi,t)=\sum_{l=0}^{+\infty}\sum_{m=-l}^{+l} c_{l m}(v)
Y_{l
m}(\theta,\phi)  e^{-D l(l+1) t/v^2}
\label{h25}
\end{eqnarray}
with
\begin{eqnarray}
c_{lm}(v)=\int_0^{2\pi} d\phi \int_0^{\pi} d\theta\, \sin\theta
\overline{f}_0(v,\theta,\phi)Y_{lm}^*(\theta,\phi),
\label{h26}
\end{eqnarray}
where $Y_{lm}(\theta,\phi)$ are the spherical harmonics.
For $t\rightarrow +\infty$, we get
\begin{eqnarray}
\overline{f}(v,\theta,\phi,t)\rightarrow
c_{00}(v)Y_{00}(\theta,\phi)=\frac{1}{4\pi}\int_0^{2\pi} d\phi \int_0^{\pi}
d\theta\, \sin\theta
\overline{f}_0(v,\theta,\phi)=\langle
\overline{f}_0(v,\theta,\phi)
\rangle_{\theta,\phi}.
\label{h27}
\end{eqnarray}
Therefore, $\overline{f}({\bf v},t)$  tends to an isotropic DF which
is equal
to the average over the angles $\theta$ and $\phi$ of the initial DF
$\overline{f}_0({\bf v})=f_0(\theta,\phi,v)$.

\subsection{Another type of equation}

If we integrate Eq. (\ref{h4}) by parts, we find that
\begin{equation}
\beta(t) =\frac{d\int \overline{f}\, d{\bf v}}{\int 
f_2 v^2 \,
d{\bf v}}.
\label{h28}
\end{equation}
If we get rid of the integrals, we get
\begin{equation}
\beta({\bf v},t) =\frac{d \overline{f}}{f_2 v^2}.
\label{h29}
\end{equation}
Substituting this relation into Eq. (\ref{h3}) we obtain
\begin{equation}
\frac{\partial \overline{f}}{\partial t}=D\frac{\partial}{\partial {\bf
v}}\cdot \left ( \frac{\partial \overline{f}}{\partial
{\bf v}}+d\overline{f}\frac{\bf v}{v^2}\right ).
\label{h30}
\end{equation}
Again, this is a closed equation. Equation (\ref{h30}) conserves the energy
(\ref{E5}). Indeed:
\begin{equation}
\dot E=\int \frac{v^2}{2}\frac{\partial \overline{f}}{\partial
t}\, d{\bf v}=\int \frac{v^2}{2}D\frac{\partial}{\partial {\bf
v}}\cdot \left ( \frac{\partial \overline{f}}{\partial
{\bf v}}+d\overline{f}\frac{\bf v}{v^2}\right )\, d{\bf v}=-\int
D \left ( \frac{\partial \overline{f}}{\partial
{\bf v}}+d\overline{f}\frac{\bf v}{v^2}\right )\cdot {\bf v}\, d{\bf v}=-\int
D \left ( \frac{\partial \overline{f}}{\partial
{\bf v}}\cdot {\bf v}+d\overline{f}\right )\, d{\bf v}=0,
\label{h31}
\end{equation}
where the last equality is obtained after performing an integration by parts.
By contrast, nothing general can be said about the sign of $\dot H$.

{\it Remark:} Equation (\ref{h31}) can be interpreted as a Smoluchowski equation
describing the evolution of a Brownian particle coupled to a thermal bath of
unit temperature $T_c=1$ and submitted to an attractive logarithmic potential
$U({\bf v})=d\ln |{\bf v}|$,
where ${\bf v}$ plays the role of the position ${\bf r}$. The stationary
solution $\overline{f}_e=Ae^{-U({\bf v})}=A/v^d$ is not normalizable
(the normalization factor diverges logarithmically at both small and large
velocities). From the general theory of Fokker-Planck equations, we know that
Eq. (\ref{h31}) satisfies an
$H$-theorem ($\dot F\le 0$) for the free energy $F=\int \overline{f}U({\bf
v})\, d{\bf v}+T_c\int \overline{f}\ln \overline{f}\, d{\bf v}$. At the
critical temperature $T_c=1$, Eq. (\ref{h30}) has the particularity to conserve
the energy $E=(1/2)\int \overline{f}v^2\, d{\bf v}$ which is analogous to the
moment of inertia if we
make the correspondance ${\bf r}\leftrightarrow {\bf v}$ (see above).
In this respect, Eq. (\ref{h31}) can be interpreted as a form of virial
theorem. The study of Eq. (\ref{h30}) at $T\le T_c$ is
subtle because it displays a
form of ``collapse'' or a form of Bose-Einstein condensation in the state ${\bf
v}={\bf 0}$ leading to a Dirac peak $\delta({\bf v})$. This is an example of
Bessel process that has been studied in, e.g., Ref. \cite{neq2}.

\eject

\end{document}